\documentclass[aps,pre,twocolumn,amsmath,amssymb,superscriptaddress,reprint,longbibliography]{revtex4-1}
\usepackage{amsfonts,amsmath,amssymb,bm}
\usepackage{graphicx,graphics,float}
\usepackage{bm}
\usepackage{amssymb}
\usepackage{colordvi}
\usepackage{graphicx}
\usepackage{color}

\usepackage[colorlinks=true,linkcolor=blue,citecolor=blue,urlcolor=blue]{hyperref}%
\usepackage{hyperref}
\usepackage{comment}
\usepackage{harpoon}

\begin{document}

\title{Energy cooperation in quantum thermoelectric systems with multiple  electric currents}

\author{Yefeng Liu}
\affiliation{School of physical science and technology \&
Collaborative Innovation Center of Suzhou Nano Science and Technology, Soochow University, Suzhou 215006, China.}

\author{Jincheng Lu}\email{jincheng.lu1993@gmail.com}
\affiliation{School of physical science and technology \&
Collaborative Innovation Center of Suzhou Nano Science and Technology, Soochow University, Suzhou 215006, China.}

\author{Rongqian Wang}
\affiliation{School of physical science and technology \&
Collaborative Innovation Center of Suzhou Nano Science and Technology, Soochow University, Suzhou 215006, China.}

\author{Chen Wang}
\address{Department of Physics, Zhejiang Normal University, Jinhua, Zhejiang 321004, China}

\author{Jian-Hua Jiang}\email{jianhuajiang@suda.edu.cn}
\affiliation{School of physical science and technology \&
Collaborative Innovation Center of Suzhou Nano Science and Technology, Soochow University, Suzhou 215006, China.}

\date{\today}

\begin{abstract}
The energy efficiency and power of a quantum thermoelectric system with multiple electric currents and only one heat currents are studied. The system is connected to the hot heat bath with one terminal but the cold bath with multiple terminals or vice versal. We find that the cooperative effects can be a potentially useful tool in improving the energy efficiency and output power in multi-terminal mesoscopic thermoelectric systems. As an example, we show that the cooperation between the two thermoelectric effects in three-terminal thermoelectric systems leads to markedly improved performance of heat engine within the linear response regime using the Landauer-B\"{u}tiker formalism. Such improvement also emerge in four-terminal thermoelectric heat engines with three output electric currents. Cooperative effects in these multi-terminal thermoelectric systems can significantly enlarge the physical parameter region with high efficiency and power. For refrigeration, we find that the energy efficiency can also be substantially improved if multi-terminal configurations are considered, suggesting a useful scheme toward electronic cooling. Our study illustrates cooperative effects as a convenient approach toward high-performance thermoelectric energy conversion in multi-terminal mesoscopic systems.
\end{abstract}

\maketitle

\section{Introduction}
Thermoelectric phenomena at nanoscales have been attracting broad research interest for their relevance to fundamental physics, material science and renewable energy~\cite{DubiRMP,RenRMP,Nanotechnology,JiangCRP,SanchezCPR,PhyRep}. Enhancing the efficiency of thermoelectric materials is one of the main challenges of great necessity for various technological applications. Understanding and harnessing thermoelectric transport at the nanoscales may lead to important applications, including, for instance, heat and energy detectors~\cite{David2011PRB,Sothmann-QW,Jiang2013,JiangNJP}, heat rectifiers~\cite{Jiangtransistors,MyPRB,SanchezAPL,WangPRE} and refrigerators~\cite{Rongqian,David-refrigerator} and energy transduction~\cite{Lena2012,JiangPRX}. Lots of efforts have been devoted to the optimization of mesoscopic heat engines and refrigerations in both theories~\cite{OraPRB2010,Rafael,Jiang2012,WhitneyPRL,JiangOra,Mazza-separation,SanchezPRL,Ora2015,BijayJiang,Yamamoto,Naoto,JiangBijayPRB17,Brandner2018,JiangNearfield,brandner2019,wangPRApplied} and experiments~\cite{hwang,Exper,Thier2015,cui2018,hartmanNP,josefsson2018,jaliel-exper,Josefsson,prete} for the reason that they may reach the ultimate efficiency and power that limited by fundamental laws of physics. Such systems may also give insight in the search of high-performance macroscopic thermoelectric materials and systems which, however, is yet to be realized.

The performance of thermoelectric materials is characterized by a single dimensionless parameter called the figure of merit, $ZT$, which is a combination of the main transport properties of a material, $ZT=\sigma S^2/(T\kappa)$~\cite{Mahan}, where $\sigma$ is electric conductivity, $\kappa$ is the thermal conductivity, and $S$ is the Seebeck coefficient, $T$ is the temperature. In the linear transport regime, the maximum energy efficiency is given by
\begin{equation}
\eta_{\max}=\eta_C\frac{\sqrt{ZT+1}-1}{\sqrt{ZT+1}+1}.
\label{eq:eta_ZT}
\end{equation}
Here, $\eta_C$ is the Carnot efficiency which is only for ideal machines operating in the reversible limit, i.e., when $ZT\rightarrow\infty$. In the linear transport regime, the dependences of the figure of merit show that the efficiency of a thermoelectric heat engine is influenced by the electronic and vibrational properties of a material.

The conventional approaches toward high-performance thermoelectrics is to tune the electronic and vibrational properties in various ways to increase the $ZT$ and the power density (characterized by the power factor $PF\equiv \sigma S^2$). However, in most situations the electric conductivity, the Seebeck coefficient, and the thermal conductivity are correlated and difficult to optimize together. Apart from the insight based on Eq.~\eqref{eq:eta_ZT}, recently it was proposed that in thermoelectric systems with multiple terminals, the energy efficiency and power can be improved through the cooperative effects. The existence of multiple thermoelectric effects in those systems enable a cooperation between various thermoelectric transport channels, which gives rise to enhancement of the thermoelectric performance. This effect was proposed in Ref.~[\onlinecite{JiangJAP}] where two coexisting thermoelectric effects induced by the inelastic hopping transport can have cooperative effects, leading to enhanced energy efficiency and output power in the linear-response regime. Generalized thermoelectric figure of merit and power factor for three-terminal systems with two heat currents and one electric current were introduced as the theoretical framework for the cooperative effect~\cite{JiangJAP}. This study was later generalized to three-terminal thermoelectric systems with elastic transport between three electrodes~\cite{MyJAP}. However, in these studies as well as in others, the use of multiple heat baths with three different temperatures are unfavorable for applications.

In this work, we show that the cooperative effects can play an important role in improving thermoelectric efficiency and power even in the set-ups with only two heat baths. Two prototypes are illustrated in Fig.~\ref{fig1} where multiple electrodes are mounted on two heat baths, the hot bath with temperature $T_h$ and the cold bath with temperature $T_c$. Due to the multi-terminal geometry, there are more than one thermoelectric effects, described by the multiple Seebeck coefficients for the thermoelectric transport between various pairs of terminals with different temperatures. By exploiting the cooperative effects, the thermoelectric performance can be improved, as compared with the situations where only one of them is used for energy conversion. Moreover, since such cooperative effects exist in a wide parameter region, the physical parameter space with high thermoelectric performance can be substantially increased. Thus, it becomes less demanding to achieve high-performance thermoelectrics in multi-terminal systems, when the cooperative effect is exploited. We use three-terminal heat engines and refrigerators as well as four-terminal heat engines to demonstrate such effects.

The main part of the paper is organized as follows. In Sec.~\ref{sec:3terminals}, we introduce the three-terminal thermoelectric system and its transport properties, give the expressions for the maximum efficiency and maximum power in the linear-response regime. In Sec.~\ref{Co_eff}, we show that the cooperative effect can enhance the efficiency and power in the linear-response regime. In Sec.~\ref{ref-3T} we study cooperative effects in the refrigeration set-up. In Sec.~\ref{4T}, we compare the optimal efficiency and power of the four-terminal thermoelectric heat engine with that of the three-terminal heat engine. We conclude our study in Sec.~\ref{conclusion} along with our future outlook.

\section{Linear-response thermoelectric transport in a quantum three-terminal heat engine}\label{sec:3terminals}

\begin{figure}
\begin{center}
\centering \includegraphics[width=8.7cm]{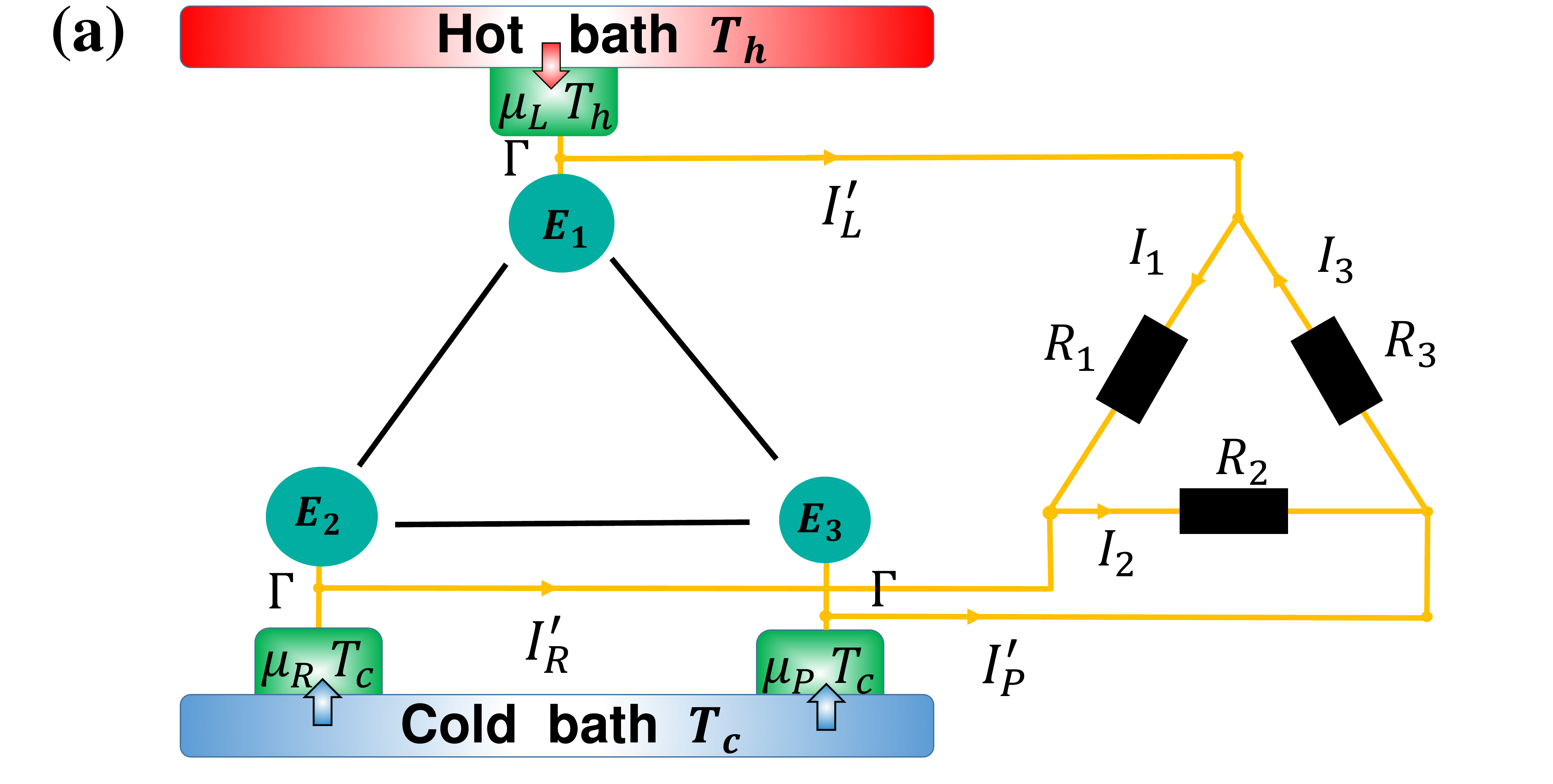}\hspace{2.2cm}\includegraphics[width=8.2cm]{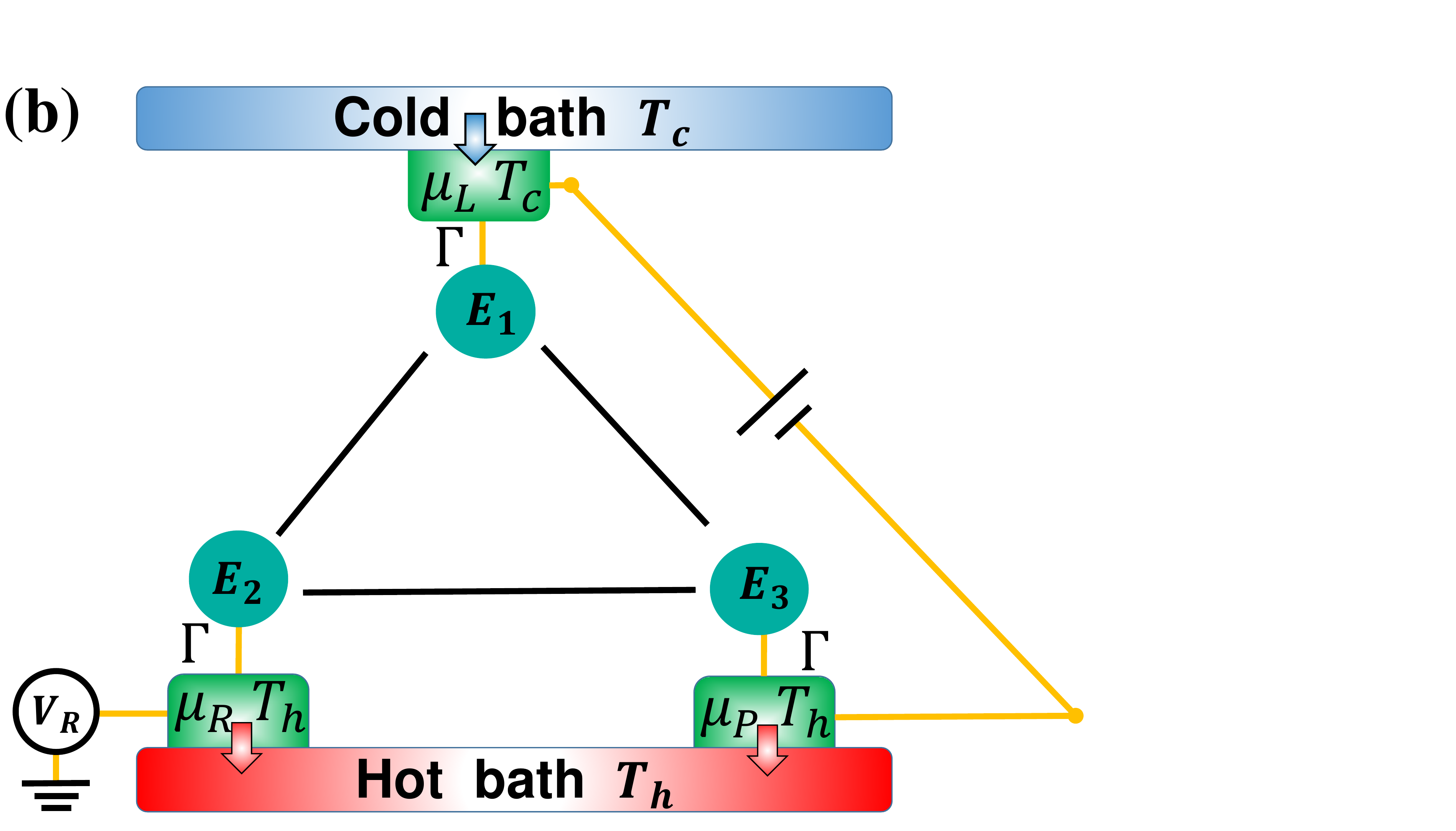}
\caption{(a) Schematic figure of a three-terminal triple-QD thermoelectric heat engine. Three quantum dots with a single energy level $E_i$ ($i=L,R,P$) are connected in series to three fermionic reservoirs $i$. The chemical potential and temperature of reservoirs are $\mu_i$ and $T_i$. The constant $\Gamma$ represents the coupling between the quantum dot and electrode. We consider this set-up that the $L$ terminal connects with hot bath and $R$ and $P$ terminal connect with cold bath. In addition, the three-terminal thermoelectric system is connected with three-resistors circuit, the resistors (denoted as $R_i$, $i=1,2,3$) are connected to three electrodes, $I_i$ represents the current following through each resistor. (b) Schematic figure of a three-terminal triple-QD refrigerator, where $L$ terminal connects with cold bath and $R$ and $P$ terminal connect with hot bath. The $L$ terminal can be cooled.}
\label{fig1}
\end{center}
\end{figure}

We consider a nanoscale thermoelectric system consisting of triple quantum dots (QDs) coupled to three electrodes [see Fig.~\ref{fig1}]. One electrode is mounted onto one heat bath, the other two electrodes are mounted onto another heat bath. The system can function as a heat engine or a refrigerator depends on the voltage and temperature configurations. In the high-temperature regime, the inter-QD Coulomb interaction can be ignored~\cite{Saito,Buttiker}. The intra-QD Coulomb interaction, which is not relevant for the essential physics to be revealed in this work, is neglected. Each QD is coupled to the nearby reservoir. We employ the indices $1/2/3$ to identify the leads $L/R/P$ thus, respectively.

The system is described by the Hamiltonian~\cite{JiangPRL}
\begin{align}
\hat{H}=\hat{H}_{\rm QD}+\hat{H}_{\rm lead}+\hat{H}_{\rm tun},
\end{align}
where
\begin{align}
\hat{H}_{\rm QD}=\sum_{i=1,2,3} E_i d_i^\dagger d_i,
\end{align}
\begin{align}
\hat{H}_{\rm lead}=\sum_{i=1,2,3}\sum_k\varepsilon_{k} c^\dagger_{ik} c_{ik},
\end{align}
\begin{align}
\hat{H}_{\rm tun}=\sum_{i,k}V_{k}d^\dagger_ic_{ik}+{\rm H.c.}.
\end{align}
Here, $d_i^{\dag}$ and $d_i$ create and annihilate an electron in the $i$th QD with an energy $E_i$, $t$ is the tunneling between the QDs.
$c_{ik}^{\dag}$ and $c_{ik}$ create and annihilate an electron in the $i$th electrode with an energy $\epsilon_k$.

We take the temperature and electrochemical potential of reservoir $R$ as a reference, we restrict our study to the linear-response regime. We thus define the following thermodynamic forces,
\begin{equation}
F^i_e=\frac{\mu_i-\mu_R}{e}, \quad F^i_Q=\frac{T_i-T_R}{T_R} \quad  (i=L,P),
\end{equation}
where $e<0$ is the electron charge. We focus on the set-up where $L$ reservoir is connected to the hot bath and the $R$ and $P$ reservoirs are connected to the cold bath, i.e., $T_L\equiv T_h$ and $T_R=T_P\equiv T_c$. Therefore, there are two independent output charge currents, $I_e^L$ and $I_e^P$, whereas there is only one input heat current $I_Q$, with the corresponding fore $F_Q\equiv F_Q^L$.

\begin{figure*}[htb]
\centering
\includegraphics[width=4.8cm]{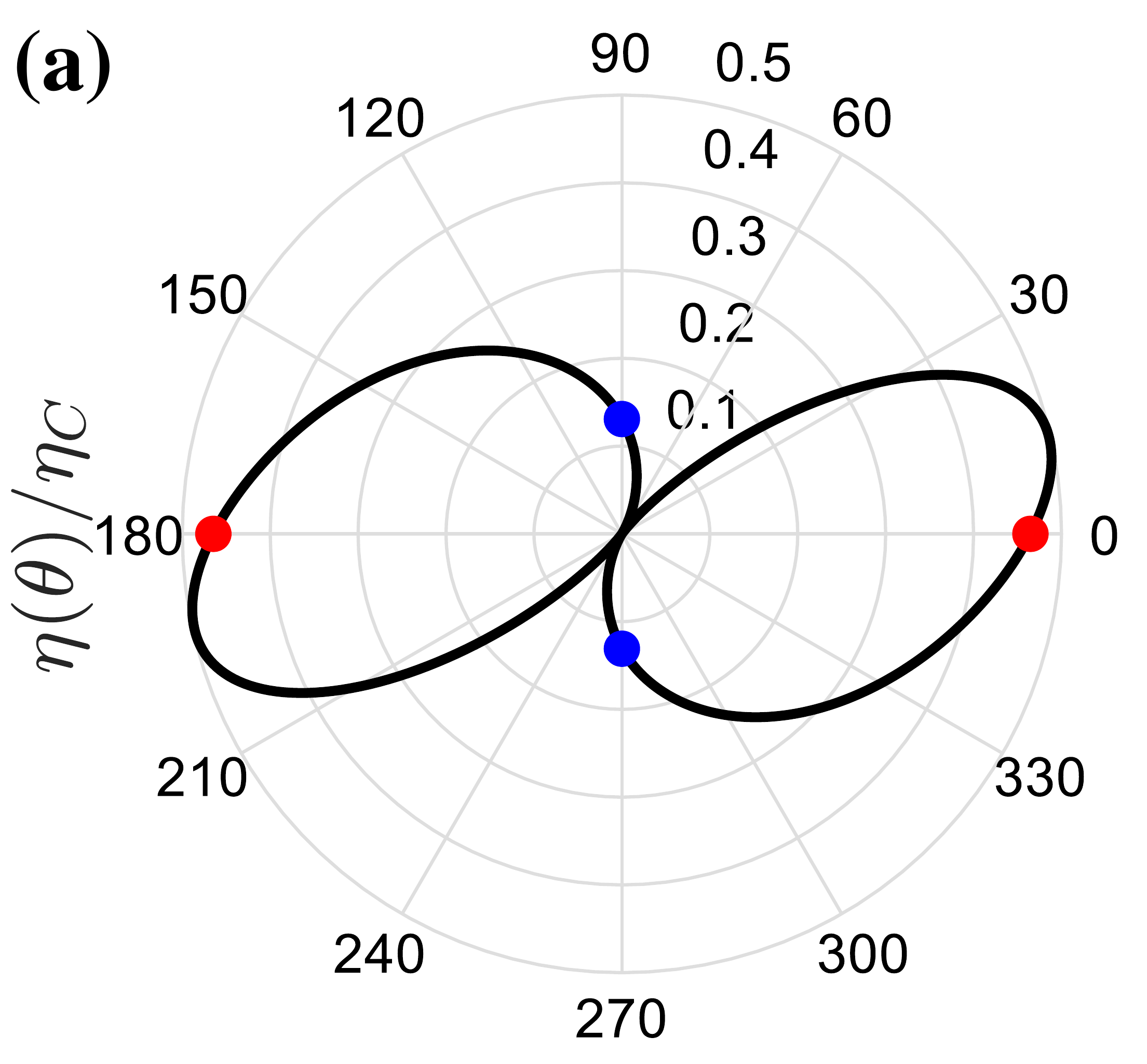}\hspace{2.2cm}\includegraphics[width=4.8cm]{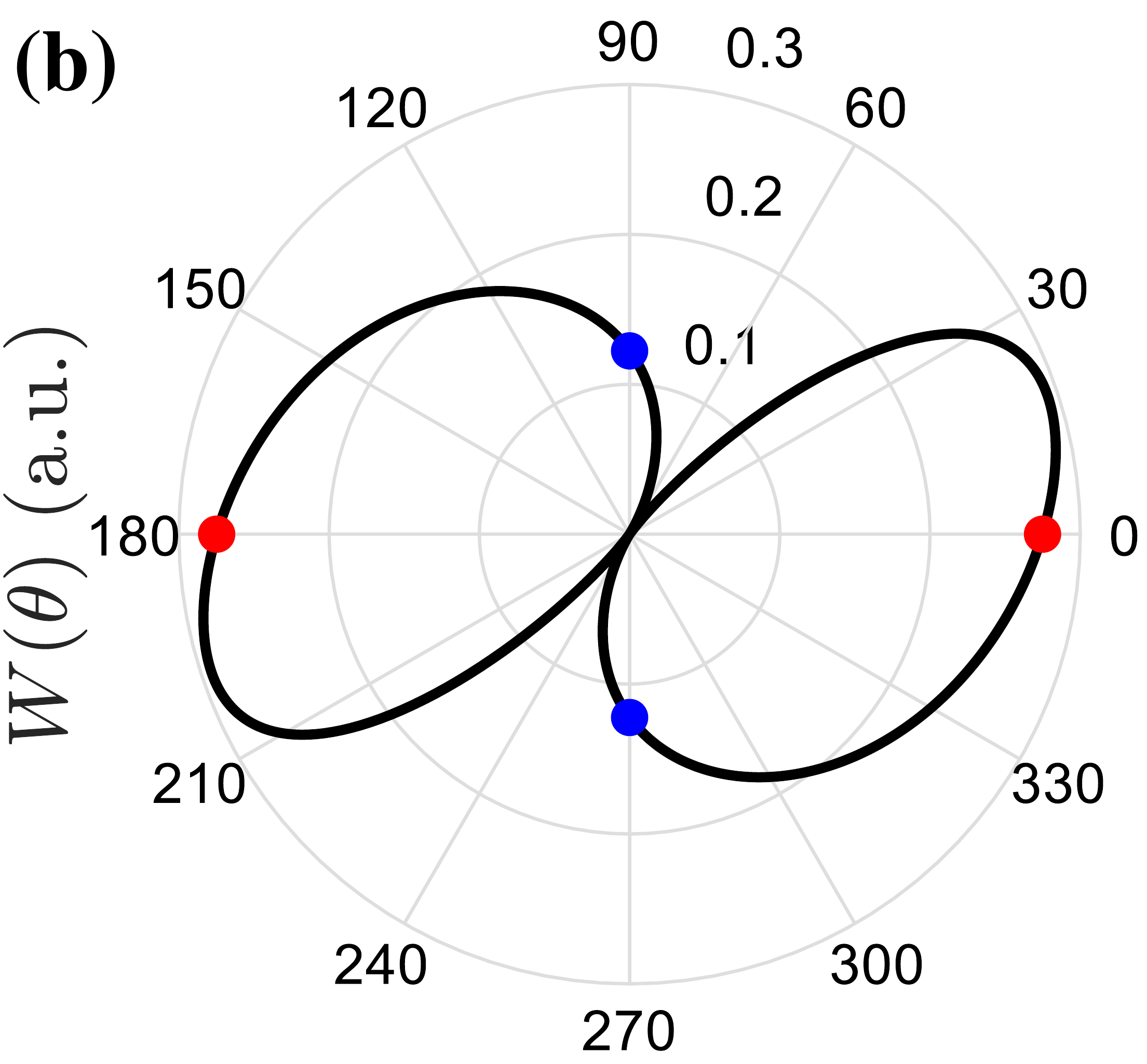}\hspace{1.8cm}
\includegraphics[width=4.8cm]{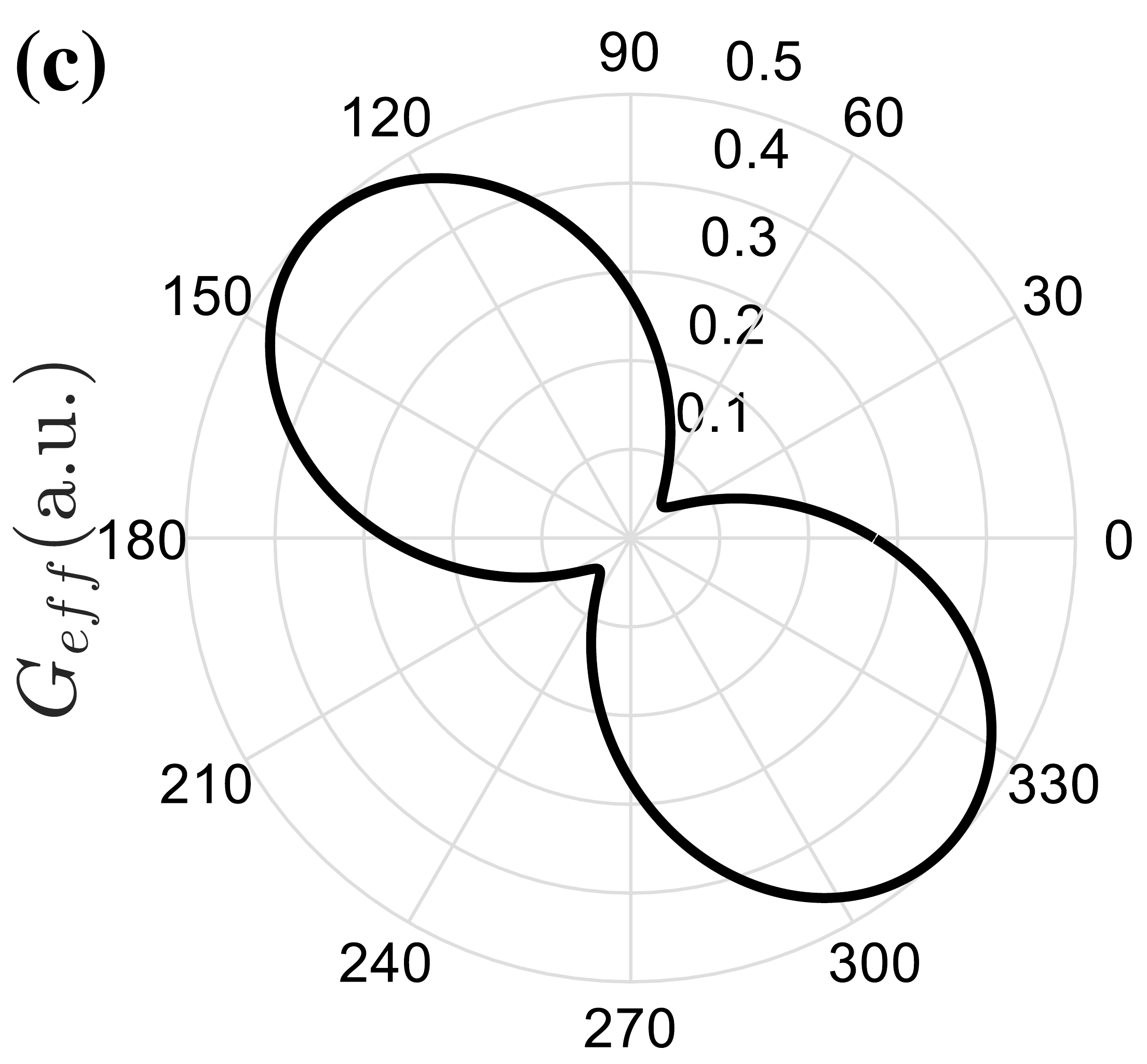}\hspace{2.2cm}\includegraphics[width=4.8cm]{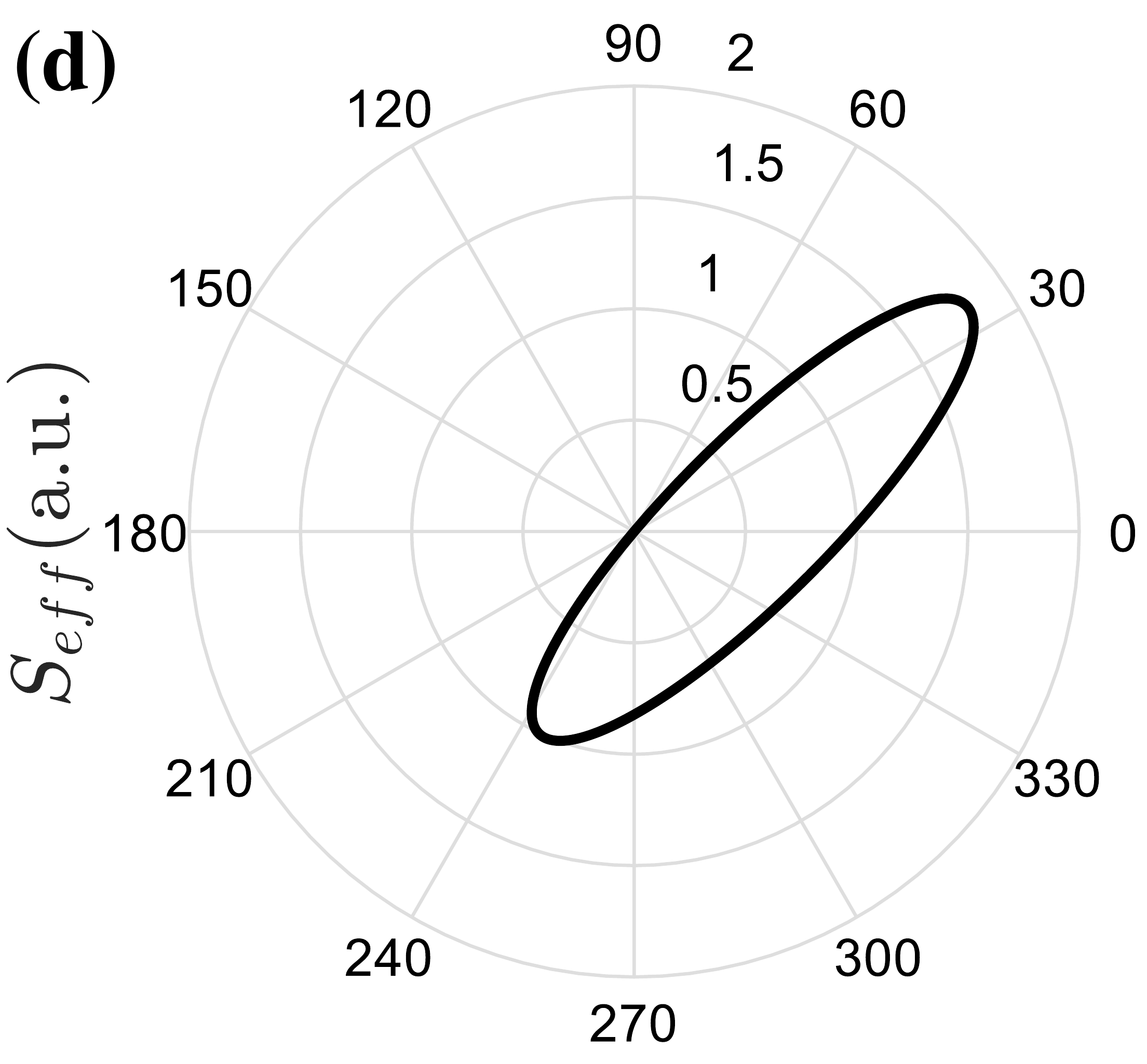}
\caption{Polar plot of (a) the efficiency $\eta(\theta)$ [Eq.~\eqref{eq:eta-theta}], (b) power $W(\theta)$ [Eq.~\eqref{eq:P}], versus angle $\theta$. At $\theta=0$ or $\pi$, $\eta$ and $W$ recover the values for the ``$L$-$R$ thermoelectric effect`` (red dots), while at $\theta=\pi/2$ or $3\pi/2$ they go back to those of the ``$P$-$R$ thermoelectric effect`` (blue dots). (c) Electronic conductance $G_{eff}$ [Eq.~\eqref{eq:G-theta}], (d) Seebeck efficient $S_{eff}$ [Eq.~\eqref{eq:S-theta}] as a function of $\theta$. The parameters are $t=-0.2k_BT$, $\Gamma=0.5k_BT$, $\mu=0$, $E_1=1.0k_BT$, $E_2=2.0k_BT$, $E_3=1.0k_BT$. }
~\label{fig:P_eta_theta}
\end{figure*}

In the linear-response regime, the charge and heat transports are described as~\cite{JiangPRE,Jiang2017}
\begin{equation}
\begin{pmatrix}
I^L_e \\
I^P_e \\
I_Q \\
\end{pmatrix}
=
\begin{pmatrix}
M_{11} & M_{12} & M_{13}  \\
M_{12} & M_{22} & M_{23}  \\
M_{13} & M_{23} & M_{33}  \\
\end{pmatrix}
\begin{pmatrix}
F^L_e \\
F^P_e \\
F_Q \\
\end{pmatrix},\label{eq:matrix}
\end{equation}
The coherent flow of charge and heat through a noninteracting system can be readily characterized by the Landauer-B\"{u}tiker theory~\cite{Sivan,butcher1990}.
The charge and heat currents following out of the left reservoir are given by
{\small{
\begin{equation}
I^L_e = \frac{2e}{h}\int_{-\infty}^{\infty}dE\sum_i{\mathcal T}_{iL}(E)[f_L(E)-f_i(E)],
~\label{eq:Ie}
\end{equation}

\begin{equation}
I_Q = \frac{2}{h}\int_{-\infty}^{\infty}dE\sum_i(E-\mu_L){\mathcal T}_{iL}(E)[f_L(E)-f_i(E)].
\label{eq:IQ}
\end{equation}
}}
Here $f_i(E)=\{\exp[(E-\mu_i)/k_BT_i]+1\}^{-1}$ is the Fermi distribution function and $h$ is the Planck's constant. The factor of two comes from the spin degeneracy of electrons. ${\mathcal T}_{ij}(E)$ is the transmission probability from terminal $j$ to terminal $i$\cite{3T-QTM}
\begin{equation}
{\mathcal T}_{ij}={\rm Tr}[\Gamma_i(E)G(E)\Gamma_j(E)G^\dagger(E)],
\label{eq:Tpq}
\end{equation}
where the (retarded) Green's function $G(E)\equiv [E-H_{\rm QD}-i\Gamma/2]^{-1}$, the dot-electrode coupling $\Gamma=2\pi\sum_k|V_{ik}|^2\delta(\omega-\varepsilon_{ik})$ is assumed to be a energy-independent constant for all three electrodes. Analogous expressions can be written for $I^P_e$, provided the index $L$ in Eq.~\eqref{eq:Ie} is substituted by $P$. The Onsager coefficients $M_{ij}$ ($i,j=1,2,3$) are calculated in details in Appendix.~\ref{coefficients}.

The entropy production accompanying this transport process reads~\cite{OraPRB2010,Iyynonlin}
\begin{equation}
T\dot{\mathcal{S}}=I_e^L F_e^L+I_e^P F_e^P+I_QF_Q,
\end{equation}
the second law of thermodynamics $\dot{\mathcal{S}}\ge0$ requires that
\begin{equation}
\begin{aligned}
&M_{11}, M_{22}, M_{33}\ge0,\quad  M_{11}M_{22}\ge M_{12}^2, \\
&\quad M_{11}M_{33}\ge M_{13}^2, \quad M_{22}M_{33}\ge M_{23}^2,
~\label{eq:Bound1}
\end{aligned}
\end{equation}
as well as that the determinant of the transport matrix in Eq.~\eqref{eq:matrix} to be non-negative.

\section{Cooperative Thermoelectric Effects: A Geometric Interpretation} \label{Co_eff}
In this section, we consider the cooperative effects in a thermoelectric engine and elucidate such cooperative effects using a geometric interpretation~\cite{JiangJAP,MyJAP}
\begin{equation}
F^L_{e} = F_e\cos{\theta}, \quad F^P_e = F_e\sin{\theta},
~\label{theta}
\end{equation}
where $F_e=\sqrt{(F^L_e)^2+(F^P_e)^2}$ is the total ``magnitude`` of the electrical affinities. To facilitate the discussion, we defined the effective electrical conductance as a function of the angle $\theta$~\cite{MyJAP},
\begin{equation}
G_{\rm{eff}}(\theta)=M_{11}\cos^2\theta+2M_{12}\sin\theta\cos\theta+M_{22}\sin^2\theta.
\label{eq:G-theta}
\end{equation}
The effective thermoelectric coefficient and the thermal conductance are
\begin{equation}
L_{\rm{eff}}(\theta)=M_{13}\cos\theta+M_{23}\sin\theta, \quad K=M_{33},
\end{equation}
respectively. Meanwhile the Seebeck coefficient is given by~\cite{3T-QTM}
\begin{equation}
S_{\rm{eff}}(\theta)=\frac{L_{eff}(\theta)}{TG_{eff}(\theta)}.
\label{eq:S-theta}
\end{equation}
Each angle $\theta$ corresponding to a particular configuration between the two electric affinities. By tuning $\theta$, we can obtain various configurations to explore the interference between the two thermoelectric effects.

Using the above relations, we can write the figure of merit $ZT$ for a given angle $\theta$ as~\cite{Saito2011}
\begin{equation}
ZT = \frac{L^2_{eff}(\theta)}{G_{eff}(\theta)K-L^2_{eff}(\theta)}.
\end{equation}

The energy efficiency of the thermoelectric engine is given by~\cite{3T-QTM}
\begin{equation}
\eta=-\frac{W}{I_Q}=-\frac{I_e^LF^L_e+I_e^PF^P_e}{I_Q}\le\eta_{\max}=\eta_C\frac{\sqrt{ZT+1}-1}{\sqrt{ZT+1}+1},
\label{eq:eta-theta}
\end{equation}
which is bounded by the Carnot efficiency, $\eta_C\equiv{(T_h-T_c)}/{T_h}$.

The maximum output power is found to be~\cite{JiangPRE,entropy-tradeoff}
\begin{equation}
\begin{aligned}
& W(\theta) = \frac{1}{4}\frac{L_{\rm{eff}}^2(\theta)}{G_{\rm{eff}}(\theta)}F^2_Q \\
&=\frac{(M_{13}\cos\theta+M_{23}\sin\theta)^2}{M_{33}(M_{11}\cos^2\theta+2M_{12}\sin\theta\cos\theta+M_{22}\sin^2\theta)}W_3,
\label{eq:P}
\end{aligned}
\end{equation}
Here we have defined $W_3=\frac{1}{4}M_{33}F^2_Q$. We shall term the thermoelectric effect associated with $F^L_e$ as the ``$L$-$R$ thermoelectric effect``, while that associated with $F^P_e$ as the ``$P$-$R$ thermoelectric effect``.

When $\theta=0$ or $\pi$, Eq.~\eqref{eq:P} gives the well-known efficiency and power for the $L$-$R$ thermoelectric effect
\begin{subequations}
\begin{align}
\eta_L=\,&\frac{\sqrt{M_{11}M_{33}}-\sqrt{M_{11}M_{33}-M^2_{13}}}{\sqrt{M_{11}M_{33}}+\sqrt{M_{11}M_{33}-M^2_{13}}}\eta_C,\\
 W_L =\,& \frac{M_{13}^2}{M_{11}M_{33}}W_3.
\label{eq:eta_l}
\end{align}
\end{subequations}

The $P$-$R$ thermoelectric efficiency and power, i.e., $\theta=\pi/2$ or $3\pi/2$, are given by
\begin{subequations}
\begin{align}
\eta_P =\, & \frac{\sqrt{M_{22}M_{33}}-\sqrt{M_{33}M_{22}-M^2_{23}}}{\sqrt{M_{22}M_{33}}+\sqrt{M_{33}M_{22}-M^2_{23}}}\eta_C,\\
 W_P =\, & \frac{M_{23}^2}{M_{22}M_{33}}W_3.
\label{eq:eta_t}
\end{align}
\end{subequations}

In the linear-response regime, the calculation of three-terminal heat engine performance with multiple electric currents and one heat current has been obtained in Ref.~[\onlinecite{trade-off}]. The maximum energy efficiency can be expressed as
\begin{equation}
\eta_{\max}=\frac{1-\sqrt{1-\lambda}}{1+\sqrt{1-\lambda}}\eta_C.
~\label{eq:eta_max}
\end{equation}
The maximum output power is
\begin{equation}
W_{\max}=\lambda W_3.
~\label{eq:W_max}
\end{equation}
Meanwhile, the efficiency at maximum output power is~\cite{Van2005,AI-PRL2012,Proesmans}
\begin{equation}
\eta(W_{\max})=\frac{\lambda}{4-2\lambda}\eta_C.
~\label{eq:eta_Wmax}
\end{equation}
and the output power at maximum efficiency is
\begin{equation}
W(\eta_{\max})=\lambda\left[1-\left(\frac{\eta_{\max}}{\eta_C}\right)^2\right]W_3.
~\label{eq:W_etamax}
\end{equation}
Here,
\begin{equation}
\lambda\equiv \frac{M_{13}^2M_{22}+M_{11}M_{23}^2-2M_{12}M_{13}M_{23}}{M_{33}(M_{11}M_{22}-M_{12}^2)}.
\end{equation}\label{eq:lambda}
is dimensionless parameter that characterize the thermoelectric performance of the system.



To concretely show how the two thermoelectric powers cooperate with each other, we plot in Fig.~\ref{fig:P_eta_theta}(a) and \ref{fig:P_eta_theta}(b) the efficiency $\eta$ and output power $W$ as a function of angle $\theta$ in a polar plot for a set of physical parameters specified in the caption. Obviously, the efficiency $\eta$ is greater than both $\eta_L$ and $\eta_P$ for $0<\theta<\pi/6$. In the same region, the output power $W$ can also be greater than both $W_L$ and $W_P$. In fact, the efficiency and output power follow the same changing trend because both of them are only dependent on the dimensionless parameter $\lambda$. Meanwhile, we find that the optimal performance of the $L$-$R$ thermoelectric effect is also better than that of the $P$-$R$ thermoelectric effect.

The cooperative effect can also induce the reduction of the Joule heat, which is related to the effective electrical conductance as $G_{eff}$. As shown in Fig.~\ref{fig:P_eta_theta}(c), the effective conductance $G_{eff}$ is reduced for $6/\pi$. This is the reason that the output power is enhanced despite the fact that the effective thermoelectric power $S_{eff}$ is reduced [see Fig.~\ref{fig:P_eta_theta}(d)], since the total power is given by $W(\theta)=L_{eff}^2W_3/G_{eff}$. The increase of the output power then leads to the enhancement of energy efficiency.

\begin{figure}[htb]
\begin{center}
\centering \includegraphics[width=4.2cm]{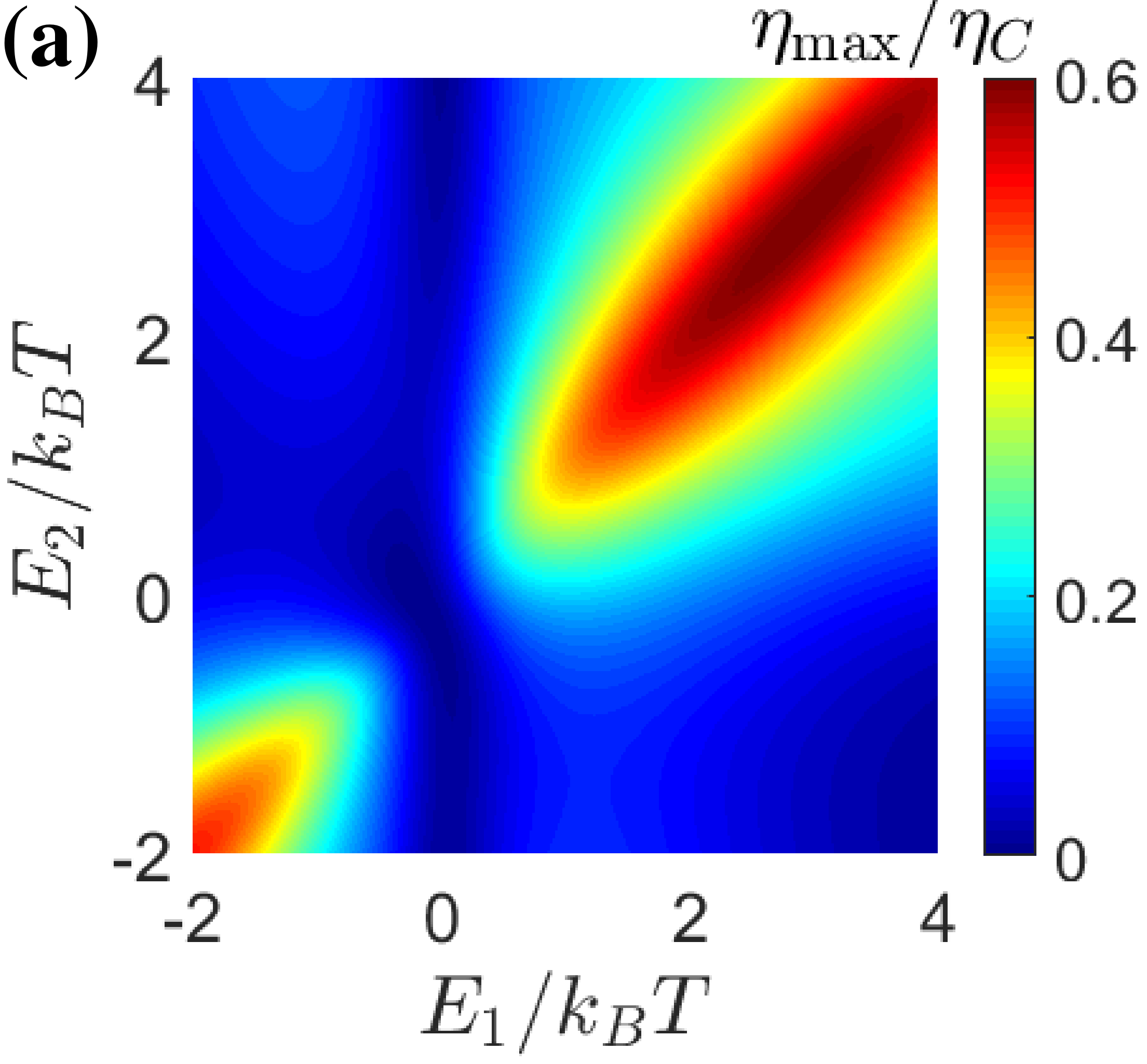}\hspace{0.2cm}\includegraphics[width=4.2cm]{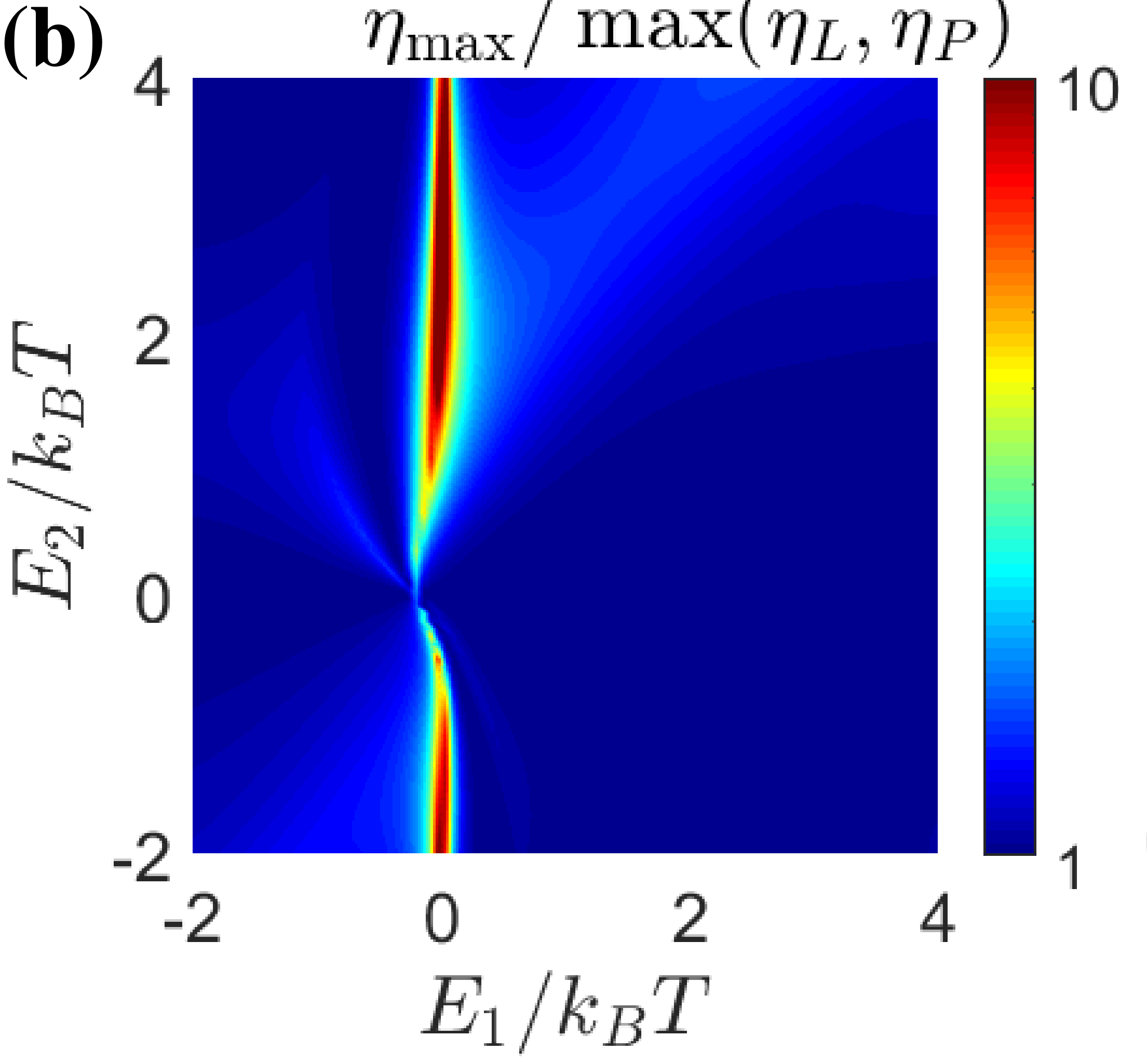}
\centering \includegraphics[width=4.2cm]{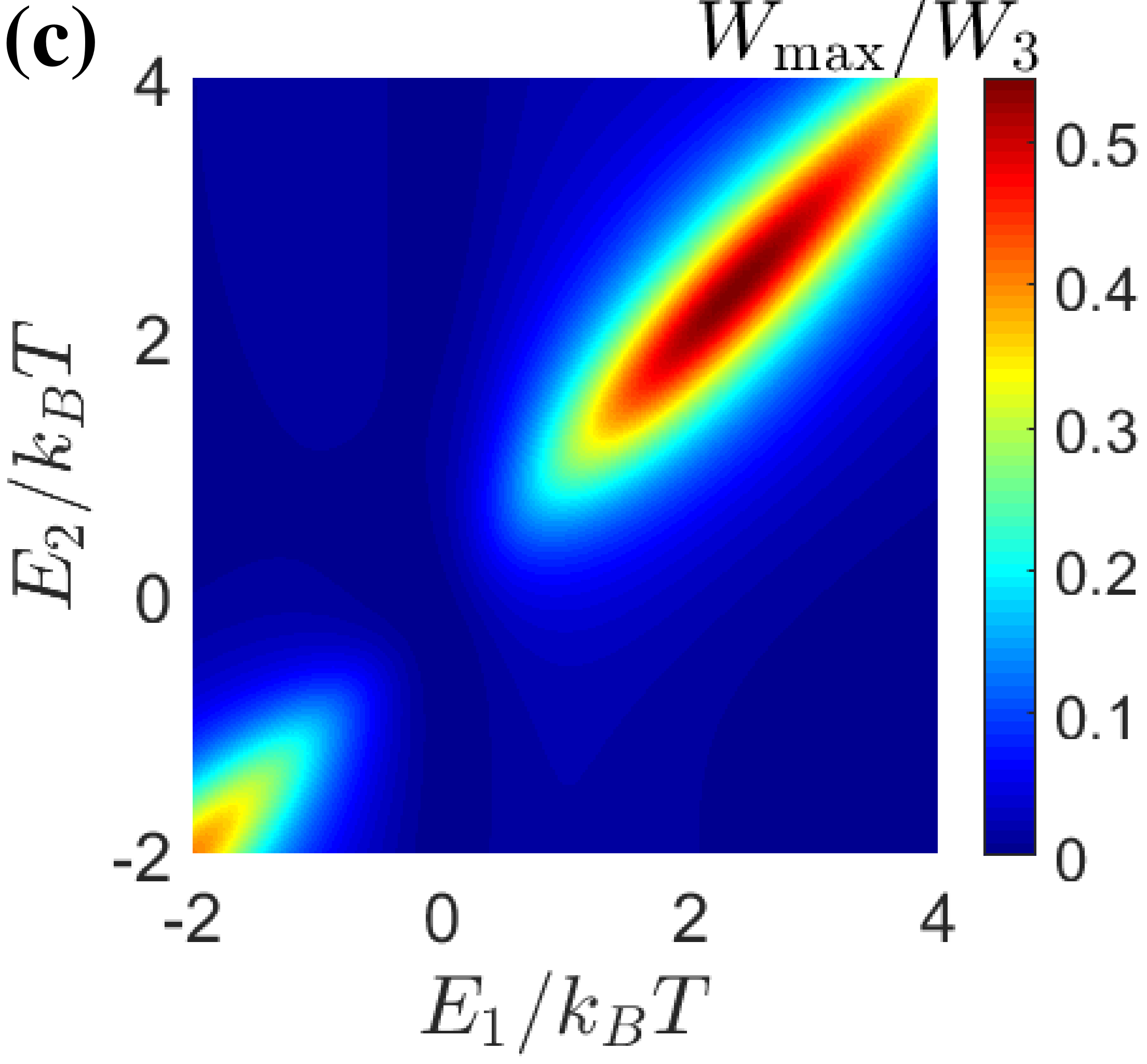}\hspace{0.2cm}\includegraphics[width=4.2cm]{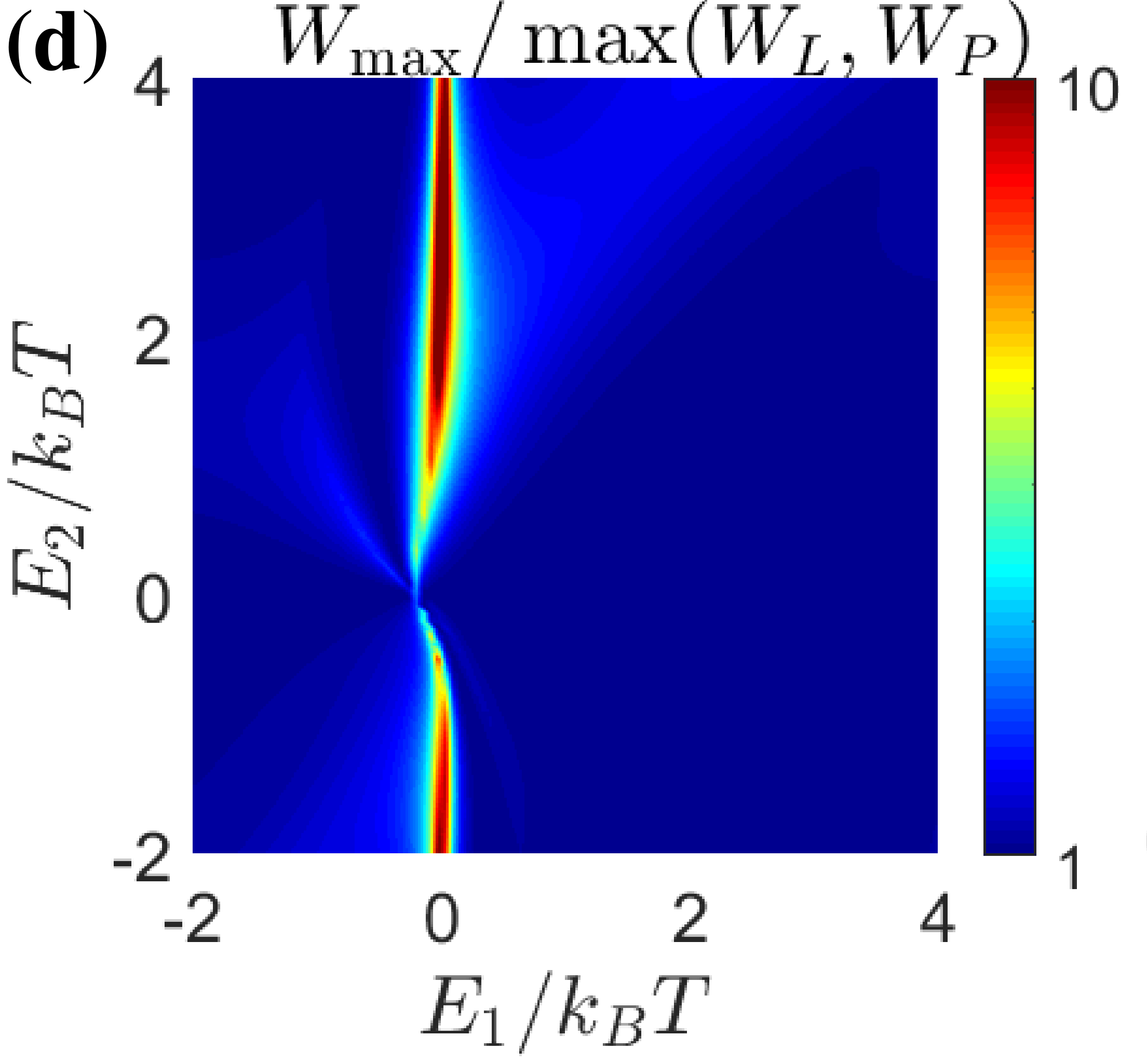}
\caption{(a) The maximum efficiency $\eta_{\max}$ [Eq.~\eqref{eq:eta_max}], (b) the enhancement of efficiency $\eta_{\max}/\max{(\eta_L,\eta_P)}$, (c) the maximum output power $W_{\max}$ [Eq.~\eqref{eq:W_max}], (d) the enhancement of power $W_{\max}/\max{(W_L,W_P)}$ as functions of $E_1$ and $E_2$. The parameters are $t=-0.2k_BT$, $\Gamma=0.5k_BT$, $\mu=0$, $E_3=0$. } \label{fig:eta_E1E2}
\end{center}
\end{figure}

To further demonstrate that the cooperation between the two thermoelectric effects leads to improved efficiency and output power, the underlying physical figures are shown by the numerical calculation to confirm using QD energies. Figs.~\ref{fig:eta_E1E2}(a) and \ref{fig:eta_E1E2}(c) indicate the maximum efficiency $\eta_{\max}$ and maximum output power $W_{\max}$ as functions of QD energy $E_1$ and $E_2$. We find that the $\eta_{\max}$ and $W_{\max}$ are very sensitive to the QDs energies which can be controlled easily via gate-voltage in experiments~\cite{Jiangtransistors}. The optimal values for $\eta_{\max}$ and $W_{\max}$ are achieved when $E_1\approx E_2>1.0k_BT$ and the $\eta_{\max}$ can reach $0.6\eta_C$, as shown by the hot-spots regions. In particular,
both of the efficiency and power are symmetric around the line of $E_1=E_2$ when $E_3=0$. Figs.~\ref{fig:eta_E1E2}(b) and \ref{fig:eta_E1E2}(d) show the enhancement of efficiency and power are large for the situation that $E_1\approx 0$ and even the enhancements can go beyond 10. Remarkably, the maximum efficiency $\eta_{\max}$ always is greater than both $\eta_L$ and $\eta_P$, i.e., the enhancements of $\eta_{\max}/\max(\eta_L,\eta_P)$ is always greater than 1. 

From the above results we conclude that cooperative effects can significantly improve the efficiency and output power for thermoelectrics in three-terminal geometry. Such improvements are helpful for optimizing the electronic systems especially. Hence, cooperative effects provide a convenient and efficient way to design and improve the performance of thermoelectrics which are potentially useful for realistic systems~\cite{Sothmann-Re,Sothmann-QW,UdoPRX,sothmann}.

Following the discussion in Ref.~[\onlinecite{trade-off}], we now discuss the practical problem of connecting the three-terminal thermoelectric heat engine to a resistor circuit which receives the electric power. We use a triangular resistor to receive the electric power to achieve the best performance, as shown in Fig.~\ref{fig1}(a). We connect a triangular resistance circuit with the three-terminal system, each resistance (resistances are denoted as $R_i$, $i=1,2,3$) is connected to the nearby reservoir. The current-force response matrix for the resistor circuit is
\begin{equation}
\vec{I}_e^{\prime} = \hat{M}^{3T} \vec{F}_e.
\label{eq:Ie2}
\end{equation}
where $\vec{I}_e^{\prime}=(I_L^{\prime},I_P^{\prime})^T$, $I_L^{\prime}$ and $I_P^{\prime}$ are charge currents following from $L$ and $P$ reservoirs into the resistor circuit, respectively. The current following through each resistor $R_i$ ($i=1,2,3$) are
\begin{subequations}
\begin{align}
I_1&=\frac{\mu_L-\mu_R}{eR_1}=\frac{F_e^L}{R_1},\\
I_2&=\frac{\mu_R-\mu_P}{eR_2}=-\frac{F_e^P}{R_2},\\
I_3&=\frac{\mu_P-\mu_L}{eR_3}=\frac{F_e^P-F_e^L}{R_3}.
\end{align}\label{eq:Ii}
\end{subequations}
On the one hand, using Kirchhoff's current law, we obtain
\begin{equation}
\begin{aligned}
I_L^{\prime}-I_1+I_3=0,\\
I_P^{\prime}+I_2-I_3=0.
\end{aligned}~\label{eq:ILP}
\end{equation}
Combining Eqs.~\eqref{eq:Ie2}, \eqref{eq:Ii} and \eqref{eq:ILP}, we can get the expression of $\hat{M}^{3T}$,
\begin{equation}
\hat{M}^{3T}=
\begin{pmatrix}
\frac{1}{R_1}+\frac{1}{R_3} & -\frac{1}{R_3}   \\
-\frac{1}{R_3} & \frac{1}{R_2}+\frac{1}{R_3}  \\
\end{pmatrix}.
\end{equation}
According to Ref.~[\onlinecite{trade-off}], we find that the maximum output power is reached at
\begin{equation}
\hat{M}^{3T}=
\begin{pmatrix}
M_{11} & M_{12}   \\
M_{12} & M_{22}  \\
\end{pmatrix},
\end{equation}
whereas the maximum energy efficiency is reached at
\begin{equation}
\hat{M}^{3T}=\sqrt{1-\lambda}
\begin{pmatrix}
M_{11} & M_{12}   \\
M_{12} & M_{22}  \\
\end{pmatrix}.
\end{equation}
The energy efficiency and output power for the two conditions are the same with Eqs.~\eqref{eq:eta_max}-\eqref{eq:W_etamax}.

\section{Thermoelectric cooperative effects on a three-terminal refrigerator}\label{ref-3T}
The three-terminal device can be tuned to be a refrigerator\cite{JiangPRE,Rongqian}, by exchanging temperatures of the three electrodes, i.e., $T_{R(P)}=T_h$, $T_L=T_c$, with $T_h>T_c$, as shown in Fig.~\ref{fig1}(b). Then, the $L$ terminal can be cooled and the heat $I_Q$ following from the $L$ terminal is transferred to the system. Here, we use the invested work as the chief power supplier,
\begin{equation}
W^{\rm in} = I_e^LF^L_e+I_e^PF^P_e.
\end{equation}
The cooling efficiency is defined by the ratio of the cooling heat $I_Q^{\rm out}$ and the input power $W^{\rm in}$:
\begin{equation}
\eta^{\rm ref} = \frac{I_Q^{\rm out}}{W^{\rm in}}=\frac{I_Q^{\rm out}}{I_e^LF^L_e+I_e^PF^P_e}\le\eta_C,
\end{equation}
and the Carnot efficiency for which the process is reversible is given by
\begin{equation}
\eta_C = \frac{T_c}{T_h-T_c}.
\end{equation}
According to Eq.~\eqref{eq:matrix}, the cooling heat $I_Q^{\rm out}$ can be expressed as
\begin{equation}
I_Q^{\rm out} = I_{Q,1} + I_{Q,2} + I_{Q,3},
\end{equation}
which consists of three parts: thermal conduction $I_{Q,3}=M_{33}F_Q$, $L$-$R$ Peltier cooling $I_{Q,1}=M_{13}F_e^L$, and $P$-$R$ Peltier cooling $I_{Q,2}=M_{23}F_e^P$. The cooling is achieved when the sum of the heat current $I_{Q,1}$ and heat current $I_{Q,2}$ exceeds the thermal conduction current $I_{Q,1}$.

\begin{figure}[htb]
\begin{center}
\centering \includegraphics[width=4.2cm]{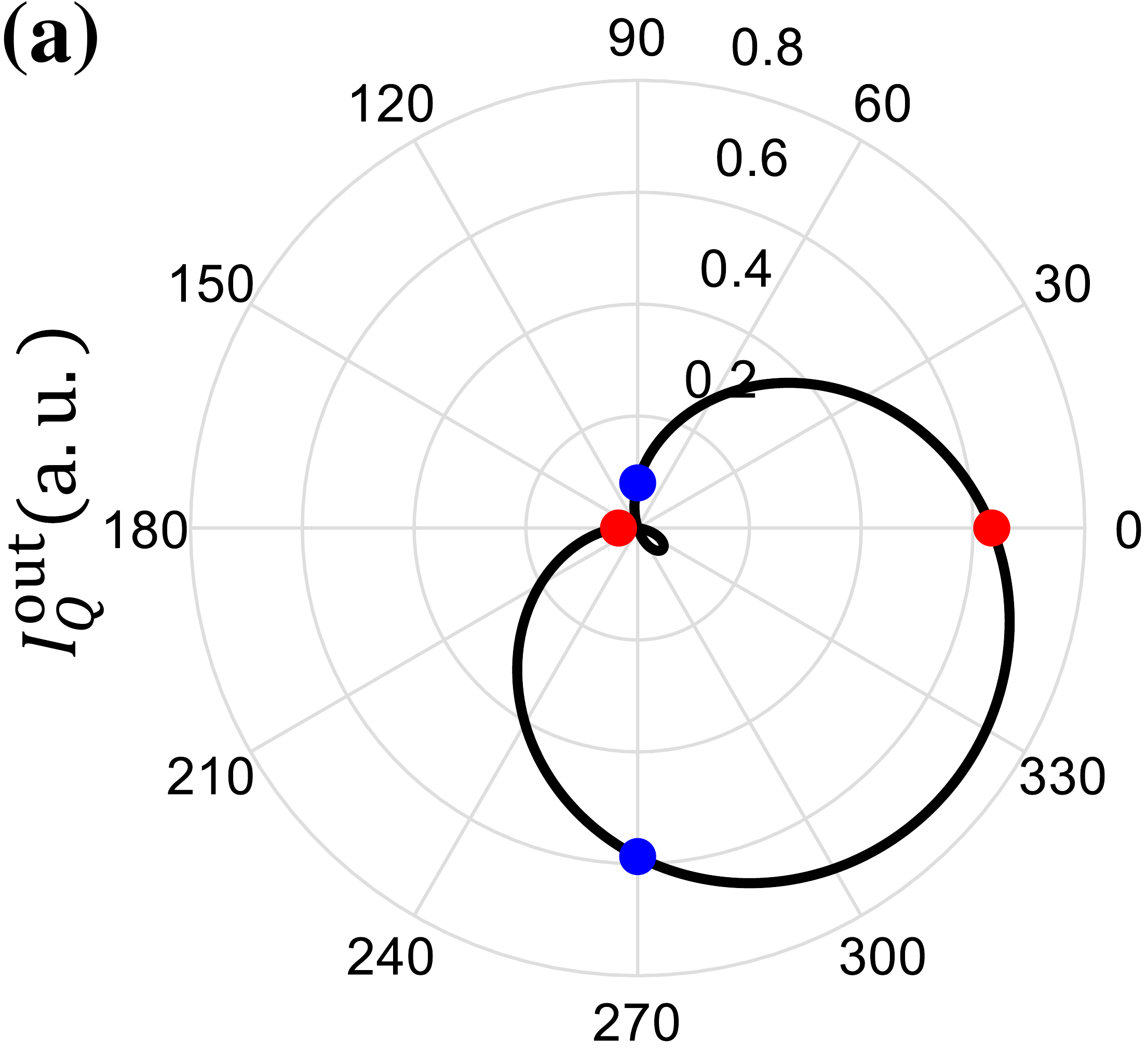}\hspace{0.21cm}\includegraphics[width=4.2cm]{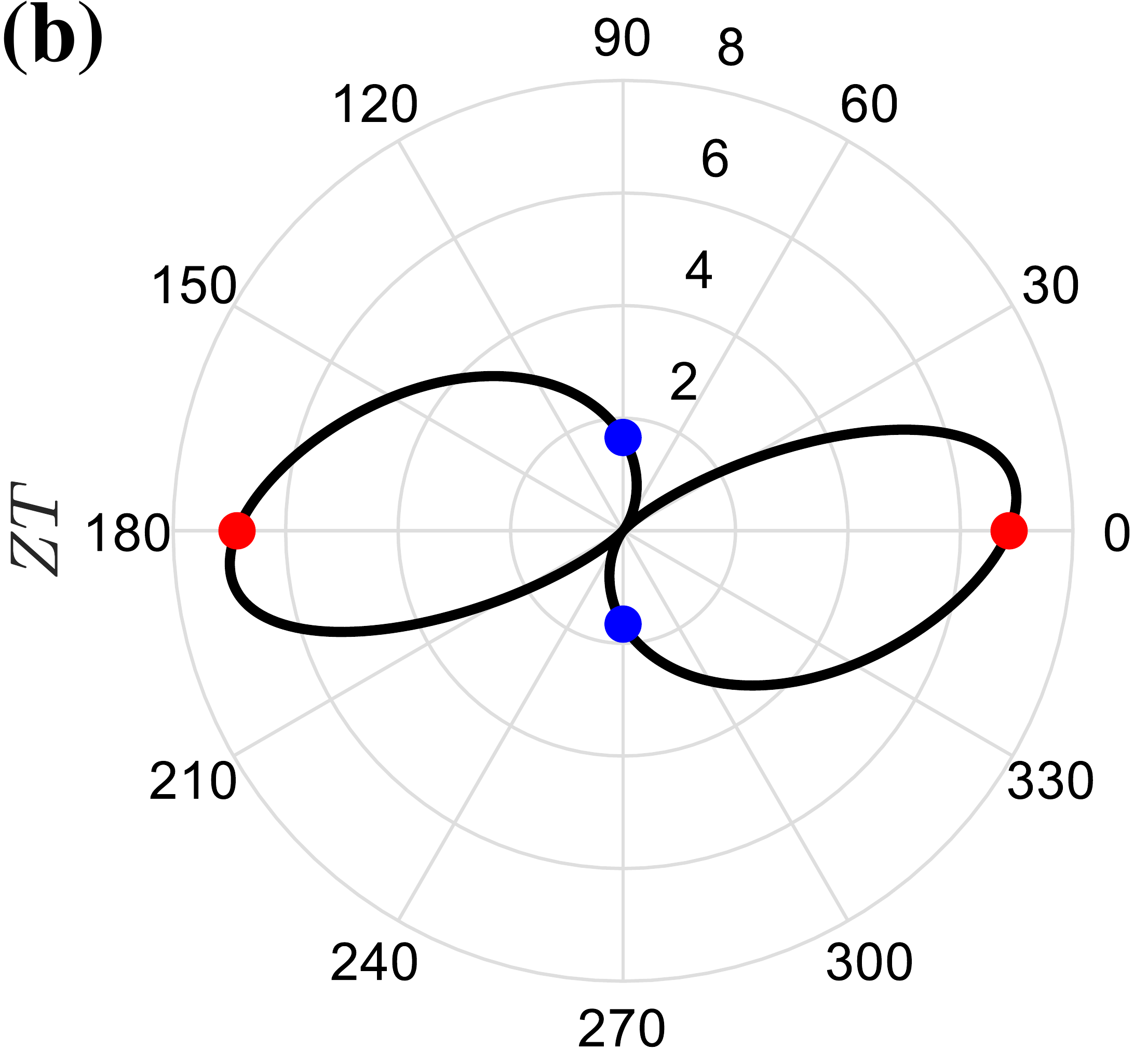}
\caption{Polar plot of (a) the effective cooling heat current $I_Q^{\rm out}$, (b) figure of merit $ZT$ as a function of $\theta$. The red dots represent the ``$L$-$R$ thermoelectric effect``, and blue dots represent the ``$P$-$R$ thermoelectric effect``. The parameters are $t=-0.2k_BT$, $\Gamma=0.5k_BT$, $\mu=0$, $E_1=1.0k_BT$, $E_2=2.0k_BT$, $E_3=1.0k_BT$, $F_e=1$ and $F_Q=1$.}\label{fig:IQ-theta}
\end{center}
\end{figure}

Combining Eq.~\eqref{theta}, the effective heat current can be expressed as a function of the angle $\theta$,
\begin{equation}
I_Q^{\rm out} = M_{13}F_e\cos\theta + M_{23}F_e\sin\theta + M_{33}F_Q.
\label{eq:IQ-theta}
\end{equation}
The Eq.~\eqref{eq:IQ-theta} indicates that tuning the angle $\theta$ changes the relative amplitude of the heat current $I_{Q,1}$ and heat current $I_{Q,2}$. As shown in Fig.~\ref{fig:IQ-theta}(a), the two heat currents can be same sign, or the opposite sign, depending on $\theta$. From Fig.~\ref{fig:IQ-theta}(b), when $I_{Q,1}$ and $I_{Q,2}$ have the same sign, the cooling is enhanced, leading high efficiency. However, when $I_{Q,1}$ and $I_{Q,2}$ have opposite sign, the cooling is suppressed and the efficiency is reduced. Our analytical results demonstrate that the cooperative effects between the two cooling mechanisms can substantially improve the efficiency of the refrigerator.

%

\section{The optimal efficiency and power of four-terminal heat engine}~\label{4T}

\begin{figure}[htb]
\begin{center}
\centering \includegraphics[width=8.5cm]{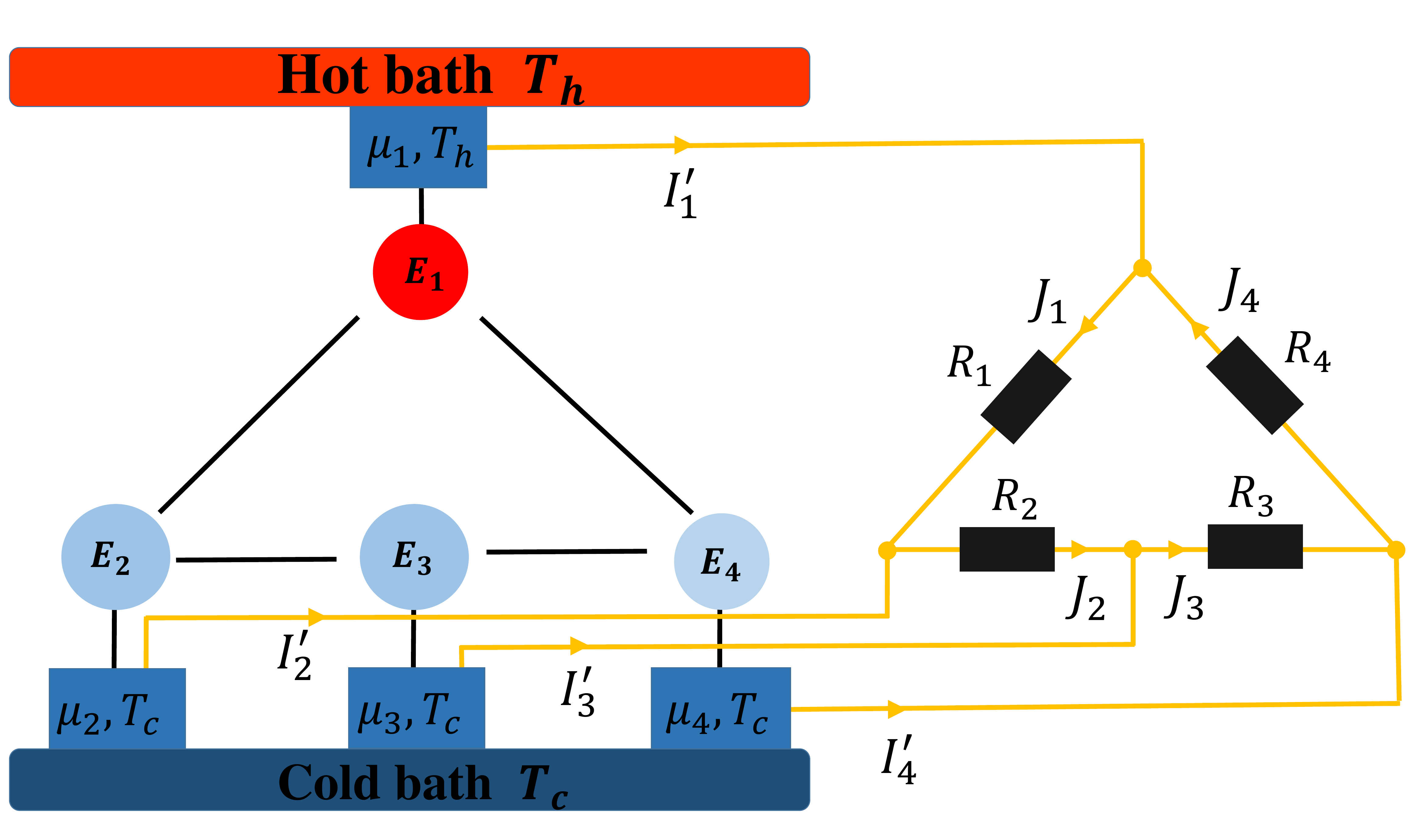}
\caption{The four-terminal thermoelectric system is connected with four-resistors circuit, the resistors (denoted as $R_i$, $i=1,2,3,4$) are connected to four electrodes, the electrochemical potential and temperature of the reservoir $i$ are $\mu_i$ and $T_i$, respectively, $J_i$ represents the current following through each resistor. }\label{fig:4T}
\end{center}
\end{figure}

\begin{figure}[htb]
\begin{center}
\centering \includegraphics[width=4.2cm]{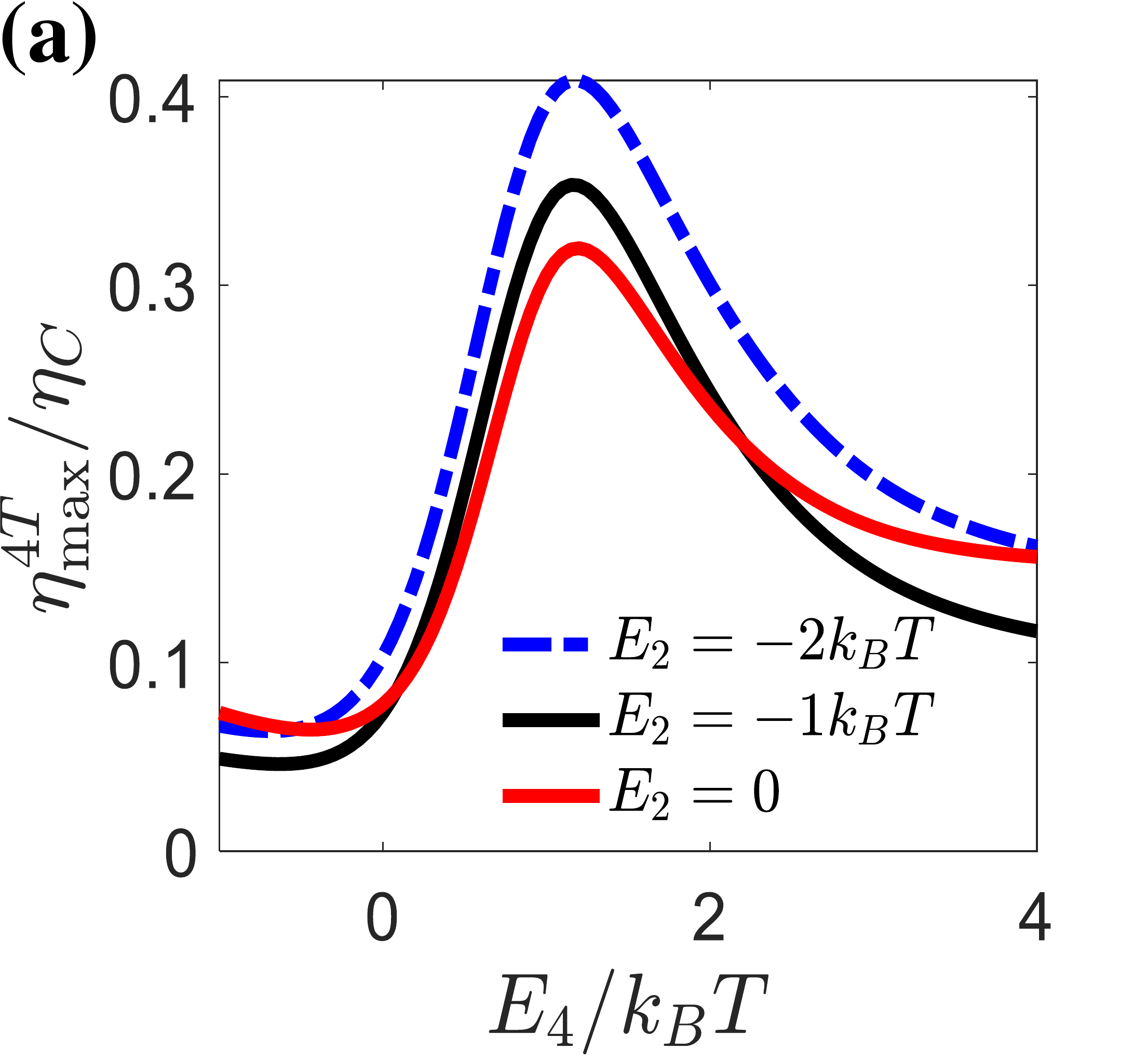}\hspace{0.2cm}\includegraphics[width=4.2cm]{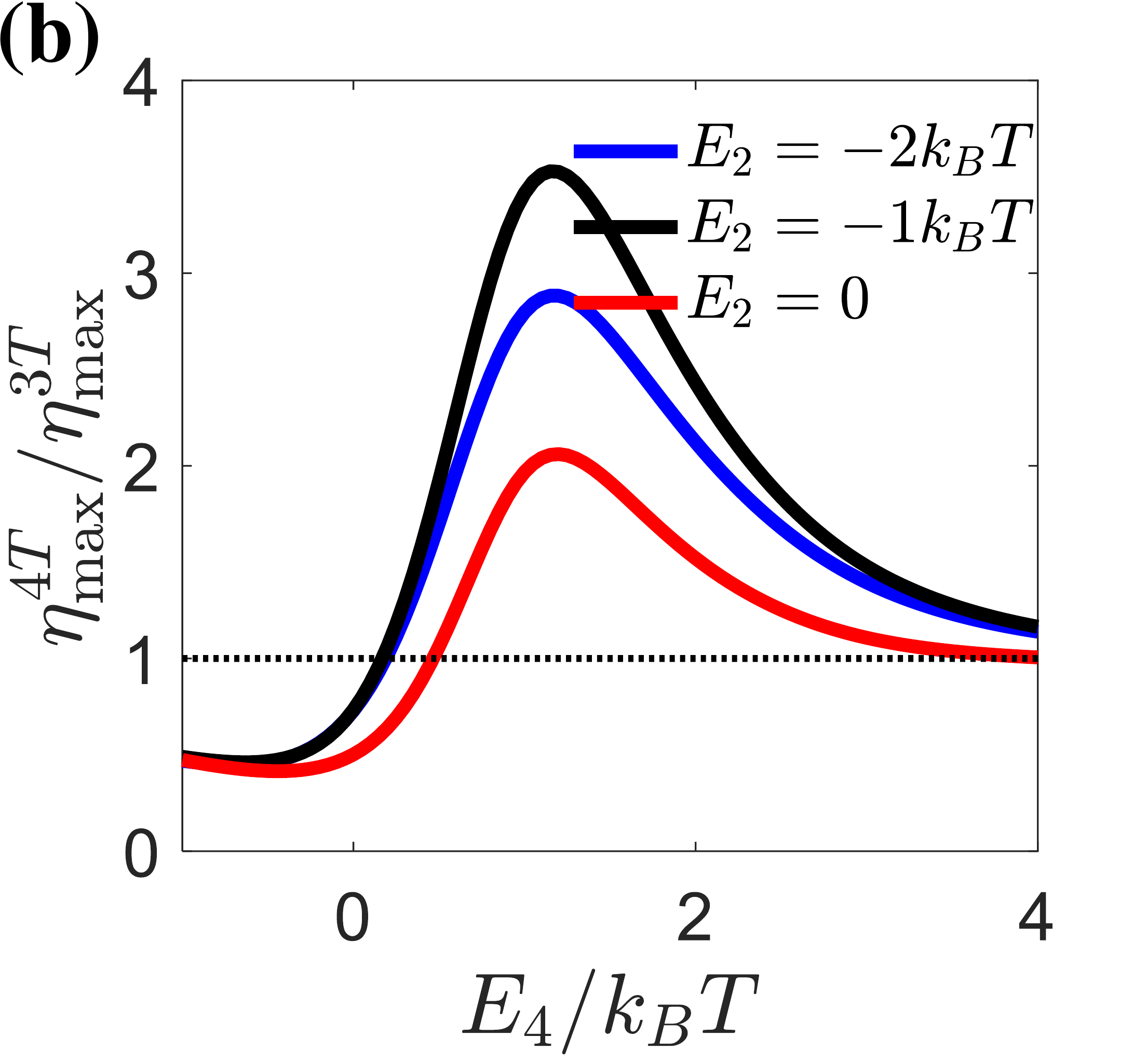}
\centering \includegraphics[width=4.2cm]{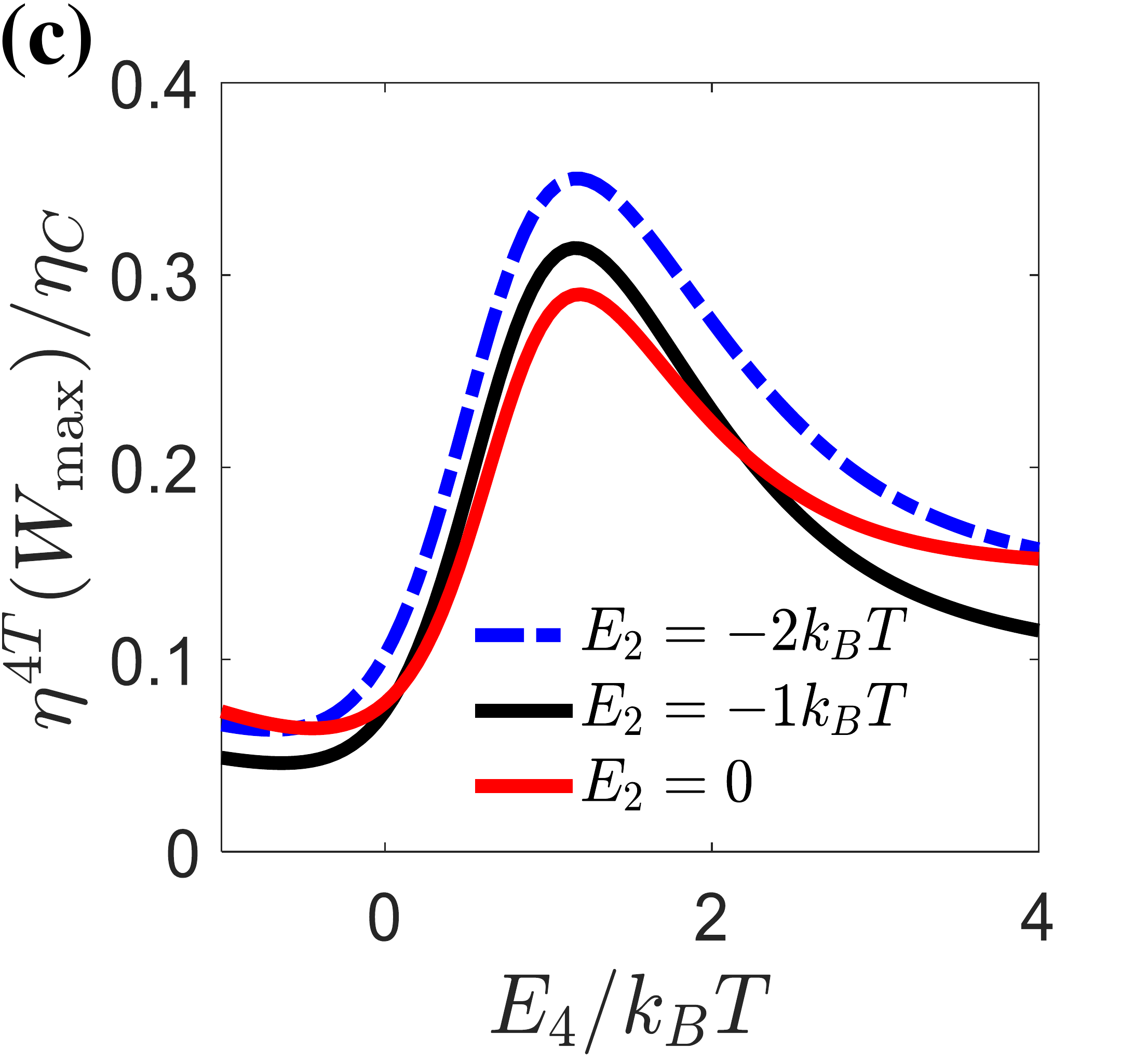}\hspace{0.2cm}\includegraphics[width=4.2cm]{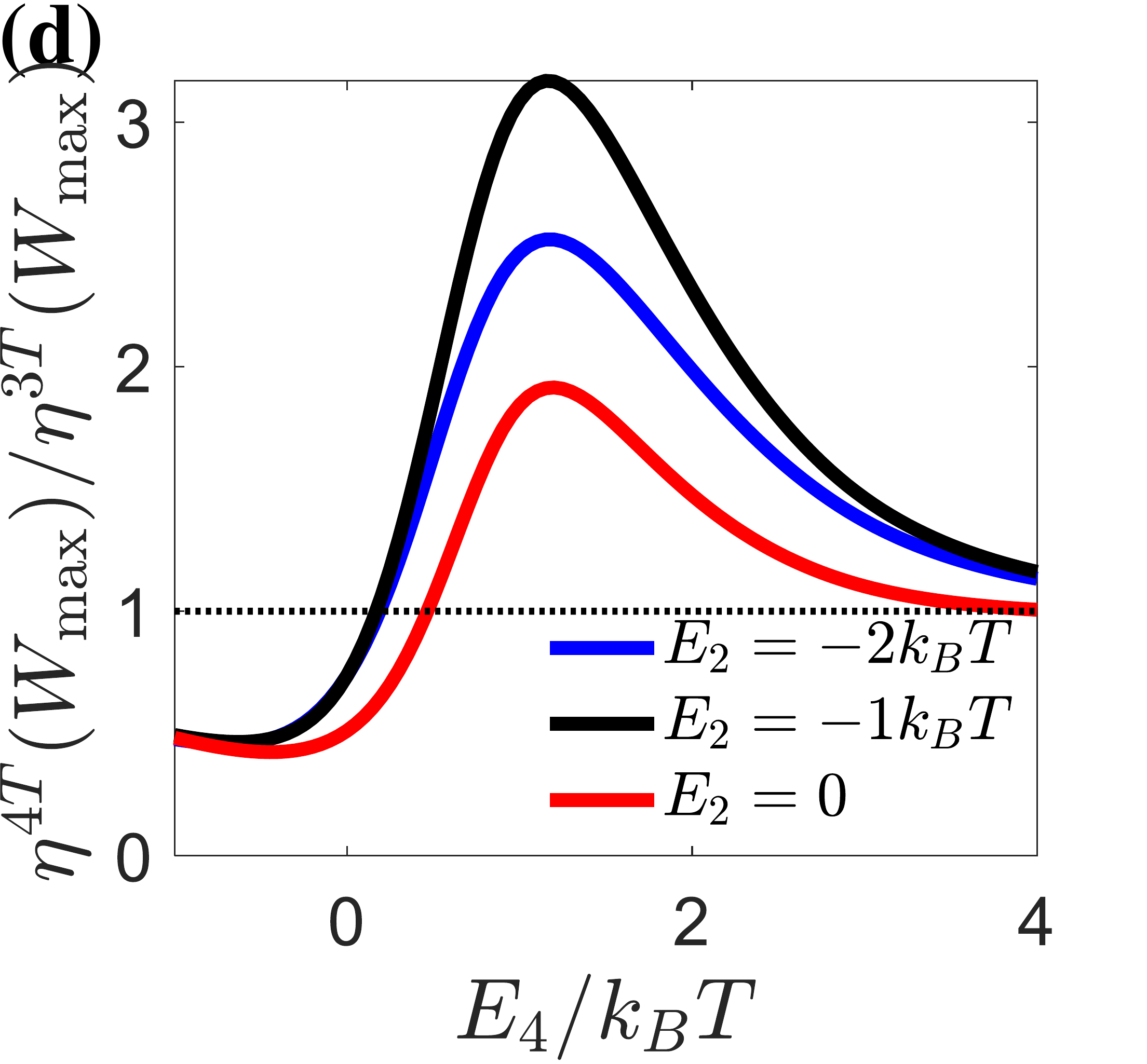}
\caption{(a) The maximum efficiency $\eta^{4T}_{\max}$ [Eq.~\eqref{eq:4T-eta}] and (c) the efficiency at maximum output power $\eta^{4T}(W_{\max})$ [Eq.~\eqref{eq:4T-W}] as a function of $E_4$ for different $E_2$. Comparing the optimal efficiency and power for the four-terminal heat engine and the case of three-terminal heat engine as a function of $E_4$ for different $E_2$: (b) the maximum efficiency, (d) efficiency at maximum output power. The parameters are $E_1=1.0k_BT$, $E_3=3.0k_BT$, $\mu=0$, and $t=-0.2k_BT$.}\label{fig:E4}
\end{center}
\end{figure}

\begin{figure}[htb]
\begin{center}
\centering \includegraphics[width=4.2cm]{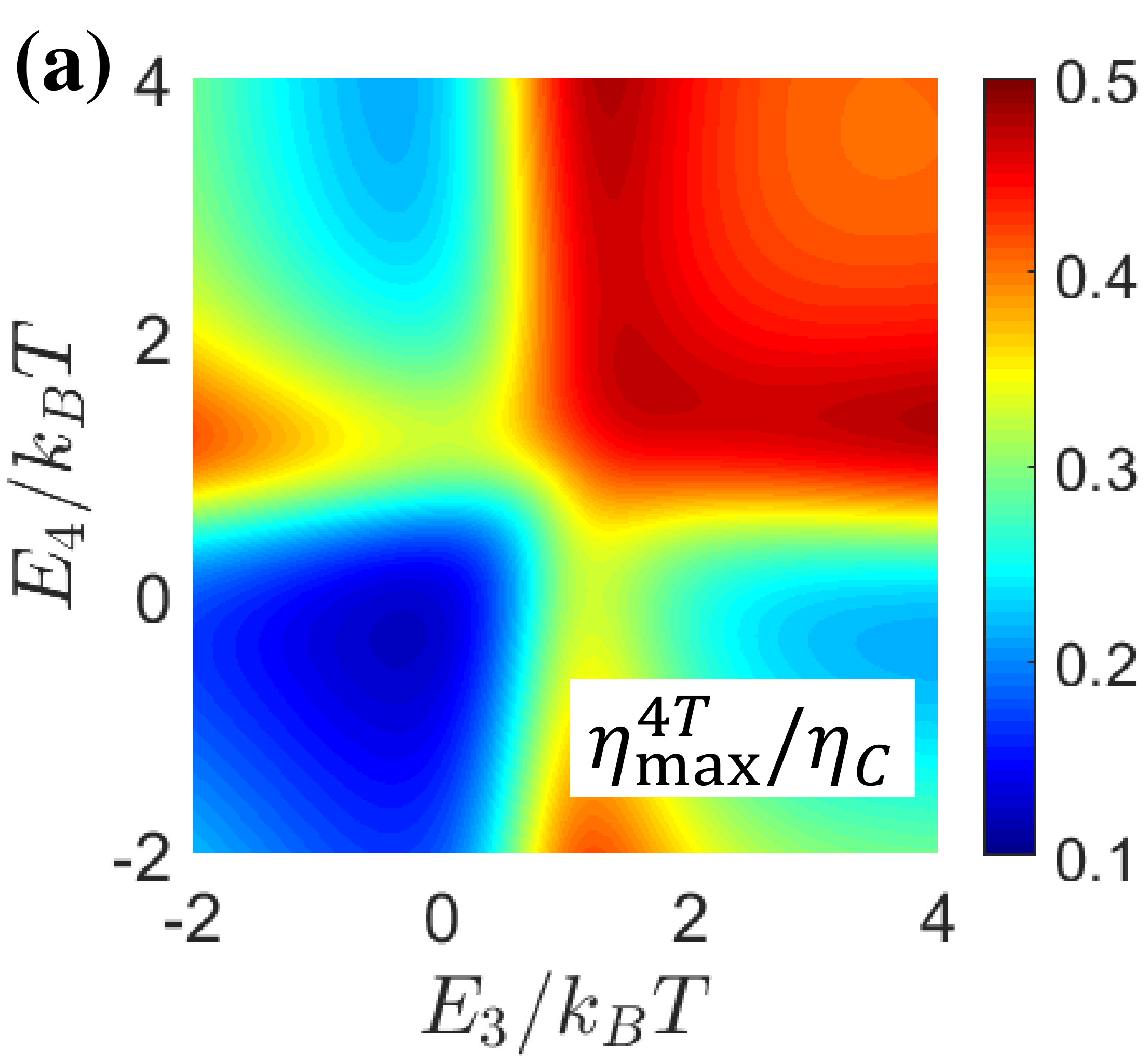}\hspace{0.2cm}\includegraphics[width=4.2cm]{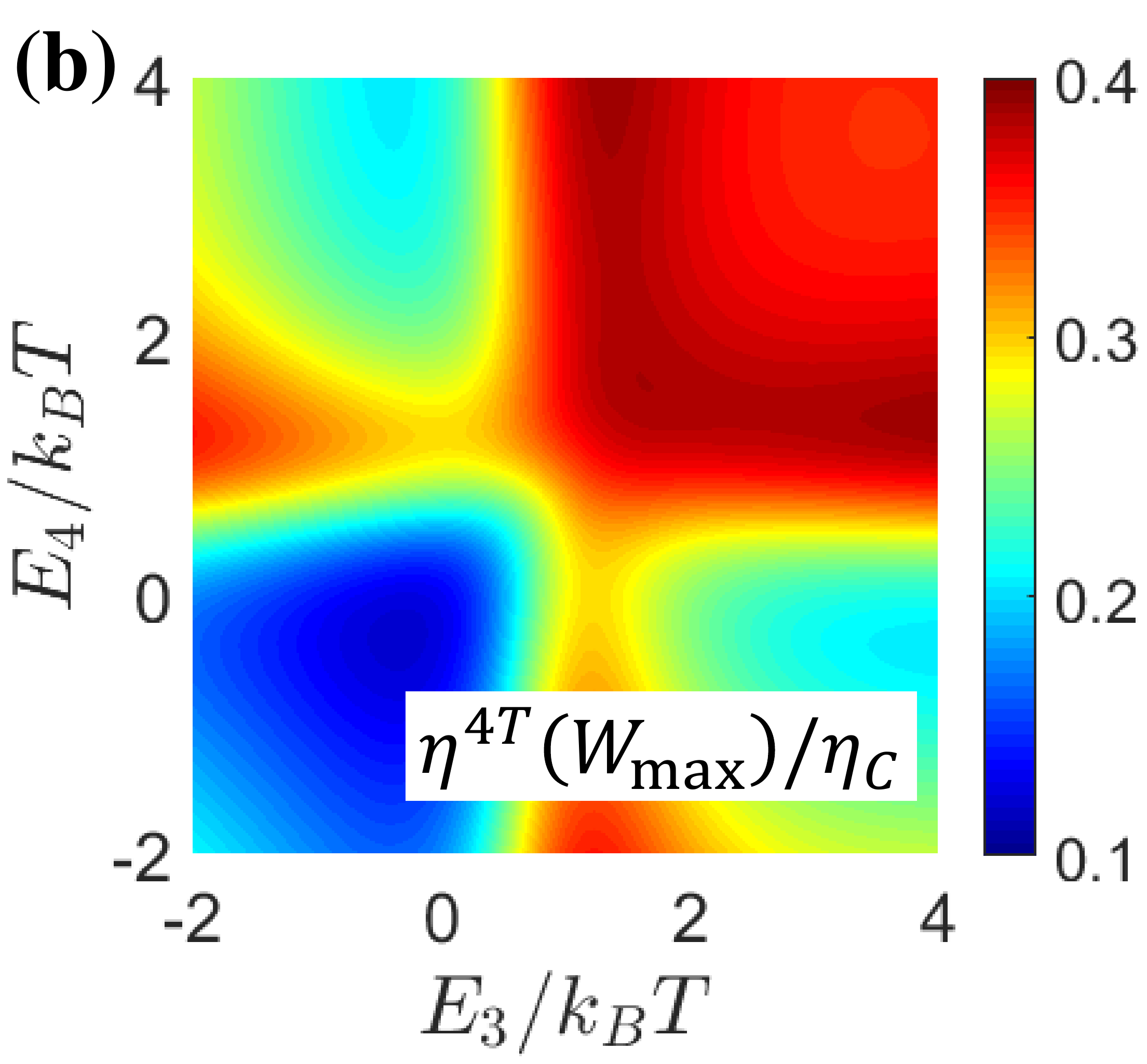}
\centering \includegraphics[width=4.2cm]{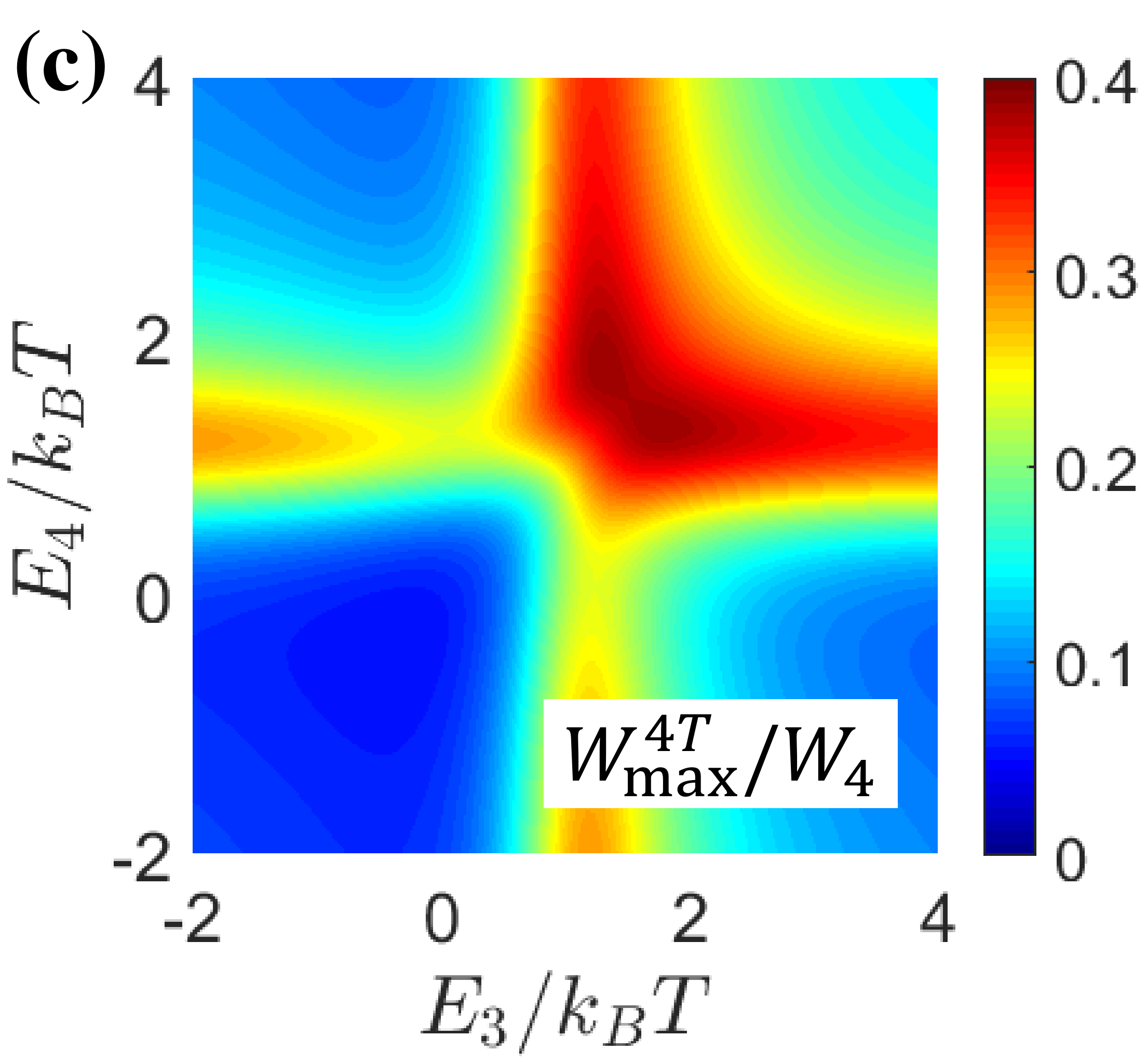}\hspace{0.2cm}\includegraphics[width=4.2cm]{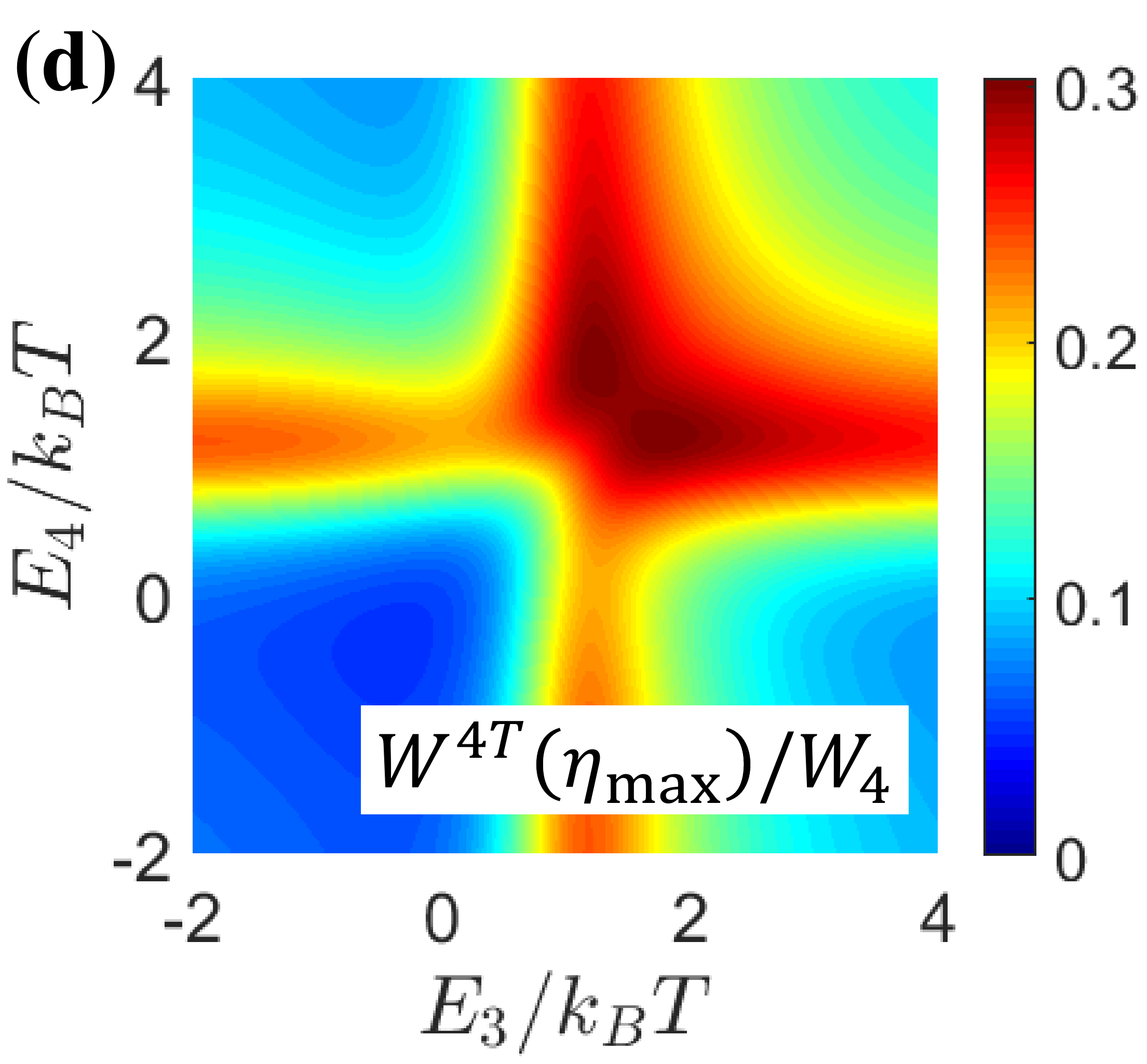}
\caption{The optimal efficiency and power of the four-terminal heat engine for various QD energies $E_3$ and $E_4$: (a) the maximum efficiency [Eq.~\eqref{eq:4T-eta}], (b) the efficiency at maximum output power [Eq.~\eqref{eq:4T-etaW}], (c) the maximum output power [Eq.~\eqref{eq:4T-W}], and (d) the output power at maximum efficiency [Eq.~\eqref{eq:4T-Weta}]. The parameters are $t=-0.2k_BT$, $\mu=0$, $E_1=1.0k_BT$, $E_2=2.0k_BT$. } \label{fig:E3E4}
\end{center}
\end{figure}

\begin{figure}[htb]
\begin{center}
\centering\includegraphics[width=4.2cm]{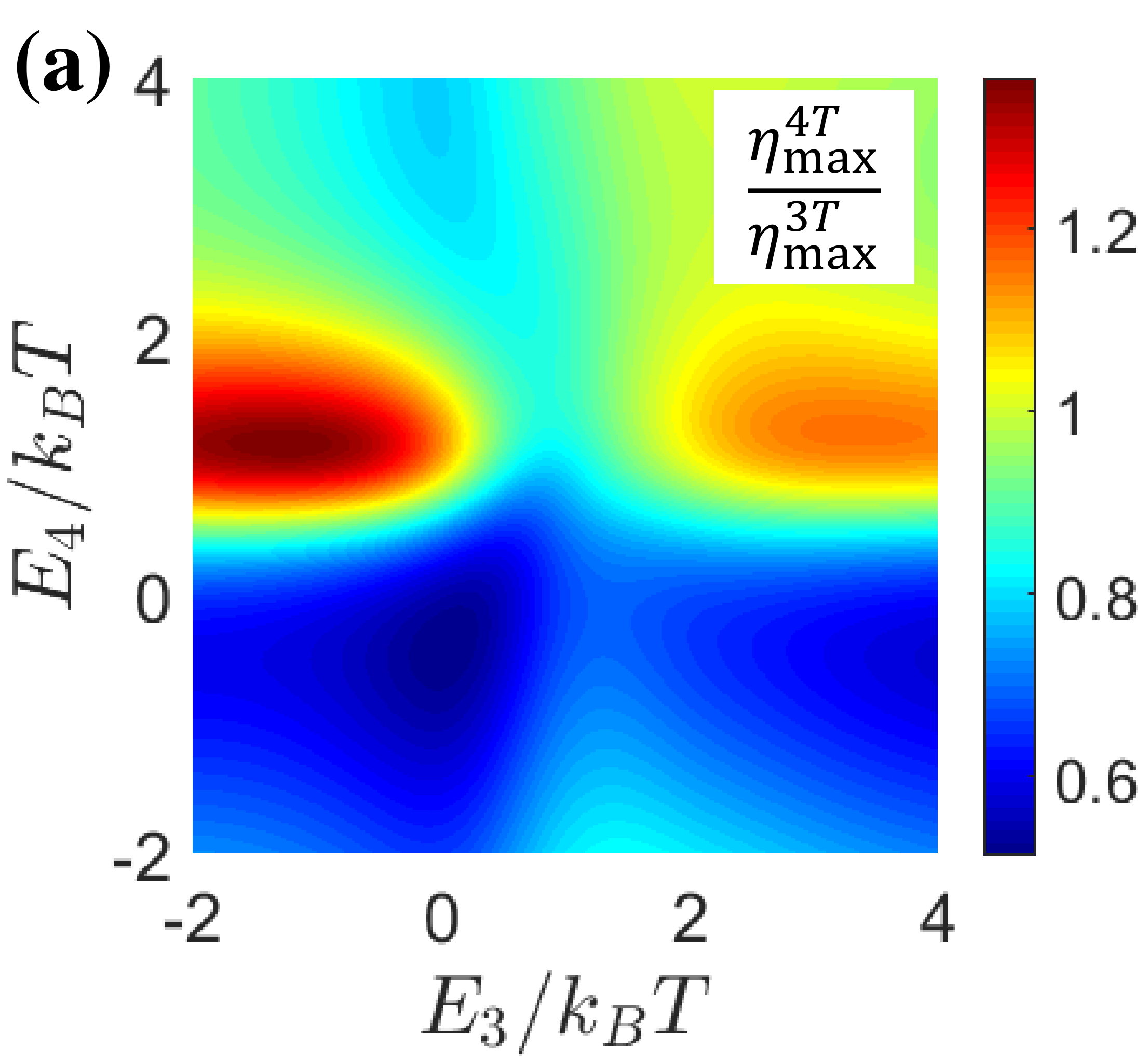}\hspace{0.2cm}\includegraphics[width=4.2cm]{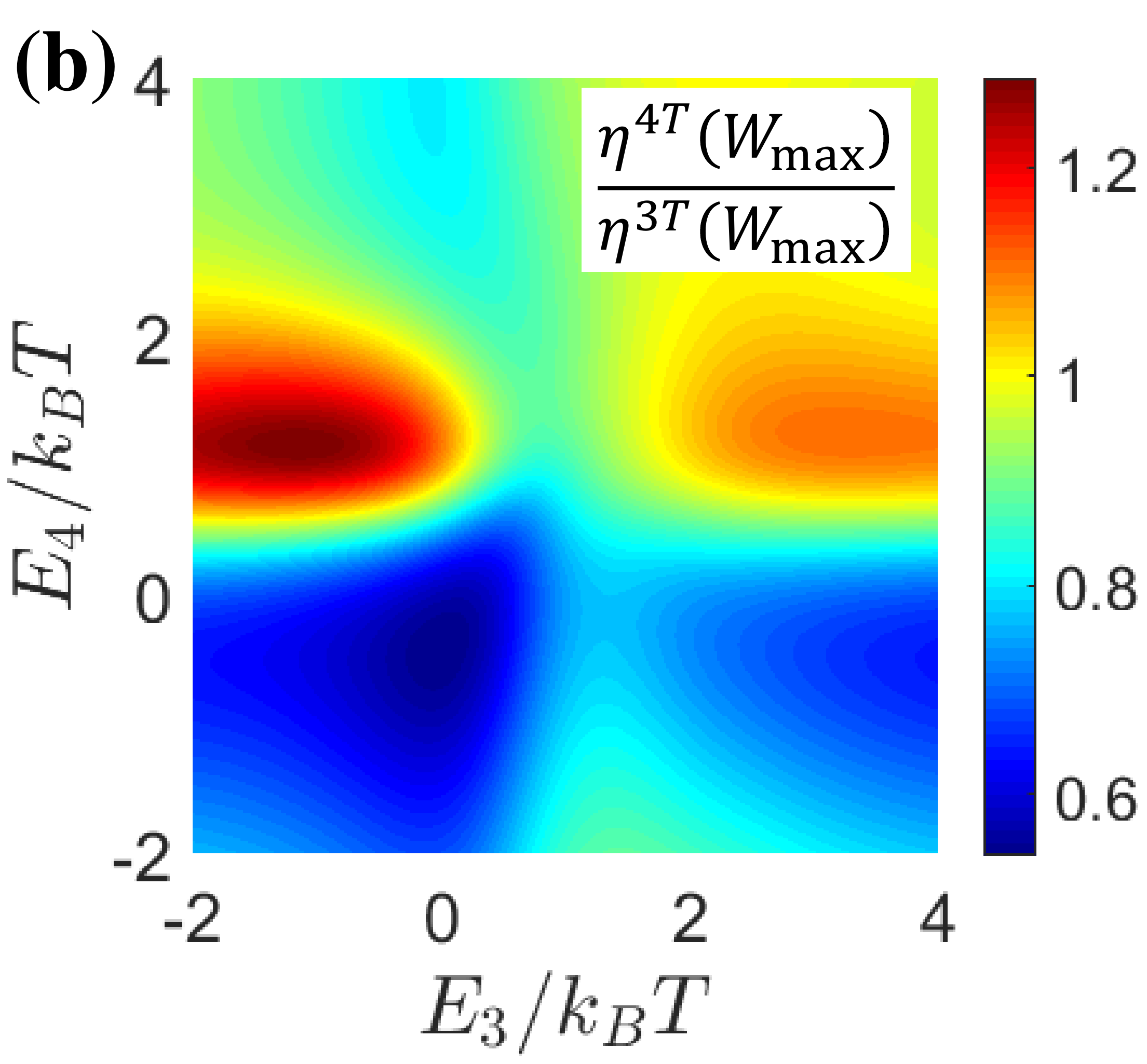}
\centering\includegraphics[width=4.2cm]{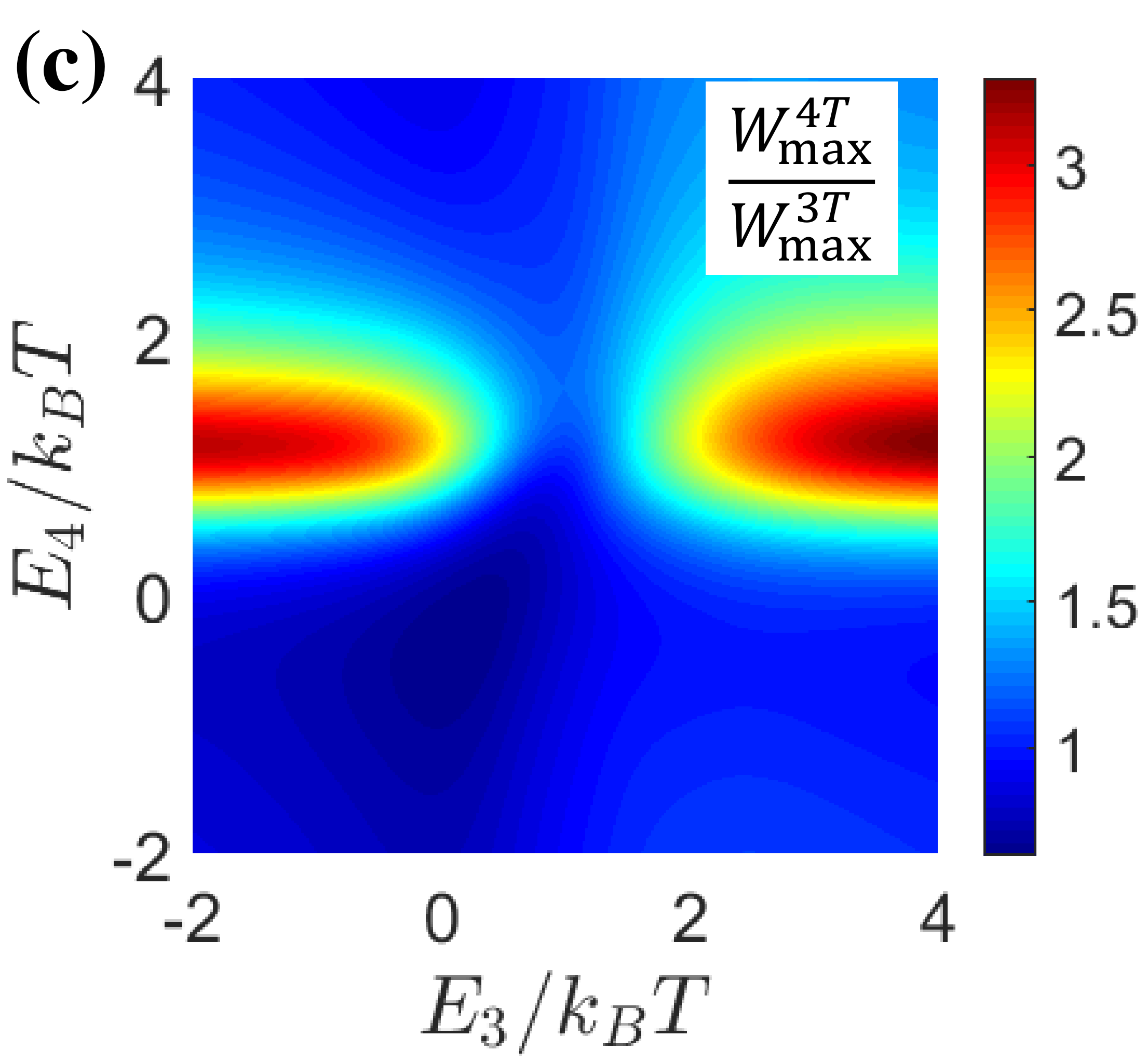}\hspace{0.2cm}\includegraphics[width=4.2cm]{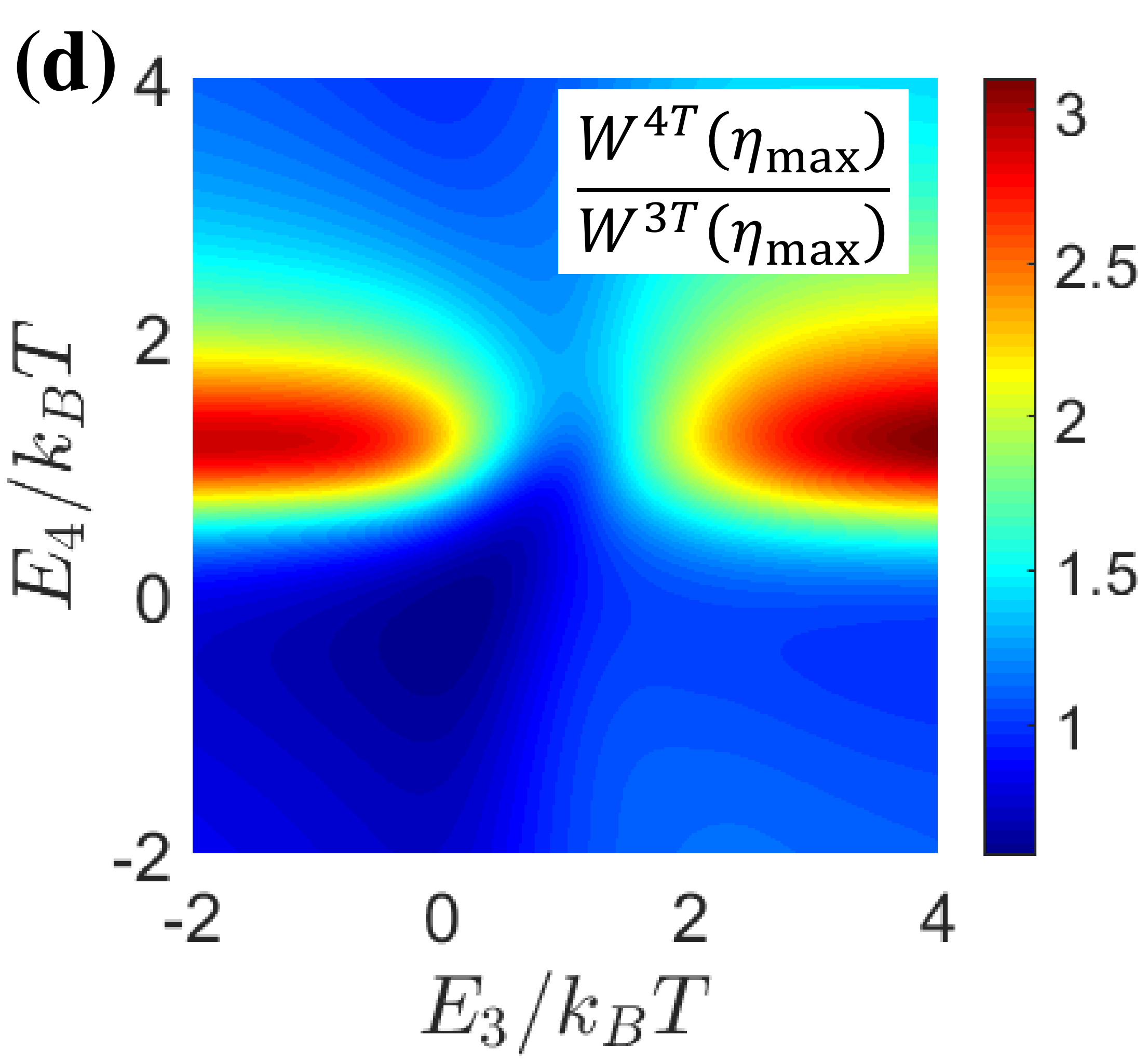}
\caption{Comparing the optimal efficiency and power for the four-terminal heat engine and the case of three-terminal heat engine for various QD energies $E_3$ and $E_4$: (a) the maximum efficiency, (b) the efficiency at maximum output power, (c) the maximum output power, and (d) the output power at maximum efficiency. The parameters are $t=-0.2k_BT$, $\mu=0$, $E_1=1.0k_BT$, $E_2=2.0k_BT$.}\label{fig:3T4TE3E4}
\end{center}
\end{figure}

In this section, we compare the optimal efficiency and output power of three-terminal thermoelectric heat engine with the four-terminal heat engine~\cite{Buttiker-4T,4T,Udo-MT,SanchezPhyE,Brandner4T}. As shown in Fig.~\ref{fig:4T}, we focus on this set-up where reservoir 1 is connected to the hot bath and the reservoirs 2, 3 and 4 are connected to the cold bath, respectively. Therefore, there are three independent output electric currents $I_e^i$ ($i=1,2,3$), whereas there is only one input heat current $I^{4T}_Q$ following from reservoir 1. In the linear-response regime, the currents and forces are described as~\cite{JiangPRE}
\begin{equation}
\begin{pmatrix}
I^1_e  \\
I^2_e  \\
I^3_e  \\
I^{4T}_Q  \\
\end{pmatrix}=
\begin{pmatrix}
M_{11} & M_{12} & M_{13} & M_{14}  \\
M_{12} & M_{22} & M_{23} & M_{24}  \\
M_{13} & M_{23} & M_{33} & M_{34}  \\
M_{14} & M_{24} & M_{34} & M_{44}  \\
\end{pmatrix}
\begin{pmatrix}
F^1_e  \\
F^2_e  \\
F^3_e  \\
F^{4T}_Q    \\
\end{pmatrix},
\end{equation}
where $I_e^i$ represents the electric current following from the electrode $i$, with $i=(1,2,3)$, the corresponding thermodynamic forces are $F_e^i=\frac{\mu_i-\mu_4}{e}$ and $F^{4T}_Q=\frac{T_1-T_4}{T_1}$. The expressions of currents and Onsager coefficients $M_{ij}$ ($i,j=1,2,3,4$) are are calculated in details in Appendix ~\ref{4T:coefficients}.

According to Ref.~[\onlinecite{trade-off}], the maximum energy efficiency can be expressed as
\begin{equation}
\eta^{4T}_{\max}=\frac{1-\sqrt{1-\Lambda}}{1+\sqrt{1-\Lambda}}\eta_C.
\label{eq:4T-eta}
\end{equation}
The maximum output power is
\begin{equation}
W^{4T}_{\max}=\Lambda W_4,\quad W_4=\frac{1}{4}M_{44}F_Q^2.
\label{eq:4T-W}
\end{equation}
Meanwhile, the efficiency at maximum output power is
\begin{equation}
\eta^{4T}(W_{\max})=\frac{\Lambda}{4-2\Lambda}\eta_C,
\label{eq:4T-etaW}
\end{equation}
and the output power at maximum efficiency is
\begin{equation}
W^{4T}(\eta_{\max})=\Lambda\left[1-\left(\frac{\eta_{\max}}{\eta_C}\right)^2\right]W_4.
\label{eq:4T-Weta}
\end{equation}
Here,
\begin{equation}
\Lambda=\hat{M}_{Qe}\hat{M}_{ee}^{-1}\hat{M}_{eQ}M_{QQ}^{-1},
\end{equation}
where $\hat{M}_{ee}$ denotes the $3\times3$ charge conductivity tensor, $\hat{M}_{Qe}$ is the $3\times1$ matrix, $\hat{M}_{eQ}$ is the $1\times3$ matrix, and $M_{QQ}$ represents the heat conductivity.

In Fig.~\ref{fig:E4} we plot the maximum efficiency and the efficiency at maximum output power as a function of QD energy $E_4$ and compare the thermoelectric performance of the four-terminal heat engine with the three-terminal heat engine in details. From Figs.~\ref{fig:E4}(a) and \ref{fig:E4}(c), we find that the both of the $\eta^{4T}_{\max}$ and $\eta^{4T}(W_{\max})$ reach their maximum values when $E_4\approx1.0k_BT$ and the maximum efficiency $\eta_{\max}$ achieves $0.4\eta_C$. Meanwhile, we also find that the efficiency of the four-terminal heat engine are larger than the efficiency of three-terminal system as shown in Figs.~\ref{fig:E4}(b) and \ref{fig:E4}(d) when $0<E_4<4k_BT$ where the four-terminal heat engine achieves its optimal performance. These results demonstrate concretely that multiple output current heat engine is more likely to achieve higher performance.

To be more careful on this conclusion, we present the maximum energy efficiency, the energy efficiency at maximum power, the maximum output power and output power at maximum efficiency as functions of $E_3$ and $E_4$. As shown in Fig.~\ref{fig:E3E4}, the maximum efficiency and the efficiency at maximum output power can be significantly improved for four-terminal set-up. In particular, both of the efficiency and power are symmetric around the line of $E_3=E_4$. The improvements are especially pronounced when $E_3>1k_BT$ and $E_4>1k_BT$, the energy efficiency achieved with four-terminal system is very high, $0.5\eta_C$. Moreover, the Fig.~\ref{fig:3T4TE3E4} shows the ratios between the two systems can be significantly improved through the QD energies $E_3$ and $E_4$, there are hot-spots for the ratios to be considerably larger than 1, particularly when $0.5k_BT<E_4<1.5k_BT$ and $E_3<0$ as well as $E_3>2k_BT$. Our analysis demonstrates the quantum heat engine with multiple output electric currents have superior efficiency and power for a large range of parameters.

Here we continue to address the practical problem of connecting the four-terminal thermoelectric heat engine to a resistor circuit. According to Ref.~[\onlinecite{JiangPRE}], for the four-terminal thermoelectric heat engine, we use a triangular resistor circuit to receive the electric power to achieve the best performance, as schematically shown in Fig.~\ref{fig:4T}. The current-force response matrix for the resistor circuit is
\begin{equation}
\vec{J}=\hat{M}^{4T} \vec{F}_e,
~\label{eq:R1}
\end{equation}
where $\vec{J}=(J_1,J_2,J_3)^T$ and $\vec{F}^{4T}_e=(F_e^1,F_e^2,F_e^3)^T$. $I_i$ ($i=1,2,3$) are electric currents following from $i$ electrodes into the resistor circuit, respectively. The electric currents following through the resistors $R_i$ are given by
\begin{equation}
\begin{aligned}
J_1 &= \frac{\mu_1-\mu_2}{eR_1}=\frac{F_e^1-F_e^2}{R_1}, \\
J_2 &= \frac{\mu_2-\mu_3}{eR_2}=\frac{F_e^2-F_e^3}{R_2}, \\
J_3 &= \frac{\mu_3-\mu_4}{eR_3}=\frac{F_e^3}{R_3}, \\
J_4 &= \frac{\mu_4-\mu_1}{eR_3}=-\frac{F_e^1}{R_4}.
\end{aligned}~\label{eq:R2}
\end{equation}
On one hand, using Kirchhoff's current law for the re- sistor circuit, we obtain
\begin{equation}
\begin{aligned}
I_1^{\prime}-J_1+J_4=0, \\
I_2^{\prime}+J_1-J_2=0, \\
I_3^{\prime}+J_2-J_3=0.
\end{aligned} ~\label{eq:R3}
\end{equation}
Combining Eqs.~\eqref{eq:R1},~\eqref{eq:R2} and \eqref{eq:R3}, we arrive at the expression of $\hat{M}^{4T}_L$,
\begin{equation}
\hat{M}^{4T}=
\begin{pmatrix}
\frac{1}{R_1}+\frac{1}{R_4} & -\frac{1}{R_1} &  0\\
-\frac{1}{R_1} & \frac{1}{R_1}+\frac{1}{R_2} & -\frac{1}{R_2}\\
0 & -\frac{1}{R_2} & \frac{1}{R_2}+\frac{1}{R_3}\\
\end{pmatrix}.
\end{equation}

On the other hand, the Kirchhoff's current law requires that
\begin{equation}
\vec{J} + \vec{I}^{4T}_{e} = 0.
\end{equation}
After some calculation, we obtain
\begin{equation}
\vec{F}^{4T}_e = -(\hat{M}_{ee}+\hat{M}^{4T})^{-1}\hat{M}_{Qe}F^{4T}_Q.
\end{equation}
The power consumed by the resistor circuit is
\begin{equation}
\begin{aligned}
&W^{4T}=-(\vec{F}^{4T}_e)^T \hat{M}^{4T} \vec{F}^{4T}_e \\
&= -\hat{M}_{Qe}(\hat{M}_{ee}+\hat{M}^{4T})^{-1}\hat{M}_L(\hat{M}_{ee}+\hat{M}^{4T})^{-1}\hat{M}_{Qe}(F^{4T}_Q)^2.
\end{aligned}
\end{equation}
The input heat currents is
\begin{equation}
I^{4T}_{Q} = [-\hat{M}_{eQ}(\hat{M}_{ee}+\hat{M}^{4T})^{-1}\hat{M}_{Qe} + {M}_{QQ}] F^{4T}_Q.
\end{equation}
The energy efficiency of this four-terminal heat engine is then given by $\eta^{4T}=W^{4T}/I^{4T}_Q$. By varying $\hat{M}^{4T}_L$, we can find that the maximum output power is reached at
\begin{equation}
\hat{M}^{4T}=\hat{M}_{ee},
\end{equation}
and the maximum energy efficiency is reached at
\begin{equation}
\hat{M}^{4T}=\sqrt{1-\Lambda}\hat{M}_{ee},
\end{equation}
The energy efficiency and output power for these two conditions are the consistent with Eqs.~\eqref{eq:4T-eta} and \eqref{eq:4T-W}.

\section{Conclusion}\label{conclusion}

In this paper, we have shown that cooperative effects can be an effective way to improve the energy efficiency and output power for the multi-terminal quantum-dot thermoelectric heat engines with multiple output electric currents. Each pair of terminals (including a hot terminal and a cold terminal) yields a thermoelectric effect. Through the calculation of the thermoelectric transport coefficients using the Landauer-B\"{u}tiker formalism, we found that both the efficiency and power can be considerably improved by the cooperative thermoelectric effect compared with using two thermoelectric effects independently. These enhancements are as effective for good thermoelectrics as that for bad thermoelectrics. Therefore, the region of physical parameters with high thermoelectric performance is considerably increased by thermoelectric effects.

After the general analysis of the transport processes in the four-terminal set-up with three output electric current and only one heat current, we investigated the optimal consequence of comparison between the four-terminal thermoelectric heat engine and the three-terminal one. We found that more output currents can improve the performance of quantum heat engine in a certain range of parameters. We demonstrated that quantum heat engine with multiple output electric currents provide superior efficiency and power.

Our results offer useful guidelines in the understanding of optimal behaviors of multiple-terminal heat engine in the linear response regime. We demonstrated that the setup can substantially enlarge the parameters region with high efficiency and power, and it thus provides an alternative route to high performance thermoelectric systems. However, the non-interacting systems which requires to go beyond the linear response regime, are not covered by our analysis. Nonlinear effects may yield interesting effects which deserve future studies.

\section*{Acknowledgment}
Y.L., J.L., R.W., and J.-H.J. acknowledge support from the National Natural Science Foundation of China (NSFC Grant No. 11675116), the Jiangsu distinguished professor funding and a Project Funded by the Priority Academic Program Development of Jiangsu Higher Education Institutions (PAPD). C.W. is supported by the National Natural Science Foundation of China under Grant No. 11704093.

\appendix

\section{Calculation of the Onsager linear-response coefficients for three-terminal model}  \label{coefficients}
For a three-terminal configuration, as in previous sections, the Onsager's coefficients are obtained from the linear expansion of the electronic currents $I_e^i$ and heat current $I_Q^i$ ($i=L,P$) given by Eqs.~\eqref{eq:Ie} and \eqref{eq:IQ}.
The Onsager coefficients $M_{ij}$ can be written in terms of the transmission function ${\mathcal T}_{ij}(E), i,j\in\{L,R,P\}$ as ~\cite{Sivan,butcher1990},
\begin{equation}
M_{11}=\frac{2e^2}{hk_BT}\int_{-\infty}^\infty dE \sum_{j\ne L} {\mathcal T}_{Lj}(E) F(E),\label{eq:M11}
\end{equation}
\begin{equation}
M_{12}=-\frac{2e^2}{hk_BT}\int_{-\infty}^\infty dE {\mathcal T}_{LP}(E) F(E),
\end{equation}
\begin{equation}
M_{13}= \frac{2e}{hk_BT}\int_{-\infty}^\infty dE (E-\mu) \sum_{j\ne L} {\mathcal T}_{Lj}(E) F(E),
\end{equation}
\begin{equation}
M_{22}=\frac{2e^2}{hk_BT}\int_{-\infty}^\infty dE \sum_{j\ne P} {\mathcal T}_{Pj}(E) F(E),
\end{equation}
\begin{equation}
M_{23}=-\frac{2e}{hk_BT}\int_{-\infty}^\infty dE(E-\mu){\mathcal T}_{PL}(E) F(E),
\end{equation}
\begin{equation}
M_{33}=\frac{2}{hk_BT}\int_{-\infty}^\infty dE (E-\mu)^2\sum_{j\ne L} {\mathcal T}_{Lj}(E) F(E).\label{eq:M33}
\end{equation}
where $F(E)\equiv\{4\cosh^2[(E-\mu)/k_BT]\}^{-1}$.

\section{Calculations of the expression of the currents and the linear-response coefficients for four-terminal system}  \label{4T:coefficients}
The system Hamiltonian for the four-terminal heat engine reads
\begin{align}
\hat{H}_{\rm QD}=\sum_{i=1,2,3,4} E_i d_i^\dagger d_i,
\end{align}
The transmission probability is given by
\begin{equation}
{\mathcal T}_{ij}={\rm Tr}[\Gamma_i(E)G(E)\Gamma_j(E)G^\dagger(E)],
\end{equation}
where the (retarded) system Green function $G(E)\equiv [E-H_{\rm QD}-i\Gamma/2]^{-1}$, the dot-lead coupling $\Gamma_i$ is assumed to be a constant for all three electrode.



The charge and heat currents following from the left reservoir are given by
{\small{
\begin{align}
I^i_e = \frac{2e}{h}\int_{-\infty}^{\infty}dE\sum_{i\ne j}{\mathcal T}_{ij}(E)[f_j(E)-f_i(E)],
\end{align}
\begin{align}
I_Q = \frac{2}{h}\int_{-\infty}^{\infty}dE(E-\mu_1)\sum_{j}{\mathcal T}_{j1}(E)[f_1(E)-f_j(E)].
\end{align}}}
The Onsager coefficients $M_{ij}$ can be written in terms of the transmission function ${\mathcal T}_{ij}(E)$ as,
\begin{equation}
M_{11}=\frac{2e^2}{hk_BT}\int_{-\infty}^\infty dE \sum_{j\ne 1} {\mathcal T}_{1j}(E) F(E),
\end{equation}
\begin{equation}
M_{12}=-\frac{2e^2}{hk_BT}\int_{-\infty}^\infty dE {\mathcal T}_{12}(E) F(E),
\end{equation}
\begin{equation}
M_{13}=-\frac{2e^2}{hk_BT}\int_{-\infty}^\infty dE {\mathcal T}_{13}(E) F(E),
\end{equation}
\begin{equation}
M_{14}=\frac{2e}{hk_BT}\int_{-\infty}^\infty dE \sum_{j\ne 1}(E-\mu) {\mathcal T}_{1k}(E) F(E);
\end{equation}
\begin{equation}
M_{22}=\frac{2e^2}{hk_BT}\int_{-\infty}^\infty dE \sum_{j\ne 2} {\mathcal T}_{2j}(E) F(E),
\end{equation}
\begin{equation}
M_{23}=-\frac{2e^2}{hk_BT}\int_{-\infty}^\infty dE {\mathcal T}_{23}(E) F(E),
\end{equation}
\begin{equation}
M_{24}=\frac{2e}{hk_BT}\int_{-\infty}^\infty dE (E-\mu) {\mathcal T}_{21}(E) F(E);
\end{equation}
\begin{equation}
M_{33}=\frac{2e^2}{hk_BT}\int_{-\infty}^\infty dE \sum_{j\ne 3}{\mathcal T}_{3j}(E) F(E),
\end{equation}
\begin{equation}
M_{34}=-\frac{2e}{hk_BT}\int_{-\infty}^\infty dE(E-\mu){\mathcal T}_{31}(E) F(E);
\end{equation}
\begin{equation}
M_{44}=\frac{2}{hk_BT}\int_{-\infty}^\infty dE(E-\mu)^2{\mathcal T}_{j\ne1}(E) F(E),
\end{equation}
where $F(E)\equiv\{4\cosh^2[(E-\mu)/k_BT]\}^{-1}$.

\bibliography{Ref_BTRS}

\begin{thebibliography}{69}%
\makeatletter
\providecommand \@ifxundefined [1]{%
 \@ifx{#1\undefined}
}%
\providecommand \@ifnum [1]{%
 \ifnum #1\expandafter \@firstoftwo
 \else \expandafter \@secondoftwo
 \fi
}%
\providecommand \@ifx [1]{%
 \ifx #1\expandafter \@firstoftwo
 \else \expandafter \@secondoftwo
 \fi
}%
\providecommand \natexlab [1]{#1}%
\providecommand \enquote  [1]{``#1''}%
\providecommand \bibnamefont  [1]{#1}%
\providecommand \bibfnamefont [1]{#1}%
\providecommand \citenamefont [1]{#1}%
\providecommand \href@noop [0]{\@secondoftwo}%
\providecommand \href [0]{\begingroup \@sanitize@url \@href}%
\providecommand \@href[1]{\@@startlink{#1}\@@href}%
\providecommand \@@href[1]{\endgroup#1\@@endlink}%
\providecommand \@sanitize@url [0]{\catcode `\\12\catcode `\$12\catcode
  `\&12\catcode `\#12\catcode `\^12\catcode `\_12\catcode `\%12\relax}%
\providecommand \@@startlink[1]{}%
\providecommand \@@endlink[0]{}%
\providecommand \url  [0]{\begingroup\@sanitize@url \@url }%
\providecommand \@url [1]{\endgroup\@href {#1}{\urlprefix }}%
\providecommand \urlprefix  [0]{URL }%
\providecommand \Eprint [0]{\href }%
\providecommand \doibase [0]{http://dx.doi.org/}%
\providecommand \selectlanguage [0]{\@gobble}%
\providecommand \bibinfo  [0]{\@secondoftwo}%
\providecommand \bibfield  [0]{\@secondoftwo}%
\providecommand \translation [1]{[#1]}%
\providecommand \BibitemOpen [0]{}%
\providecommand \bibitemStop [0]{}%
\providecommand \bibitemNoStop [0]{.\EOS\space}%
\providecommand \EOS [0]{\spacefactor3000\relax}%
\providecommand \BibitemShut  [1]{\csname bibitem#1\endcsname}%
\let\auto@bib@innerbib\@empty
\bibitem [{\citenamefont {Dubi}\ and\ \citenamefont
  {Di~Ventra}(2011)}]{DubiRMP}%
  \BibitemOpen
  \bibfield  {author} {\bibinfo {author} {\bibfnamefont {Y.}~\bibnamefont
  {Dubi}}\ and\ \bibinfo {author} {\bibfnamefont {M.}~\bibnamefont
  {Di~Ventra}},\ }\bibfield  {title} {\enquote {\bibinfo {title} {Heat flow and
  thermoelectricity in atomic and molecular junctions},}\ }\href {\doibase
  10.1103/RevModPhys.83.131} {\bibfield  {journal} {\bibinfo  {journal} {Rev.
  Mod. Phys.}\ }\textbf {\bibinfo {volume} {83}},\ \bibinfo {pages} {131--155}
  (\bibinfo {year} {2011})}\BibitemShut {NoStop}%
\bibitem [{\citenamefont {Li}\ \emph {et~al.}(2012)\citenamefont {Li},
  \citenamefont {Ren}, \citenamefont {Wang}, \citenamefont {Zhang},
  \citenamefont {H\"anggi},\ and\ \citenamefont {Li}}]{RenRMP}%
  \BibitemOpen
  \bibfield  {author} {\bibinfo {author} {\bibfnamefont {N.}~\bibnamefont
  {Li}}, \bibinfo {author} {\bibfnamefont {J.}~\bibnamefont {Ren}}, \bibinfo
  {author} {\bibfnamefont {L.}~\bibnamefont {Wang}}, \bibinfo {author}
  {\bibfnamefont {G.}~\bibnamefont {Zhang}}, \bibinfo {author} {\bibfnamefont
  {P.}~\bibnamefont {H\"anggi}}, \ and\ \bibinfo {author} {\bibfnamefont
  {B.}~\bibnamefont {Li}},\ }\bibfield  {title} {\enquote {\bibinfo {title}
  {Phononics: Manipulating heat flow with electronic analogs and beyond},}\
  }\href {\doibase 10.1103/RevModPhys.84.1045} {\bibfield  {journal} {\bibinfo
  {journal} {Rev. Mod. Phys.}\ }\textbf {\bibinfo {volume} {84}},\ \bibinfo
  {pages} {1045--1066} (\bibinfo {year} {2012})}\BibitemShut {NoStop}%
\bibitem [{\citenamefont {Sothmann}\ \emph {et~al.}(2015)\citenamefont
  {Sothmann}, \citenamefont {S{\'a}nchez},\ and\ \citenamefont
  {Jordan}}]{Nanotechnology}%
  \BibitemOpen
  \bibfield  {author} {\bibinfo {author} {\bibfnamefont {B.}~\bibnamefont
  {Sothmann}}, \bibinfo {author} {\bibfnamefont {R.}~\bibnamefont
  {S{\'a}nchez}}, \ and\ \bibinfo {author} {\bibfnamefont {A.~N}\ \bibnamefont
  {Jordan}},\ }\bibfield  {title} {\enquote {\bibinfo {title} {Thermoelectric
  energy harvesting with quantum dots},}\ }\href
  {http://stacks.iop.org/0957-4484/26/i=3/a=032001} {\bibfield  {journal}
  {\bibinfo  {journal} {Nanotechnology}\ }\textbf {\bibinfo {volume} {26}},\
  \bibinfo {pages} {032001} (\bibinfo {year} {2015})}\BibitemShut {NoStop}%
\bibitem [{\citenamefont {Jiang}\ and\ \citenamefont {Imry}(2016)}]{JiangCRP}%
  \BibitemOpen
  \bibfield  {author} {\bibinfo {author} {\bibfnamefont {J.-H.}\ \bibnamefont
  {Jiang}}\ and\ \bibinfo {author} {\bibfnamefont {Y.}~\bibnamefont {Imry}},\
  }\bibfield  {title} {\enquote {\bibinfo {title} {Linear and nonlinear
  mesoscopic thermoelectric transport with coupling with heat baths},}\ }\href
  {\doibase https://doi.org/10.1016/j.crhy.2016.08.006} {\bibfield  {journal}
  {\bibinfo  {journal} {C. R. Phys.}\ }\textbf {\bibinfo {volume} {17}},\
  \bibinfo {pages} {1047 -- 1059} (\bibinfo {year} {2016})}\BibitemShut
  {NoStop}%
\bibitem [{\citenamefont {Thierschmann}\ \emph {et~al.}(2016)\citenamefont
  {Thierschmann}, \citenamefont {S{\'{a}}nchez}, \citenamefont {Sothmann},
  \citenamefont {Buhmann},\ and\ \citenamefont {Molenkamp}}]{SanchezCPR}%
  \BibitemOpen
  \bibfield  {author} {\bibinfo {author} {\bibfnamefont {H.}~\bibnamefont
  {Thierschmann}}, \bibinfo {author} {\bibfnamefont {R.}~\bibnamefont
  {S{\'{a}}nchez}}, \bibinfo {author} {\bibfnamefont {B.}~\bibnamefont
  {Sothmann}}, \bibinfo {author} {\bibfnamefont {H.}~\bibnamefont {Buhmann}}, \
  and\ \bibinfo {author} {\bibfnamefont {L.~W.}\ \bibnamefont {Molenkamp}},\
  }\bibfield  {title} {\enquote {\bibinfo {title} {Thermoelectrics with
  coulomb-coupled quantum dots},}\ }\href {\doibase
  https://doi.org/10.1016/j.crhy.2016.08.001} {\bibfield  {journal} {\bibinfo
  {journal} {C. R. Phys.}\ }\textbf {\bibinfo {volume} {17}},\ \bibinfo {pages}
  {1109 -- 1122} (\bibinfo {year} {2016})}\BibitemShut {NoStop}%
\bibitem [{\citenamefont {Benenti}\ \emph {et~al.}(2017)\citenamefont
  {Benenti}, \citenamefont {Casati}, \citenamefont {Saito},\ and\ \citenamefont
  {Whitney}}]{PhyRep}%
  \BibitemOpen
  \bibfield  {author} {\bibinfo {author} {\bibfnamefont {G.}~\bibnamefont
  {Benenti}}, \bibinfo {author} {\bibfnamefont {G.}~\bibnamefont {Casati}},
  \bibinfo {author} {\bibfnamefont {K.}~\bibnamefont {Saito}}, \ and\ \bibinfo
  {author} {\bibfnamefont {R.~S.}\ \bibnamefont {Whitney}},\ }\bibfield
  {title} {\enquote {\bibinfo {title} {Fundamental aspects of steady-state
  conversion of heat to work at the nanoscale},}\ }\href {\doibase
  10.1016/j.physrep.2017.05.008} {\bibfield  {journal} {\bibinfo  {journal}
  {Phys. Rep.}\ }\textbf {\bibinfo {volume} {694}},\ \bibinfo {pages} {1 --
  124} (\bibinfo {year} {2017})}\BibitemShut {NoStop}%
\bibitem [{\citenamefont {S\'anchez}\ and\ \citenamefont
  {Serra}(2011)}]{David2011PRB}%
  \BibitemOpen
  \bibfield  {author} {\bibinfo {author} {\bibfnamefont {D.}~\bibnamefont
  {S\'anchez}}\ and\ \bibinfo {author} {\bibfnamefont {L.}~\bibnamefont
  {Serra}},\ }\bibfield  {title} {\enquote {\bibinfo {title} {Thermoelectric
  transport of mesoscopic conductors coupled to voltage and thermal probes},}\
  }\href {\doibase 10.1103/PhysRevB.84.201307} {\bibfield  {journal} {\bibinfo
  {journal} {Phys. Rev. B}\ }\textbf {\bibinfo {volume} {84}},\ \bibinfo
  {pages} {201307} (\bibinfo {year} {2011})}\BibitemShut {NoStop}%
\bibitem [{\citenamefont {Sothmann}\ \emph {et~al.}(2013)\citenamefont
  {Sothmann}, \citenamefont {S{\'a}nchez}, \citenamefont {Jordan},\ and\
  \citenamefont {B{\"u}ttiker}}]{Sothmann-QW}%
  \BibitemOpen
  \bibfield  {author} {\bibinfo {author} {\bibfnamefont {B.}~\bibnamefont
  {Sothmann}}, \bibinfo {author} {\bibfnamefont {R.}~\bibnamefont
  {S{\'a}nchez}}, \bibinfo {author} {\bibfnamefont {A.~N}\ \bibnamefont
  {Jordan}}, \ and\ \bibinfo {author} {\bibfnamefont {M.}~\bibnamefont
  {B{\"u}ttiker}},\ }\bibfield  {title} {\enquote {\bibinfo {title} {Powerful
  energy harvester based on resonant-tunneling quantum wells},}\ }\href
  {http://stacks.iop.org/1367-2630/15/i=9/a=095021} {\bibfield  {journal}
  {\bibinfo  {journal} {New J. Phys.}\ }\textbf {\bibinfo {volume} {15}},\
  \bibinfo {pages} {095021} (\bibinfo {year} {2013})}\BibitemShut {NoStop}%
\bibitem [{\citenamefont {Jiang}\ \emph
  {et~al.}(2013{\natexlab{a}})\citenamefont {Jiang}, \citenamefont
  {Entin-Wohlman},\ and\ \citenamefont {Imry}}]{Jiang2013}%
  \BibitemOpen
  \bibfield  {author} {\bibinfo {author} {\bibfnamefont {J.-H.}\ \bibnamefont
  {Jiang}}, \bibinfo {author} {\bibfnamefont {O.}~\bibnamefont
  {Entin-Wohlman}}, \ and\ \bibinfo {author} {\bibfnamefont {Y.}~\bibnamefont
  {Imry}},\ }\bibfield  {title} {\enquote {\bibinfo {title} {Hopping
  thermoelectric transport in finite systems: Boundary effects},}\ }\href
  {\doibase 10.1103/PhysRevB.87.205420} {\bibfield  {journal} {\bibinfo
  {journal} {Phys. Rev. B}\ }\textbf {\bibinfo {volume} {87}},\ \bibinfo
  {pages} {205420} (\bibinfo {year} {2013}{\natexlab{a}})}\BibitemShut
  {NoStop}%
\bibitem [{\citenamefont {Jiang}\ \emph
  {et~al.}(2013{\natexlab{b}})\citenamefont {Jiang}, \citenamefont
  {Entin-Wohlman},\ and\ \citenamefont {Imry}}]{JiangNJP}%
  \BibitemOpen
  \bibfield  {author} {\bibinfo {author} {\bibfnamefont {J.-H.}\ \bibnamefont
  {Jiang}}, \bibinfo {author} {\bibfnamefont {O.}~\bibnamefont
  {Entin-Wohlman}}, \ and\ \bibinfo {author} {\bibfnamefont {Y.}~\bibnamefont
  {Imry}},\ }\bibfield  {title} {\enquote {\bibinfo {title} {Three-terminal
  semiconductor junction thermoelectric devices: improving performance},}\
  }\href {http://stacks.iop.org/1367-2630/15/i=7/a=075021} {\bibfield
  {journal} {\bibinfo  {journal} {New J. Phys.}\ }\textbf {\bibinfo {volume}
  {15}},\ \bibinfo {pages} {075021} (\bibinfo {year}
  {2013}{\natexlab{b}})}\BibitemShut {NoStop}%
\bibitem [{\citenamefont {Jiang}\ \emph
  {et~al.}(2015{\natexlab{a}})\citenamefont {Jiang}, \citenamefont {Kulkarni},
  \citenamefont {Segal},\ and\ \citenamefont {Imry}}]{Jiangtransistors}%
  \BibitemOpen
  \bibfield  {author} {\bibinfo {author} {\bibfnamefont {J.-H.}\ \bibnamefont
  {Jiang}}, \bibinfo {author} {\bibfnamefont {M.}~\bibnamefont {Kulkarni}},
  \bibinfo {author} {\bibfnamefont {D.}~\bibnamefont {Segal}}, \ and\ \bibinfo
  {author} {\bibfnamefont {Y.}~\bibnamefont {Imry}},\ }\bibfield  {title}
  {\enquote {\bibinfo {title} {Phonon thermoelectric transistors and
  rectifiers},}\ }\href {\doibase 10.1103/PhysRevB.92.045309} {\bibfield
  {journal} {\bibinfo  {journal} {Phys. Rev. B}\ }\textbf {\bibinfo {volume}
  {92}},\ \bibinfo {pages} {045309} (\bibinfo {year}
  {2015}{\natexlab{a}})}\BibitemShut {NoStop}%
\bibitem [{\citenamefont {Lu}\ \emph {et~al.}(2019{\natexlab{a}})\citenamefont
  {Lu}, \citenamefont {Wang}, \citenamefont {Ren}, \citenamefont {Kulkarni},\
  and\ \citenamefont {Jiang}}]{MyPRB}%
  \BibitemOpen
  \bibfield  {author} {\bibinfo {author} {\bibfnamefont {J.}~\bibnamefont
  {Lu}}, \bibinfo {author} {\bibfnamefont {R.}~\bibnamefont {Wang}}, \bibinfo
  {author} {\bibfnamefont {J.}~\bibnamefont {Ren}}, \bibinfo {author}
  {\bibfnamefont {M.}~\bibnamefont {Kulkarni}}, \ and\ \bibinfo {author}
  {\bibfnamefont {J.-H.}\ \bibnamefont {Jiang}},\ }\bibfield  {title} {\enquote
  {\bibinfo {title} {Quantum-dot circuit-qed thermoelectric diodes and
  transistors},}\ }\href {\doibase 10.1103/PhysRevB.99.035129} {\bibfield
  {journal} {\bibinfo  {journal} {Phys. Rev. B}\ }\textbf {\bibinfo {volume}
  {99}},\ \bibinfo {pages} {035129} (\bibinfo {year}
  {2019}{\natexlab{a}})}\BibitemShut {NoStop}%
\bibitem [{\citenamefont {Goury}\ and\ \citenamefont
  {S\'anchez}(2019)}]{SanchezAPL}%
  \BibitemOpen
  \bibfield  {author} {\bibinfo {author} {\bibfnamefont {D.}~\bibnamefont
  {Goury}}\ and\ \bibinfo {author} {\bibfnamefont {R.}~\bibnamefont
  {S\'anchez}},\ }\bibfield  {title} {\enquote {\bibinfo {title} {Reversible
  thermal diode and energy harvester with a superconducting quantum
  interference single-electron transistor},}\ }\href {\doibase
  10.1063/1.5109100} {\bibfield  {journal} {\bibinfo  {journal} {Appl. Phys.
  Lett.}\ }\textbf {\bibinfo {volume} {115}},\ \bibinfo {pages} {092601}
  (\bibinfo {year} {2019})}\BibitemShut {NoStop}%
\bibitem [{\citenamefont {Wang}\ \emph
  {et~al.}(2019{\natexlab{a}})\citenamefont {Wang}, \citenamefont {Xu},
  \citenamefont {Liu},\ and\ \citenamefont {Gao}}]{WangPRE}%
  \BibitemOpen
  \bibfield  {author} {\bibinfo {author} {\bibfnamefont {C.}~\bibnamefont
  {Wang}}, \bibinfo {author} {\bibfnamefont {D.}~\bibnamefont {Xu}}, \bibinfo
  {author} {\bibfnamefont {H.}~\bibnamefont {Liu}}, \ and\ \bibinfo {author}
  {\bibfnamefont {X.}~\bibnamefont {Gao}},\ }\bibfield  {title} {\enquote
  {\bibinfo {title} {Thermal rectification and heat amplification in a
  nonequilibrium v-type three-level system},}\ }\href {\doibase
  10.1103/PhysRevE.99.042102} {\bibfield  {journal} {\bibinfo  {journal} {Phys.
  Rev. E}\ }\textbf {\bibinfo {volume} {99}},\ \bibinfo {pages} {042102}
  (\bibinfo {year} {2019}{\natexlab{a}})}\BibitemShut {NoStop}%
\bibitem [{\citenamefont {Wang}\ \emph {et~al.}(2018)\citenamefont {Wang},
  \citenamefont {Lu}, \citenamefont {Wang},\ and\ \citenamefont
  {Jiang}}]{Rongqian}%
  \BibitemOpen
  \bibfield  {author} {\bibinfo {author} {\bibfnamefont {R.}~\bibnamefont
  {Wang}}, \bibinfo {author} {\bibfnamefont {J.}~\bibnamefont {Lu}}, \bibinfo
  {author} {\bibfnamefont {C.}~\bibnamefont {Wang}}, \ and\ \bibinfo {author}
  {\bibfnamefont {J.-H.}\ \bibnamefont {Jiang}},\ }\bibfield  {title} {\enquote
  {\bibinfo {title} {Nonlinear effects for three-terminal heat engine and
  refrigerator},}\ }\href {\doibase 10.1038/s41598-018-20757-8} {\bibfield
  {journal} {\bibinfo  {journal} {Sci. Rep.}\ }\textbf {\bibinfo {volume}
  {8}},\ \bibinfo {pages} {2607} (\bibinfo {year} {2018})}\BibitemShut
  {NoStop}%
\bibitem [{\citenamefont {S\'anchez}\ \emph {et~al.}(2019)\citenamefont
  {S\'anchez}, \citenamefont {S\'anchez}, \citenamefont {L\'opez},\ and\
  \citenamefont {Sothmann}}]{David-refrigerator}%
  \BibitemOpen
  \bibfield  {author} {\bibinfo {author} {\bibfnamefont {D.}~\bibnamefont
  {S\'anchez}}, \bibinfo {author} {\bibfnamefont {R.}~\bibnamefont
  {S\'anchez}}, \bibinfo {author} {\bibfnamefont {R.}~\bibnamefont {L\'opez}},
  \ and\ \bibinfo {author} {\bibfnamefont {B.}~\bibnamefont {Sothmann}},\
  }\bibfield  {title} {\enquote {\bibinfo {title} {Nonlinear chiral
  refrigerators},}\ }\href {\doibase 10.1103/PhysRevB.99.245304} {\bibfield
  {journal} {\bibinfo  {journal} {Phys. Rev. B}\ }\textbf {\bibinfo {volume}
  {99}},\ \bibinfo {pages} {245304} (\bibinfo {year} {2019})}\BibitemShut
  {NoStop}%
\bibitem [{\citenamefont {Simine}\ and\ \citenamefont
  {Segal}(2012)}]{Lena2012}%
  \BibitemOpen
  \bibfield  {author} {\bibinfo {author} {\bibfnamefont {L.}~\bibnamefont
  {Simine}}\ and\ \bibinfo {author} {\bibfnamefont {D.}~\bibnamefont {Segal}},\
  }\bibfield  {title} {\enquote {\bibinfo {title} {Vibrational cooling{,}
  heating{,} and instability in molecular conducting junctions: full counting
  statistics analysis},}\ }\href {\doibase 10.1039/C2CP40851A} {\bibfield
  {journal} {\bibinfo  {journal} {Phys. Chem. Chem. Phys.}\ }\textbf {\bibinfo
  {volume} {14}},\ \bibinfo {pages} {13820--13834} (\bibinfo {year}
  {2012})}\BibitemShut {NoStop}%
\bibitem [{\citenamefont {Jiang}\ and\ \citenamefont {John}(2014)}]{JiangPRX}%
  \BibitemOpen
  \bibfield  {author} {\bibinfo {author} {\bibfnamefont {J.-H.}\ \bibnamefont
  {Jiang}}\ and\ \bibinfo {author} {\bibfnamefont {S.}~\bibnamefont {John}},\
  }\bibfield  {title} {\enquote {\bibinfo {title} {Photonic crystal
  architecture for room-temperature equilibrium bose-einstein condensation of
  exciton polaritons},}\ }\href {\doibase 10.1103/PhysRevX.4.031025} {\bibfield
   {journal} {\bibinfo  {journal} {Phys. Rev. X}\ }\textbf {\bibinfo {volume}
  {4}},\ \bibinfo {pages} {031025} (\bibinfo {year} {2014})}\BibitemShut
  {NoStop}%
\bibitem [{\citenamefont {Entin-Wohlman}\ \emph {et~al.}(2010)\citenamefont
  {Entin-Wohlman}, \citenamefont {Imry},\ and\ \citenamefont
  {Aharony}}]{OraPRB2010}%
  \BibitemOpen
  \bibfield  {author} {\bibinfo {author} {\bibfnamefont {O.}~\bibnamefont
  {Entin-Wohlman}}, \bibinfo {author} {\bibfnamefont {Y.}~\bibnamefont {Imry}},
  \ and\ \bibinfo {author} {\bibfnamefont {A.}~\bibnamefont {Aharony}},\
  }\bibfield  {title} {\enquote {\bibinfo {title} {Three-terminal
  thermoelectric transport through a molecular junction},}\ }\href {\doibase
  10.1103/PhysRevB.82.115314} {\bibfield  {journal} {\bibinfo  {journal} {Phys.
  Rev. B}\ }\textbf {\bibinfo {volume} {82}},\ \bibinfo {pages} {115314}
  (\bibinfo {year} {2010})}\BibitemShut {NoStop}%
\bibitem [{\citenamefont {S\'anchez}\ and\ \citenamefont
  {B\"uttiker}(2011)}]{Rafael}%
  \BibitemOpen
  \bibfield  {author} {\bibinfo {author} {\bibfnamefont {R.}~\bibnamefont
  {S\'anchez}}\ and\ \bibinfo {author} {\bibfnamefont {M.}~\bibnamefont
  {B\"uttiker}},\ }\bibfield  {title} {\enquote {\bibinfo {title} {Optimal
  energy quanta to current conversion},}\ }\href {\doibase
  10.1103/PhysRevB.83.085428} {\bibfield  {journal} {\bibinfo  {journal} {Phys.
  Rev. B}\ }\textbf {\bibinfo {volume} {83}},\ \bibinfo {pages} {085428}
  (\bibinfo {year} {2011})}\BibitemShut {NoStop}%
\bibitem [{\citenamefont {Jiang}\ \emph {et~al.}(2012)\citenamefont {Jiang},
  \citenamefont {Entin-Wohlman},\ and\ \citenamefont {Imry}}]{Jiang2012}%
  \BibitemOpen
  \bibfield  {author} {\bibinfo {author} {\bibfnamefont {J.-H.}\ \bibnamefont
  {Jiang}}, \bibinfo {author} {\bibfnamefont {O.}~\bibnamefont
  {Entin-Wohlman}}, \ and\ \bibinfo {author} {\bibfnamefont {Y.}~\bibnamefont
  {Imry}},\ }\bibfield  {title} {\enquote {\bibinfo {title} {Thermoelectric
  three-terminal hopping transport through one-dimensional nanosystems},}\
  }\href {\doibase 10.1103/PhysRevB.85.075412} {\bibfield  {journal} {\bibinfo
  {journal} {Phys. Rev. B}\ }\textbf {\bibinfo {volume} {85}},\ \bibinfo
  {pages} {075412} (\bibinfo {year} {2012})}\BibitemShut {NoStop}%
\bibitem [{\citenamefont {Whitney}(2014)}]{WhitneyPRL}%
  \BibitemOpen
  \bibfield  {author} {\bibinfo {author} {\bibfnamefont {R.~S.}\ \bibnamefont
  {Whitney}},\ }\bibfield  {title} {\enquote {\bibinfo {title} {Most efficient
  quantum thermoelectric at finite power output},}\ }\href {\doibase
  10.1103/PhysRevLett.112.130601} {\bibfield  {journal} {\bibinfo  {journal}
  {Phys. Rev. Lett.}\ }\textbf {\bibinfo {volume} {112}},\ \bibinfo {pages}
  {130601} (\bibinfo {year} {2014})}\BibitemShut {NoStop}%
\bibitem [{\citenamefont {Entin-Wohlman}\ \emph {et~al.}(2014)\citenamefont
  {Entin-Wohlman}, \citenamefont {Jiang},\ and\ \citenamefont
  {Imry}}]{JiangOra}%
  \BibitemOpen
  \bibfield  {author} {\bibinfo {author} {\bibfnamefont {O.}~\bibnamefont
  {Entin-Wohlman}}, \bibinfo {author} {\bibfnamefont {J.-H.}\ \bibnamefont
  {Jiang}}, \ and\ \bibinfo {author} {\bibfnamefont {Y.}~\bibnamefont {Imry}},\
  }\bibfield  {title} {\enquote {\bibinfo {title} {Efficiency and dissipation
  in a two-terminal thermoelectric junction, emphasizing small dissipation},}\
  }\href {\doibase 10.1103/PhysRevE.89.012123} {\bibfield  {journal} {\bibinfo
  {journal} {Phys. Rev. E}\ }\textbf {\bibinfo {volume} {89}},\ \bibinfo
  {pages} {012123} (\bibinfo {year} {2014})}\BibitemShut {NoStop}%
\bibitem [{\citenamefont {Mazza}\ \emph {et~al.}(2015)\citenamefont {Mazza},
  \citenamefont {Valentini}, \citenamefont {Bosisio}, \citenamefont {Benenti},
  \citenamefont {Giovannetti}, \citenamefont {Fazio},\ and\ \citenamefont
  {Taddei}}]{Mazza-separation}%
  \BibitemOpen
  \bibfield  {author} {\bibinfo {author} {\bibfnamefont {F.}~\bibnamefont
  {Mazza}}, \bibinfo {author} {\bibfnamefont {S.}~\bibnamefont {Valentini}},
  \bibinfo {author} {\bibfnamefont {R.}~\bibnamefont {Bosisio}}, \bibinfo
  {author} {\bibfnamefont {G.}~\bibnamefont {Benenti}}, \bibinfo {author}
  {\bibfnamefont {V.}~\bibnamefont {Giovannetti}}, \bibinfo {author}
  {\bibfnamefont {R.}~\bibnamefont {Fazio}}, \ and\ \bibinfo {author}
  {\bibfnamefont {F.}~\bibnamefont {Taddei}},\ }\bibfield  {title} {\enquote
  {\bibinfo {title} {Separation of heat and charge currents for boosted
  thermoelectric conversion},}\ }\href {\doibase 10.1103/PhysRevB.91.245435}
  {\bibfield  {journal} {\bibinfo  {journal} {Phys. Rev. B}\ }\textbf {\bibinfo
  {volume} {91}},\ \bibinfo {pages} {245435} (\bibinfo {year}
  {2015})}\BibitemShut {NoStop}%
\bibitem [{\citenamefont {S\'anchez}\ \emph {et~al.}(2015)\citenamefont
  {S\'anchez}, \citenamefont {Sothmann},\ and\ \citenamefont
  {Jordan}}]{SanchezPRL}%
  \BibitemOpen
  \bibfield  {author} {\bibinfo {author} {\bibfnamefont {R.}~\bibnamefont
  {S\'anchez}}, \bibinfo {author} {\bibfnamefont {B.}~\bibnamefont {Sothmann}},
  \ and\ \bibinfo {author} {\bibfnamefont {A.~N.}\ \bibnamefont {Jordan}},\
  }\bibfield  {title} {\enquote {\bibinfo {title} {Chiral thermoelectrics with
  quantum hall edge states},}\ }\href {\doibase 10.1103/PhysRevLett.114.146801}
  {\bibfield  {journal} {\bibinfo  {journal} {Phys. Rev. Lett.}\ }\textbf
  {\bibinfo {volume} {114}},\ \bibinfo {pages} {146801} (\bibinfo {year}
  {2015})}\BibitemShut {NoStop}%
\bibitem [{\citenamefont {Entin-Wohlman}\ \emph {et~al.}(2015)\citenamefont
  {Entin-Wohlman}, \citenamefont {Imry},\ and\ \citenamefont
  {Aharony}}]{Ora2015}%
  \BibitemOpen
  \bibfield  {author} {\bibinfo {author} {\bibfnamefont {O.}~\bibnamefont
  {Entin-Wohlman}}, \bibinfo {author} {\bibfnamefont {Y.}~\bibnamefont {Imry}},
  \ and\ \bibinfo {author} {\bibfnamefont {A.}~\bibnamefont {Aharony}},\
  }\bibfield  {title} {\enquote {\bibinfo {title} {Enhanced performance of
  joint cooling and energy production},}\ }\href {\doibase
  10.1103/PhysRevB.91.054302} {\bibfield  {journal} {\bibinfo  {journal} {Phys.
  Rev. B}\ }\textbf {\bibinfo {volume} {91}},\ \bibinfo {pages} {054302}
  (\bibinfo {year} {2015})}\BibitemShut {NoStop}%
\bibitem [{\citenamefont {Agarwalla}\ \emph {et~al.}(2015)\citenamefont
  {Agarwalla}, \citenamefont {Jiang},\ and\ \citenamefont
  {Segal}}]{BijayJiang}%
  \BibitemOpen
  \bibfield  {author} {\bibinfo {author} {\bibfnamefont {B.~K.}\ \bibnamefont
  {Agarwalla}}, \bibinfo {author} {\bibfnamefont {J.-H.}\ \bibnamefont
  {Jiang}}, \ and\ \bibinfo {author} {\bibfnamefont {D.}~\bibnamefont
  {Segal}},\ }\bibfield  {title} {\enquote {\bibinfo {title} {Full counting
  statistics of vibrationally assisted electronic conduction: Transport and
  fluctuations of thermoelectric efficiency},}\ }\href {\doibase
  10.1103/PhysRevB.92.245418} {\bibfield  {journal} {\bibinfo  {journal} {Phys.
  Rev. B}\ }\textbf {\bibinfo {volume} {92}},\ \bibinfo {pages} {245418}
  (\bibinfo {year} {2015})}\BibitemShut {NoStop}%
\bibitem [{\citenamefont {Yamamoto}\ \emph {et~al.}(2016)\citenamefont
  {Yamamoto}, \citenamefont {Entin-Wohlman}, \citenamefont {Aharony},\ and\
  \citenamefont {Hatano}}]{Yamamoto}%
  \BibitemOpen
  \bibfield  {author} {\bibinfo {author} {\bibfnamefont {K.}~\bibnamefont
  {Yamamoto}}, \bibinfo {author} {\bibfnamefont {O.}~\bibnamefont
  {Entin-Wohlman}}, \bibinfo {author} {\bibfnamefont {A.}~\bibnamefont
  {Aharony}}, \ and\ \bibinfo {author} {\bibfnamefont {N.}~\bibnamefont
  {Hatano}},\ }\bibfield  {title} {\enquote {\bibinfo {title} {Efficiency
  bounds on thermoelectric transport in magnetic fields: The role of inelastic
  processes},}\ }\href {\doibase 10.1103/PhysRevB.94.121402} {\bibfield
  {journal} {\bibinfo  {journal} {Phys. Rev. B}\ }\textbf {\bibinfo {volume}
  {94}},\ \bibinfo {pages} {121402} (\bibinfo {year} {2016})}\BibitemShut
  {NoStop}%
\bibitem [{\citenamefont {Shiraishi}\ \emph {et~al.}(2016)\citenamefont
  {Shiraishi}, \citenamefont {Saito},\ and\ \citenamefont {Tasaki}}]{Naoto}%
  \BibitemOpen
  \bibfield  {author} {\bibinfo {author} {\bibfnamefont {N.}~\bibnamefont
  {Shiraishi}}, \bibinfo {author} {\bibfnamefont {K.}~\bibnamefont {Saito}}, \
  and\ \bibinfo {author} {\bibfnamefont {H.}~\bibnamefont {Tasaki}},\
  }\bibfield  {title} {\enquote {\bibinfo {title} {Universal trade-off relation
  between power and efficiency for heat engines},}\ }\href {\doibase
  10.1103/PhysRevLett.117.190601} {\bibfield  {journal} {\bibinfo  {journal}
  {Phys. Rev. Lett.}\ }\textbf {\bibinfo {volume} {117}},\ \bibinfo {pages}
  {190601} (\bibinfo {year} {2016})}\BibitemShut {NoStop}%
\bibitem [{\citenamefont {Agarwalla}\ \emph {et~al.}(2017)\citenamefont
  {Agarwalla}, \citenamefont {Jiang},\ and\ \citenamefont
  {Segal}}]{JiangBijayPRB17}%
  \BibitemOpen
  \bibfield  {author} {\bibinfo {author} {\bibfnamefont {B.~K.}\ \bibnamefont
  {Agarwalla}}, \bibinfo {author} {\bibfnamefont {J.-H.}\ \bibnamefont
  {Jiang}}, \ and\ \bibinfo {author} {\bibfnamefont {D.}~\bibnamefont
  {Segal}},\ }\bibfield  {title} {\enquote {\bibinfo {title} {Quantum
  efficiency bound for continuous heat engines coupled to noncanonical
  reservoirs},}\ }\href {\doibase 10.1103/PhysRevB.96.104304} {\bibfield
  {journal} {\bibinfo  {journal} {Phys. Rev. B}\ }\textbf {\bibinfo {volume}
  {96}},\ \bibinfo {pages} {104304} (\bibinfo {year} {2017})}\BibitemShut
  {NoStop}%
\bibitem [{\citenamefont {Macieszczak}\ \emph {et~al.}(2018)\citenamefont
  {Macieszczak}, \citenamefont {Brandner},\ and\ \citenamefont
  {Garrahan}}]{Brandner2018}%
  \BibitemOpen
  \bibfield  {author} {\bibinfo {author} {\bibfnamefont {K.}~\bibnamefont
  {Macieszczak}}, \bibinfo {author} {\bibfnamefont {K.}~\bibnamefont
  {Brandner}}, \ and\ \bibinfo {author} {\bibfnamefont {J.~P.}\ \bibnamefont
  {Garrahan}},\ }\bibfield  {title} {\enquote {\bibinfo {title} {Unified
  thermodynamic uncertainty relations in linear response},}\ }\href {\doibase
  10.1103/PhysRevLett.121.130601} {\bibfield  {journal} {\bibinfo  {journal}
  {Phys. Rev. Lett.}\ }\textbf {\bibinfo {volume} {121}},\ \bibinfo {pages}
  {130601} (\bibinfo {year} {2018})}\BibitemShut {NoStop}%
\bibitem [{\citenamefont {Jiang}\ and\ \citenamefont
  {Imry}(2018)}]{JiangNearfield}%
  \BibitemOpen
  \bibfield  {author} {\bibinfo {author} {\bibfnamefont {J.-H.}\ \bibnamefont
  {Jiang}}\ and\ \bibinfo {author} {\bibfnamefont {Y.}~\bibnamefont {Imry}},\
  }\bibfield  {title} {\enquote {\bibinfo {title} {Near-field three-terminal
  thermoelectric heat engine},}\ }\href {\doibase 10.1103/PhysRevB.97.125422}
  {\bibfield  {journal} {\bibinfo  {journal} {Phys. Rev. B}\ }\textbf {\bibinfo
  {volume} {97}},\ \bibinfo {pages} {125422} (\bibinfo {year}
  {2018})}\BibitemShut {NoStop}%
\bibitem [{\citenamefont {Brandner}\ and\ \citenamefont
  {Saito}(2019)}]{brandner2019}%
  \BibitemOpen
  \bibfield  {author} {\bibinfo {author} {\bibfnamefont {K.}~\bibnamefont
  {Brandner}}\ and\ \bibinfo {author} {\bibfnamefont {K.}~\bibnamefont
  {Saito}},\ }\bibfield  {title} {\enquote {\bibinfo {title} {Thermodynamic
  geometry of microscopic heat engines},}\ }\href
  {https://arxiv.org/abs/1907.06780} {\bibfield  {journal} {\bibinfo  {journal}
  {arXiv:1907.06780}\ } (\bibinfo {year} {2019})}\BibitemShut {NoStop}%
\bibitem [{\citenamefont {Wang}\ \emph
  {et~al.}(2019{\natexlab{b}})\citenamefont {Wang}, \citenamefont {Lu},\ and\
  \citenamefont {Jiang}}]{wangPRApplied}%
  \BibitemOpen
  \bibfield  {author} {\bibinfo {author} {\bibfnamefont {R.}~\bibnamefont
  {Wang}}, \bibinfo {author} {\bibfnamefont {J.}~\bibnamefont {Lu}}, \ and\
  \bibinfo {author} {\bibfnamefont {J.-H.}\ \bibnamefont {Jiang}},\ }\bibfield
  {title} {\enquote {\bibinfo {title} {Enhancing thermophotovoltaic performance
  using graphene-bn-$\mathrm{In}\mathrm{Sb}$ near-field heterostructures},}\
  }\href {\doibase 10.1103/PhysRevApplied.12.044038} {\bibfield  {journal}
  {\bibinfo  {journal} {Phys. Rev. Applied}\ }\textbf {\bibinfo {volume}
  {12}},\ \bibinfo {pages} {044038} (\bibinfo {year}
  {2019}{\natexlab{b}})}\BibitemShut {NoStop}%
\bibitem [{\citenamefont {Hwang}\ \emph {et~al.}(2013)\citenamefont {Hwang},
  \citenamefont {S{\'a}nchez}, \citenamefont {Lee},\ and\ \citenamefont
  {L{\'o}pez}}]{hwang}%
  \BibitemOpen
  \bibfield  {author} {\bibinfo {author} {\bibfnamefont {S.-Y.}\ \bibnamefont
  {Hwang}}, \bibinfo {author} {\bibfnamefont {D.}~\bibnamefont {S{\'a}nchez}},
  \bibinfo {author} {\bibfnamefont {M.}~\bibnamefont {Lee}}, \ and\ \bibinfo
  {author} {\bibfnamefont {R.}~\bibnamefont {L{\'o}pez}},\ }\bibfield  {title}
  {\enquote {\bibinfo {title} {Magnetic-field asymmetry of nonlinear
  thermoelectric and heat transport},}\ }\href
  {http://stacks.iop.org/1367-2630/15/i=10/a=105012} {\bibfield  {journal}
  {\bibinfo  {journal} {New J. Phys.}\ }\textbf {\bibinfo {volume} {15}},\
  \bibinfo {pages} {105012} (\bibinfo {year} {2013})}\BibitemShut {NoStop}%
\bibitem [{\citenamefont {Matthews}\ \emph {et~al.}(2014)\citenamefont
  {Matthews}, \citenamefont {Battista}, \citenamefont {S\'anchez},
  \citenamefont {Samuelsson},\ and\ \citenamefont {Linke}}]{Exper}%
  \BibitemOpen
  \bibfield  {author} {\bibinfo {author} {\bibfnamefont {J.}~\bibnamefont
  {Matthews}}, \bibinfo {author} {\bibfnamefont {F.}~\bibnamefont {Battista}},
  \bibinfo {author} {\bibfnamefont {D.}~\bibnamefont {S\'anchez}}, \bibinfo
  {author} {\bibfnamefont {P.}~\bibnamefont {Samuelsson}}, \ and\ \bibinfo
  {author} {\bibfnamefont {H.}~\bibnamefont {Linke}},\ }\bibfield  {title}
  {\enquote {\bibinfo {title} {Experimental verification of reciprocity
  relations in quantum thermoelectric transport},}\ }\href {\doibase
  10.1103/PhysRevB.90.165428} {\bibfield  {journal} {\bibinfo  {journal} {Phys.
  Rev. B}\ }\textbf {\bibinfo {volume} {90}},\ \bibinfo {pages} {165428}
  (\bibinfo {year} {2014})}\BibitemShut {NoStop}%
\bibitem [{\citenamefont {Thierschmann}\ \emph {et~al.}(2015)\citenamefont
  {Thierschmann}, \citenamefont {S{\'a}nchez}, \citenamefont {Sothmann},
  \citenamefont {Arnold}, \citenamefont {Heyn}, \citenamefont {Hansen},
  \citenamefont {Buhmann},\ and\ \citenamefont {Molenkamp}}]{Thier2015}%
  \BibitemOpen
  \bibfield  {author} {\bibinfo {author} {\bibfnamefont {H.}~\bibnamefont
  {Thierschmann}}, \bibinfo {author} {\bibfnamefont {R.}~\bibnamefont
  {S{\'a}nchez}}, \bibinfo {author} {\bibfnamefont {B.}~\bibnamefont
  {Sothmann}}, \bibinfo {author} {\bibfnamefont {F.}~\bibnamefont {Arnold}},
  \bibinfo {author} {\bibfnamefont {C.}~\bibnamefont {Heyn}}, \bibinfo {author}
  {\bibfnamefont {W.}~\bibnamefont {Hansen}}, \bibinfo {author} {\bibfnamefont
  {H.}~\bibnamefont {Buhmann}}, \ and\ \bibinfo {author} {\bibfnamefont
  {L.~W.}\ \bibnamefont {Molenkamp}},\ }\bibfield  {title} {\enquote {\bibinfo
  {title} {Three-terminal energy harvester with coupled quantum dots},}\ }\href
  {\doibase 10.1038/nnano.2015.176} {\bibfield  {journal} {\bibinfo  {journal}
  {Nat. Nanotech.}\ }\textbf {\bibinfo {volume} {10}},\ \bibinfo {pages} {854}
  (\bibinfo {year} {2015})}\BibitemShut {NoStop}%
\bibitem [{\citenamefont {Cui}\ \emph {et~al.}(2018)\citenamefont {Cui},
  \citenamefont {Miao}, \citenamefont {Wang}, \citenamefont {Thompson},
  \citenamefont {Zotti}, \citenamefont {Cuevas}, \citenamefont {Meyhofer},\
  and\ \citenamefont {Reddy}}]{cui2018}%
  \BibitemOpen
  \bibfield  {author} {\bibinfo {author} {\bibfnamefont {L.}~\bibnamefont
  {Cui}}, \bibinfo {author} {\bibfnamefont {R.}~\bibnamefont {Miao}}, \bibinfo
  {author} {\bibfnamefont {K.}~\bibnamefont {Wang}}, \bibinfo {author}
  {\bibfnamefont {D.}~\bibnamefont {Thompson}}, \bibinfo {author}
  {\bibfnamefont {L.~A.}\ \bibnamefont {Zotti}}, \bibinfo {author}
  {\bibfnamefont {J.~C.}\ \bibnamefont {Cuevas}}, \bibinfo {author}
  {\bibfnamefont {E.}~\bibnamefont {Meyhofer}}, \ and\ \bibinfo {author}
  {\bibfnamefont {P.}~\bibnamefont {Reddy}},\ }\bibfield  {title} {\enquote
  {\bibinfo {title} {Peltier cooling in molecular junctions},}\ }\href
  {\doibase 10.1038/s41565-017-0020-z} {\bibfield  {journal} {\bibinfo
  {journal} {Nat. Nanotech.}\ }\textbf {\bibinfo {volume} {13}},\ \bibinfo
  {pages} {122} (\bibinfo {year} {2018})}\BibitemShut {NoStop}%
\bibitem [{\citenamefont {Hartman}\ \emph {et~al.}(2018)\citenamefont
  {Hartman}, \citenamefont {Olsen}, \citenamefont {L{\"u}scher}, \citenamefont
  {Samani}, \citenamefont {Fallahi}, \citenamefont {Gardner}, \citenamefont
  {Manfra},\ and\ \citenamefont {Folk}}]{hartmanNP}%
  \BibitemOpen
  \bibfield  {author} {\bibinfo {author} {\bibfnamefont {N.}~\bibnamefont
  {Hartman}}, \bibinfo {author} {\bibfnamefont {C.}~\bibnamefont {Olsen}},
  \bibinfo {author} {\bibfnamefont {S.}~\bibnamefont {L{\"u}scher}}, \bibinfo
  {author} {\bibfnamefont {M.}~\bibnamefont {Samani}}, \bibinfo {author}
  {\bibfnamefont {S.}~\bibnamefont {Fallahi}}, \bibinfo {author} {\bibfnamefont
  {G.~C}\ \bibnamefont {Gardner}}, \bibinfo {author} {\bibfnamefont
  {M.}~\bibnamefont {Manfra}}, \ and\ \bibinfo {author} {\bibfnamefont
  {J.}~\bibnamefont {Folk}},\ }\bibfield  {title} {\enquote {\bibinfo {title}
  {Direct entropy measurement in a mesoscopic quantum system},}\ }\href
  {\doibase https://doi.org/10.1038/s41567-018-0250-5} {\bibfield  {journal}
  {\bibinfo  {journal} {Nat. Phys.}\ }\textbf {\bibinfo {volume} {14}},\
  \bibinfo {pages} {1083--1086} (\bibinfo {year} {2018})}\BibitemShut {NoStop}%
\bibitem [{\citenamefont {Josefsson}\ \emph {et~al.}(2018)\citenamefont
  {Josefsson}, \citenamefont {Svilans}, \citenamefont {Burke}, \citenamefont
  {Hoffmann}, \citenamefont {Fahlvik}, \citenamefont {Thelander}, \citenamefont
  {Leijnse},\ and\ \citenamefont {Linke}}]{josefsson2018}%
  \BibitemOpen
  \bibfield  {author} {\bibinfo {author} {\bibfnamefont {M.}~\bibnamefont
  {Josefsson}}, \bibinfo {author} {\bibfnamefont {A.}~\bibnamefont {Svilans}},
  \bibinfo {author} {\bibfnamefont {A.~M}\ \bibnamefont {Burke}}, \bibinfo
  {author} {\bibfnamefont {E.~A}\ \bibnamefont {Hoffmann}}, \bibinfo {author}
  {\bibfnamefont {S.}~\bibnamefont {Fahlvik}}, \bibinfo {author} {\bibfnamefont
  {C.}~\bibnamefont {Thelander}}, \bibinfo {author} {\bibfnamefont
  {M.}~\bibnamefont {Leijnse}}, \ and\ \bibinfo {author} {\bibfnamefont
  {H.}~\bibnamefont {Linke}},\ }\bibfield  {title} {\enquote {\bibinfo {title}
  {A quantum-dot heat engine operating close to the thermodynamic efficiency
  limits},}\ }\href {\doibase https://doi.org/10.1038/s41565-018-0200-5}
  {\bibfield  {journal} {\bibinfo  {journal} {Nat. Nanotech.}\ }\textbf
  {\bibinfo {volume} {13}},\ \bibinfo {pages} {920} (\bibinfo {year}
  {2018})}\BibitemShut {NoStop}%
\bibitem [{\citenamefont {Jaliel}\ \emph {et~al.}(2019)\citenamefont {Jaliel},
  \citenamefont {Puddy}, \citenamefont {S\'anchez}, \citenamefont {Jordan},
  \citenamefont {Sothmann}, \citenamefont {Farrer}, \citenamefont {Griffiths},
  \citenamefont {Ritchie},\ and\ \citenamefont {Smith}}]{jaliel-exper}%
  \BibitemOpen
  \bibfield  {author} {\bibinfo {author} {\bibfnamefont {G.}~\bibnamefont
  {Jaliel}}, \bibinfo {author} {\bibfnamefont {R.~K.}\ \bibnamefont {Puddy}},
  \bibinfo {author} {\bibfnamefont {R.}~\bibnamefont {S\'anchez}}, \bibinfo
  {author} {\bibfnamefont {A.~N.}\ \bibnamefont {Jordan}}, \bibinfo {author}
  {\bibfnamefont {B.}~\bibnamefont {Sothmann}}, \bibinfo {author}
  {\bibfnamefont {I.}~\bibnamefont {Farrer}}, \bibinfo {author} {\bibfnamefont
  {J.~P.}\ \bibnamefont {Griffiths}}, \bibinfo {author} {\bibfnamefont {D.~A.}\
  \bibnamefont {Ritchie}}, \ and\ \bibinfo {author} {\bibfnamefont {C.~G.}\
  \bibnamefont {Smith}},\ }\bibfield  {title} {\enquote {\bibinfo {title}
  {Experimental realization of a quantum dot energy harvester},}\ }\href
  {\doibase 10.1103/PhysRevLett.123.117701} {\bibfield  {journal} {\bibinfo
  {journal} {Phys. Rev. Lett.}\ }\textbf {\bibinfo {volume} {123}},\ \bibinfo
  {pages} {117701} (\bibinfo {year} {2019})}\BibitemShut {NoStop}%
\bibitem [{\citenamefont {Josefsson}\ \emph {et~al.}(2019)\citenamefont
  {Josefsson}, \citenamefont {Svilans}, \citenamefont {Linke},\ and\
  \citenamefont {Leijnse}}]{Josefsson}%
  \BibitemOpen
  \bibfield  {author} {\bibinfo {author} {\bibfnamefont {M.}~\bibnamefont
  {Josefsson}}, \bibinfo {author} {\bibfnamefont {A.}~\bibnamefont {Svilans}},
  \bibinfo {author} {\bibfnamefont {H.}~\bibnamefont {Linke}}, \ and\ \bibinfo
  {author} {\bibfnamefont {M.}~\bibnamefont {Leijnse}},\ }\bibfield  {title}
  {\enquote {\bibinfo {title} {Optimal power and efficiency of single quantum
  dot heat engines: Theory and experiment},}\ }\href {\doibase
  10.1103/PhysRevB.99.235432} {\bibfield  {journal} {\bibinfo  {journal} {Phys.
  Rev. B}\ }\textbf {\bibinfo {volume} {99}},\ \bibinfo {pages} {235432}
  (\bibinfo {year} {2019})}\BibitemShut {NoStop}%
\bibitem [{\citenamefont {Prete}\ \emph {et~al.}(2019)\citenamefont {Prete},
  \citenamefont {Erdman}, \citenamefont {Demontis}, \citenamefont {Zannier},
  \citenamefont {Ercolani}, \citenamefont {Sorba}, \citenamefont {Beltram},
  \citenamefont {Rossella}, \citenamefont {Taddei},\ and\ \citenamefont
  {Roddaro}}]{prete}%
  \BibitemOpen
  \bibfield  {author} {\bibinfo {author} {\bibfnamefont {D.}~\bibnamefont
  {Prete}}, \bibinfo {author} {\bibfnamefont {P.~A.}\ \bibnamefont {Erdman}},
  \bibinfo {author} {\bibfnamefont {V.}~\bibnamefont {Demontis}}, \bibinfo
  {author} {\bibfnamefont {V.}~\bibnamefont {Zannier}}, \bibinfo {author}
  {\bibfnamefont {D.}~\bibnamefont {Ercolani}}, \bibinfo {author}
  {\bibfnamefont {L.}~\bibnamefont {Sorba}}, \bibinfo {author} {\bibfnamefont
  {F.}~\bibnamefont {Beltram}}, \bibinfo {author} {\bibfnamefont
  {F.}~\bibnamefont {Rossella}}, \bibinfo {author} {\bibfnamefont
  {F.}~\bibnamefont {Taddei}}, \ and\ \bibinfo {author} {\bibfnamefont
  {S.}~\bibnamefont {Roddaro}},\ }\bibfield  {title} {\enquote {\bibinfo
  {title} {Thermoelectric conversion at $\rm{30K}$ in $\rm{InAs/InP}$ nanowire
  quantum dots},}\ }\href {\doibase 10.1021/acs.nanolett.9b00276} {\bibfield
  {journal} {\bibinfo  {journal} {Nano Lett.}\ }\textbf {\bibinfo {volume}
  {19}},\ \bibinfo {pages} {3033--3039} (\bibinfo {year} {2019})}\BibitemShut
  {NoStop}%
\bibitem [{\citenamefont {Mahan}\ and\ \citenamefont {Sofo}(1996)}]{Mahan}%
  \BibitemOpen
  \bibfield  {author} {\bibinfo {author} {\bibfnamefont {G.~D.}\ \bibnamefont
  {Mahan}}\ and\ \bibinfo {author} {\bibfnamefont {J.~O.}\ \bibnamefont
  {Sofo}},\ }\bibfield  {title} {\enquote {\bibinfo {title} {The best
  thermoelectric},}\ }\href {\doibase 10.1073/pnas.93.15.7436} {\bibfield
  {journal} {\bibinfo  {journal} {Proc. Natl. Acad. Sci. USA}\ }\textbf
  {\bibinfo {volume} {93}},\ \bibinfo {pages} {7436--7439} (\bibinfo {year}
  {1996})}\BibitemShut {NoStop}%
\bibitem [{\citenamefont {Jiang}(2014{\natexlab{a}})}]{JiangJAP}%
  \BibitemOpen
  \bibfield  {author} {\bibinfo {author} {\bibfnamefont {J.-H.}\ \bibnamefont
  {Jiang}},\ }\bibfield  {title} {\enquote {\bibinfo {title} {Enhancing
  efficiency and power of quantum-dots resonant tunneling thermoelectrics in
  three-terminal geometry by cooperative effects},}\ }\href {\doibase
  10.1063/1.4901120} {\bibfield  {journal} {\bibinfo  {journal} {J. Appl.
  Phys.}\ }\textbf {\bibinfo {volume} {116}},\ \bibinfo {pages} {194303}
  (\bibinfo {year} {2014}{\natexlab{a}})}\BibitemShut {NoStop}%
\bibitem [{\citenamefont {Lu}\ \emph {et~al.}(2017)\citenamefont {Lu},
  \citenamefont {Wang}, \citenamefont {Liu},\ and\ \citenamefont
  {Jiang}}]{MyJAP}%
  \BibitemOpen
  \bibfield  {author} {\bibinfo {author} {\bibfnamefont {J.}~\bibnamefont
  {Lu}}, \bibinfo {author} {\bibfnamefont {R.}~\bibnamefont {Wang}}, \bibinfo
  {author} {\bibfnamefont {Y.}~\bibnamefont {Liu}}, \ and\ \bibinfo {author}
  {\bibfnamefont {J.-H.}\ \bibnamefont {Jiang}},\ }\bibfield  {title} {\enquote
  {\bibinfo {title} {Thermoelectric cooperative effect in three-terminal
  elastic transport through a quantum dot},}\ }\href {\doibase
  10.1063/1.4995532} {\bibfield  {journal} {\bibinfo  {journal} {J. Appl.
  Phys.}\ }\textbf {\bibinfo {volume} {122}},\ \bibinfo {pages} {044301}
  (\bibinfo {year} {2017})}\BibitemShut {NoStop}%
\bibitem [{\citenamefont {Saito}\ \emph {et~al.}(2011)\citenamefont {Saito},
  \citenamefont {Benenti}, \citenamefont {Casati},\ and\ \citenamefont
  {Prosen}}]{Saito}%
  \BibitemOpen
  \bibfield  {author} {\bibinfo {author} {\bibfnamefont {K.}~\bibnamefont
  {Saito}}, \bibinfo {author} {\bibfnamefont {G.}~\bibnamefont {Benenti}},
  \bibinfo {author} {\bibfnamefont {G.}~\bibnamefont {Casati}}, \ and\ \bibinfo
  {author} {\bibfnamefont {T.}~\bibnamefont {Prosen}},\ }\bibfield  {title}
  {\enquote {\bibinfo {title} {Thermopower with broken time-reversal
  symmetry},}\ }\href {\doibase 10.1103/PhysRevB.84.201306} {\bibfield
  {journal} {\bibinfo  {journal} {Phys. Rev. B}\ }\textbf {\bibinfo {volume}
  {84}},\ \bibinfo {pages} {201306} (\bibinfo {year} {2011})}\BibitemShut
  {NoStop}%
\bibitem [{\citenamefont {Buttiker}(1988)}]{Buttiker}%
  \BibitemOpen
  \bibfield  {author} {\bibinfo {author} {\bibfnamefont {M.}~\bibnamefont
  {Buttiker}},\ }\bibfield  {title} {\enquote {\bibinfo {title} {Coherent and
  sequential tunneling in series barriers},}\ }\href {\doibase
  10.1147/rd.321.0063} {\bibfield  {journal} {\bibinfo  {journal} {IBM J. Res.
  Dev.}\ }\textbf {\bibinfo {volume} {32}},\ \bibinfo {pages} {63--75}
  (\bibinfo {year} {1988})}\BibitemShut {NoStop}%
\bibitem [{\citenamefont {Jiang}\ \emph
  {et~al.}(2015{\natexlab{b}})\citenamefont {Jiang}, \citenamefont
  {Agarwalla},\ and\ \citenamefont {Segal}}]{JiangPRL}%
  \BibitemOpen
  \bibfield  {author} {\bibinfo {author} {\bibfnamefont {J.-H.}\ \bibnamefont
  {Jiang}}, \bibinfo {author} {\bibfnamefont {B.~K.}\ \bibnamefont
  {Agarwalla}}, \ and\ \bibinfo {author} {\bibfnamefont {D.}~\bibnamefont
  {Segal}},\ }\bibfield  {title} {\enquote {\bibinfo {title} {Efficiency
  statistics and bounds for systems with broken time-reversal symmetry},}\
  }\href {\doibase 10.1103/PhysRevLett.115.040601} {\bibfield  {journal}
  {\bibinfo  {journal} {Phys. Rev. Lett.}\ }\textbf {\bibinfo {volume} {115}},\
  \bibinfo {pages} {040601} (\bibinfo {year} {2015}{\natexlab{b}})}\BibitemShut
  {NoStop}%
\bibitem [{\citenamefont {Jiang}(2014{\natexlab{b}})}]{JiangPRE}%
  \BibitemOpen
  \bibfield  {author} {\bibinfo {author} {\bibfnamefont {J.-H.}\ \bibnamefont
  {Jiang}},\ }\bibfield  {title} {\enquote {\bibinfo {title} {Thermodynamic
  bounds and general properties of optimal efficiency and power in linear
  responses},}\ }\href {\doibase 10.1103/PhysRevE.90.042126} {\bibfield
  {journal} {\bibinfo  {journal} {Phys. Rev. E}\ }\textbf {\bibinfo {volume}
  {90}},\ \bibinfo {pages} {042126} (\bibinfo {year}
  {2014}{\natexlab{b}})}\BibitemShut {NoStop}%
\bibitem [{\citenamefont {Jiang}\ and\ \citenamefont {Imry}(2017)}]{Jiang2017}%
  \BibitemOpen
  \bibfield  {author} {\bibinfo {author} {\bibfnamefont {J.-H.}\ \bibnamefont
  {Jiang}}\ and\ \bibinfo {author} {\bibfnamefont {Y.}~\bibnamefont {Imry}},\
  }\bibfield  {title} {\enquote {\bibinfo {title} {Enhancing thermoelectric
  performance using nonlinear transport effects},}\ }\href {\doibase
  10.1103/PhysRevApplied.7.064001} {\bibfield  {journal} {\bibinfo  {journal}
  {Phys. Rev. Applied}\ }\textbf {\bibinfo {volume} {7}},\ \bibinfo {pages}
  {064001} (\bibinfo {year} {2017})}\BibitemShut {NoStop}%
\bibitem [{\citenamefont {Sivan}\ and\ \citenamefont {Imry}(1986)}]{Sivan}%
  \BibitemOpen
  \bibfield  {author} {\bibinfo {author} {\bibfnamefont {U.}~\bibnamefont
  {Sivan}}\ and\ \bibinfo {author} {\bibfnamefont {Y.}~\bibnamefont {Imry}},\
  }\bibfield  {title} {\enquote {\bibinfo {title} {Multichannel landauer
  formula for thermoelectric transport with application to thermopower near the
  mobility edge},}\ }\href {\doibase 10.1103/PhysRevB.33.551} {\bibfield
  {journal} {\bibinfo  {journal} {Phys. Rev. B}\ }\textbf {\bibinfo {volume}
  {33}},\ \bibinfo {pages} {551--558} (\bibinfo {year} {1986})}\BibitemShut
  {NoStop}%
\bibitem [{\citenamefont {Butcher}(1990)}]{butcher1990}%
  \BibitemOpen
  \bibfield  {author} {\bibinfo {author} {\bibfnamefont {P.~N.}\ \bibnamefont
  {Butcher}},\ }\bibfield  {title} {\enquote {\bibinfo {title} {Thermal and
  electrical transport formalism for electronic microstructures with many
  terminals},}\ }\href {http://stacks.iop.org/0953-8984/2/i=22/a=008}
  {\bibfield  {journal} {\bibinfo  {journal} {J. Phys.: Condens. Matter}\
  }\textbf {\bibinfo {volume} {2}},\ \bibinfo {pages} {4869} (\bibinfo {year}
  {1990})}\BibitemShut {NoStop}%
\bibitem [{\citenamefont {Mazza}\ \emph {et~al.}(2014)\citenamefont {Mazza},
  \citenamefont {Bosisio}, \citenamefont {Benenti}, \citenamefont
  {Giovannetti}, \citenamefont {Fazio},\ and\ \citenamefont {Taddei}}]{3T-QTM}%
  \BibitemOpen
  \bibfield  {author} {\bibinfo {author} {\bibfnamefont {F.}~\bibnamefont
  {Mazza}}, \bibinfo {author} {\bibfnamefont {R.}~\bibnamefont {Bosisio}},
  \bibinfo {author} {\bibfnamefont {G.}~\bibnamefont {Benenti}}, \bibinfo
  {author} {\bibfnamefont {V.}~\bibnamefont {Giovannetti}}, \bibinfo {author}
  {\bibfnamefont {R.}~\bibnamefont {Fazio}}, \ and\ \bibinfo {author}
  {\bibfnamefont {F.}~\bibnamefont {Taddei}},\ }\bibfield  {title} {\enquote
  {\bibinfo {title} {Thermoelectric efficiency of three-terminal quantum
  thermal machines},}\ }\href {http://stacks.iop.org/1367-2630/16/i=8/a=085001}
  {\bibfield  {journal} {\bibinfo  {journal} {New J. Phys.}\ }\textbf {\bibinfo
  {volume} {16}},\ \bibinfo {pages} {085001} (\bibinfo {year}
  {2014})}\BibitemShut {NoStop}%
\bibitem [{\citenamefont {Iyyappan}\ and\ \citenamefont
  {Ponmurugan}(2018)}]{Iyynonlin}%
  \BibitemOpen
  \bibfield  {author} {\bibinfo {author} {\bibfnamefont {I.}~\bibnamefont
  {Iyyappan}}\ and\ \bibinfo {author} {\bibfnamefont {M.}~\bibnamefont
  {Ponmurugan}},\ }\bibfield  {title} {\enquote {\bibinfo {title} {General
  relations between the power, efficiency, and dissipation for the irreversible
  heat engines in the nonlinear response regime},}\ }\href {\doibase
  10.1103/PhysRevE.97.012141} {\bibfield  {journal} {\bibinfo  {journal} {Phys.
  Rev. E}\ }\textbf {\bibinfo {volume} {97}},\ \bibinfo {pages} {012141}
  (\bibinfo {year} {2018})}\BibitemShut {NoStop}%
\bibitem [{\citenamefont {Benenti}\ \emph {et~al.}(2011)\citenamefont
  {Benenti}, \citenamefont {Saito},\ and\ \citenamefont {Casati}}]{Saito2011}%
  \BibitemOpen
  \bibfield  {author} {\bibinfo {author} {\bibfnamefont {G.}~\bibnamefont
  {Benenti}}, \bibinfo {author} {\bibfnamefont {K.}~\bibnamefont {Saito}}, \
  and\ \bibinfo {author} {\bibfnamefont {G.}~\bibnamefont {Casati}},\
  }\bibfield  {title} {\enquote {\bibinfo {title} {Thermodynamic bounds on
  efficiency for systems with broken time-reversal symmetry},}\ }\href
  {\doibase 10.1103/PhysRevLett.106.230602} {\bibfield  {journal} {\bibinfo
  {journal} {Phys. Rev. Lett.}\ }\textbf {\bibinfo {volume} {106}},\ \bibinfo
  {pages} {230602} (\bibinfo {year} {2011})}\BibitemShut {NoStop}%
\bibitem [{\citenamefont {Kheradsoud}\ \emph {et~al.}(2019)\citenamefont
  {Kheradsoud}, \citenamefont {Dashti}, \citenamefont {Misiorny}, \citenamefont
  {Potts}, \citenamefont {Splettstoesser},\ and\ \citenamefont
  {Samuelsson}}]{entropy-tradeoff}%
  \BibitemOpen
  \bibfield  {author} {\bibinfo {author} {\bibfnamefont {S.}~\bibnamefont
  {Kheradsoud}}, \bibinfo {author} {\bibfnamefont {N.}~\bibnamefont {Dashti}},
  \bibinfo {author} {\bibfnamefont {M.}~\bibnamefont {Misiorny}}, \bibinfo
  {author} {\bibfnamefont {P.~P.}\ \bibnamefont {Potts}}, \bibinfo {author}
  {\bibfnamefont {J.}~\bibnamefont {Splettstoesser}}, \ and\ \bibinfo {author}
  {\bibfnamefont {P.}~\bibnamefont {Samuelsson}},\ }\bibfield  {title}
  {\enquote {\bibinfo {title} {Power, efficiency and fluctuations in a quantum
  point contact as steady-state thermoelectric heat engine},}\ }\href {\doibase
  10.3390/e21080777} {\bibfield  {journal} {\bibinfo  {journal} {Entropy}\
  }\textbf {\bibinfo {volume} {21}} (\bibinfo {year} {2019}),\
  10.3390/e21080777}\BibitemShut {NoStop}%
\bibitem [{\citenamefont {Lu}\ \emph {et~al.}(2019{\natexlab{b}})\citenamefont
  {Lu}, \citenamefont {Liu}, \citenamefont {Wang}, \citenamefont {Wang},\ and\
  \citenamefont {Jiang}}]{trade-off}%
  \BibitemOpen
  \bibfield  {author} {\bibinfo {author} {\bibfnamefont {J.}~\bibnamefont
  {Lu}}, \bibinfo {author} {\bibfnamefont {Y.}~\bibnamefont {Liu}}, \bibinfo
  {author} {\bibfnamefont {R.}~\bibnamefont {Wang}}, \bibinfo {author}
  {\bibfnamefont {C.}~\bibnamefont {Wang}}, \ and\ \bibinfo {author}
  {\bibfnamefont {J.-H.}\ \bibnamefont {Jiang}},\ }\bibfield  {title} {\enquote
  {\bibinfo {title} {Optimal efficiency and power, and their trade-off in
  three-terminal quantum thermoelectric engines with two output electric
  currents},}\ }\href {\doibase 10.1103/PhysRevB.100.115438} {\bibfield
  {journal} {\bibinfo  {journal} {Phys. Rev. B}\ }\textbf {\bibinfo {volume}
  {100}},\ \bibinfo {pages} {115438} (\bibinfo {year}
  {2019}{\natexlab{b}})}\BibitemShut {NoStop}%
\bibitem [{\citenamefont {Van~den Broeck}(2005)}]{Van2005}%
  \BibitemOpen
  \bibfield  {author} {\bibinfo {author} {\bibfnamefont {C.}~\bibnamefont
  {Van~den Broeck}},\ }\bibfield  {title} {\enquote {\bibinfo {title}
  {Thermodynamic efficiency at maximum power},}\ }\href {\doibase
  10.1103/PhysRevLett.95.190602} {\bibfield  {journal} {\bibinfo  {journal}
  {Phys. Rev. Lett.}\ }\textbf {\bibinfo {volume} {95}},\ \bibinfo {pages}
  {190602} (\bibinfo {year} {2005})}\BibitemShut {NoStop}%
\bibitem [{\citenamefont {Golubeva}\ and\ \citenamefont
  {Imparato}(2012)}]{AI-PRL2012}%
  \BibitemOpen
  \bibfield  {author} {\bibinfo {author} {\bibfnamefont {N.}~\bibnamefont
  {Golubeva}}\ and\ \bibinfo {author} {\bibfnamefont {A.}~\bibnamefont
  {Imparato}},\ }\bibfield  {title} {\enquote {\bibinfo {title} {Efficiency at
  maximum power of interacting molecular machines},}\ }\href {\doibase
  10.1103/PhysRevLett.109.190602} {\bibfield  {journal} {\bibinfo  {journal}
  {Phys. Rev. Lett.}\ }\textbf {\bibinfo {volume} {109}},\ \bibinfo {pages}
  {190602} (\bibinfo {year} {2012})}\BibitemShut {NoStop}%
\bibitem [{\citenamefont {Proesmans}\ \emph {et~al.}(2016)\citenamefont
  {Proesmans}, \citenamefont {Cleuren},\ and\ \citenamefont {Van~den
  Broeck}}]{Proesmans}%
  \BibitemOpen
  \bibfield  {author} {\bibinfo {author} {\bibfnamefont {K.}~\bibnamefont
  {Proesmans}}, \bibinfo {author} {\bibfnamefont {B.}~\bibnamefont {Cleuren}},
  \ and\ \bibinfo {author} {\bibfnamefont {C.}~\bibnamefont {Van~den Broeck}},\
  }\bibfield  {title} {\enquote {\bibinfo {title} {Power-efficiency-dissipation
  relations in linear thermodynamics},}\ }\href {\doibase
  10.1103/PhysRevLett.116.220601} {\bibfield  {journal} {\bibinfo  {journal}
  {Phys. Rev. Lett.}\ }\textbf {\bibinfo {volume} {116}},\ \bibinfo {pages}
  {220601} (\bibinfo {year} {2016})}\BibitemShut {NoStop}%
\bibitem [{\citenamefont {Sothmann}\ \emph {et~al.}(2012)\citenamefont
  {Sothmann}, \citenamefont {S\'anchez}, \citenamefont {Jordan},\ and\
  \citenamefont {B\"uttiker}}]{Sothmann-Re}%
  \BibitemOpen
  \bibfield  {author} {\bibinfo {author} {\bibfnamefont {B.}~\bibnamefont
  {Sothmann}}, \bibinfo {author} {\bibfnamefont {R.}~\bibnamefont {S\'anchez}},
  \bibinfo {author} {\bibfnamefont {A.~N.}\ \bibnamefont {Jordan}}, \ and\
  \bibinfo {author} {\bibfnamefont {M.}~\bibnamefont {B\"uttiker}},\ }\bibfield
   {title} {\enquote {\bibinfo {title} {Rectification of thermal fluctuations
  in a chaotic cavity heat engine},}\ }\href {\doibase
  10.1103/PhysRevB.85.205301} {\bibfield  {journal} {\bibinfo  {journal} {Phys.
  Rev. B}\ }\textbf {\bibinfo {volume} {85}},\ \bibinfo {pages} {205301}
  (\bibinfo {year} {2012})}\BibitemShut {NoStop}%
\bibitem [{\citenamefont {Stark}\ \emph {et~al.}(2014)\citenamefont {Stark},
  \citenamefont {Brandner}, \citenamefont {Saito},\ and\ \citenamefont
  {Seifert}}]{UdoPRX}%
  \BibitemOpen
  \bibfield  {author} {\bibinfo {author} {\bibfnamefont {J.}~\bibnamefont
  {Stark}}, \bibinfo {author} {\bibfnamefont {K.}~\bibnamefont {Brandner}},
  \bibinfo {author} {\bibfnamefont {K.}~\bibnamefont {Saito}}, \ and\ \bibinfo
  {author} {\bibfnamefont {U.}~\bibnamefont {Seifert}},\ }\bibfield  {title}
  {\enquote {\bibinfo {title} {Classical nernst engine},}\ }\href {\doibase
  10.1103/PhysRevLett.112.140601} {\bibfield  {journal} {\bibinfo  {journal}
  {Phys. Rev. Lett.}\ }\textbf {\bibinfo {volume} {112}},\ \bibinfo {pages}
  {140601} (\bibinfo {year} {2014})}\BibitemShut {NoStop}%
\bibitem [{\citenamefont {Sothmann}\ \emph {et~al.}(2014)\citenamefont
  {Sothmann}, \citenamefont {S{\'a}nchez},\ and\ \citenamefont
  {Jordan}}]{sothmann}%
  \BibitemOpen
  \bibfield  {author} {\bibinfo {author} {\bibfnamefont {B.}~\bibnamefont
  {Sothmann}}, \bibinfo {author} {\bibfnamefont {R.}~\bibnamefont
  {S{\'a}nchez}}, \ and\ \bibinfo {author} {\bibfnamefont {A.~N}\ \bibnamefont
  {Jordan}},\ }\bibfield  {title} {\enquote {\bibinfo {title} {Quantum nernst
  engines},}\ }\href {http://stacks.iop.org/0295-5075/107/i=4/a=47003}
  {\bibfield  {journal} {\bibinfo  {journal} {Europhys. Lett.}\ }\textbf
  {\bibinfo {volume} {107}},\ \bibinfo {pages} {47003} (\bibinfo {year}
  {2014})}\BibitemShut {NoStop}%
\bibitem [{\citenamefont {B\"uttiker}(1986)}]{Buttiker-4T}%
  \BibitemOpen
  \bibfield  {author} {\bibinfo {author} {\bibfnamefont {M.}~\bibnamefont
  {B\"uttiker}},\ }\bibfield  {title} {\enquote {\bibinfo {title}
  {Four-terminal phase-coherent conductance},}\ }\href {\doibase
  10.1103/PhysRevLett.57.1761} {\bibfield  {journal} {\bibinfo  {journal}
  {Phys. Rev. Lett.}\ }\textbf {\bibinfo {volume} {57}},\ \bibinfo {pages}
  {1761--1764} (\bibinfo {year} {1986})}\BibitemShut {NoStop}%
\bibitem [{\citenamefont {Hofer}\ and\ \citenamefont {Sothmann}(2015)}]{4T}%
  \BibitemOpen
  \bibfield  {author} {\bibinfo {author} {\bibfnamefont {P.~P.}\ \bibnamefont
  {Hofer}}\ and\ \bibinfo {author} {\bibfnamefont {B.}~\bibnamefont
  {Sothmann}},\ }\bibfield  {title} {\enquote {\bibinfo {title} {Quantum heat
  engines based on electronic mach-zehnder interferometers},}\ }\href {\doibase
  10.1103/PhysRevB.91.195406} {\bibfield  {journal} {\bibinfo  {journal} {Phys.
  Rev. B}\ }\textbf {\bibinfo {volume} {91}},\ \bibinfo {pages} {195406}
  (\bibinfo {year} {2015})}\BibitemShut {NoStop}%
\bibitem [{\citenamefont {Brandner}\ and\ \citenamefont
  {Seifert}(2013)}]{Udo-MT}%
  \BibitemOpen
  \bibfield  {author} {\bibinfo {author} {\bibfnamefont {K.}~\bibnamefont
  {Brandner}}\ and\ \bibinfo {author} {\bibfnamefont {U.}~\bibnamefont
  {Seifert}},\ }\bibfield  {title} {\enquote {\bibinfo {title} {Multi-terminal
  thermoelectric transport in a magnetic field: bounds on onsager coefficients
  and efficiency},}\ }\href {http://stacks.iop.org/1367-2630/15/i=10/a=105003}
  {\bibfield  {journal} {\bibinfo  {journal} {New J. Phys.}\ }\textbf {\bibinfo
  {volume} {15}},\ \bibinfo {pages} {105003} (\bibinfo {year}
  {2013})}\BibitemShut {NoStop}%
\bibitem [{\citenamefont {S{\'{a}}nchez}\ \emph {et~al.}(2016)\citenamefont
  {S{\'{a}}nchez}, \citenamefont {Sothmann},\ and\ \citenamefont
  {Jordan}}]{SanchezPhyE}%
  \BibitemOpen
  \bibfield  {author} {\bibinfo {author} {\bibfnamefont {R.}~\bibnamefont
  {S{\'{a}}nchez}}, \bibinfo {author} {\bibfnamefont {B.}~\bibnamefont
  {Sothmann}}, \ and\ \bibinfo {author} {\bibfnamefont {A.~N.}\ \bibnamefont
  {Jordan}},\ }\bibfield  {title} {\enquote {\bibinfo {title} {Effect of
  incoherent scattering on three-terminal quantum hall thermoelectrics},}\
  }\href {\doibase https://doi.org/10.1016/j.physe.2015.09.004} {\bibfield
  {journal} {\bibinfo  {journal} {Physica E: Low-dimensional Systems and
  Nanostructures}\ }\textbf {\bibinfo {volume} {75}},\ \bibinfo {pages} {86 --
  92} (\bibinfo {year} {2016})}\BibitemShut {NoStop}%
\bibitem [{\citenamefont {Brandner}\ \emph {et~al.}(2018)\citenamefont
  {Brandner}, \citenamefont {Hanazato},\ and\ \citenamefont
  {Saito}}]{Brandner4T}%
  \BibitemOpen
  \bibfield  {author} {\bibinfo {author} {\bibfnamefont {K.}~\bibnamefont
  {Brandner}}, \bibinfo {author} {\bibfnamefont {T.}~\bibnamefont {Hanazato}},
  \ and\ \bibinfo {author} {\bibfnamefont {K.}~\bibnamefont {Saito}},\
  }\bibfield  {title} {\enquote {\bibinfo {title} {Thermodynamic bounds on
  precision in ballistic multiterminal transport},}\ }\href {\doibase
  10.1103/PhysRevLett.120.090601} {\bibfield  {journal} {\bibinfo  {journal}
  {Phys. Rev. Lett.}\ }\textbf {\bibinfo {volume} {120}},\ \bibinfo {pages}
  {090601} (\bibinfo {year} {2018})}\BibitemShut {NoStop}%
\end{thebibliography}%

\end{document}